\documentclass[11pt]{article}
\usepackage[T1]{fontenc}
\usepackage{lmodern}
\usepackage[margin=1in]{geometry}
\usepackage{amsmath,amssymb,amsthm,mathtools}
\numberwithin{equation}{section}
\usepackage{bm}
\usepackage{algorithm,algpseudocode}
\usepackage{microtype}
\usepackage{booktabs,tabularx,array}
\newlength{\comparisontablewidth}
\usepackage[numbers,square]{natbib}
\setcitestyle{semicolon}
\usepackage[pagebackref,colorlinks=true,allcolors=blue]{hyperref}
\renewcommand*{\backref}[1]{}
\renewcommand*{\backrefalt}[4]{%
  \ifcase#1\relax
  \or\space(Cited on page~#2.)%
  \else\space(Cited on pages~#2.)%
  \fi}

\newtheorem{theorem}{Theorem}[section]
\newtheorem{lemma}[theorem]{Lemma}
\newtheorem{corollary}[theorem]{Corollary}
\newtheorem{proposition}[theorem]{Proposition}

\theoremstyle{remark}

\newcommand{\Tr}{\operatorname{Tr}}
\newcommand{\id}{\mathrm{id}}
\newcommand{\ad}{\operatorname{ad}}
\newcommand{\poly}{\operatorname{poly}}

\newcommand{\norm}[1]{\left\lVert#1\right\rVert}
\newcommand{\cL}{\mathcal L}
\newcommand{\tO}{\widetilde{O}}
\DeclareMathOperator{\EP}{EP}
\DeclareMathOperator{\supp}{supp}
\title{Near-Optimal Learning of Quantum Many-Body Systems from Gibbs and Thermal Metastable States via the Stationarity Test}
\author{Tai-Hsuan Yang\thanks{\href{mailto:ty39@illinois.edu}{ty39@illinois.edu}}\\[0.5em]
\small Department of Physics, Illinois Quantum Information Science and Technology Center,\\
\small and Institute of Condensed Matter Theory, The Grainger College of Engineering,\\
\small University of Illinois at Urbana-Champaign, Urbana, Illinois 61801, USA}
\date{}

\begin{document}
\maketitle

\begin{abstract}
We study the problem of learning a finite-range quantum Hamiltonian on an $n$-qubit $D$-dimensional lattice from copies of its Gibbs states at a known inverse temperature $\beta$ or its thermal metastable states. Existing efficient algorithms at arbitrary temperature have sample guarantees with an $e^{\poly(\beta)}$ dependence, leaving open whether the exponent can be made linear in $\beta$. 
We achieve this improvement: for $\beta\ge1$, our algorithm learns all Hamiltonian coefficients to $\ell_\infty$-error $\epsilon$ with probability at least $1-\delta$ using $\tO\bigl(e^{O(\beta)}\epsilon^{-2}\log(n/\delta)\bigr)$ copies and total time $\tO\bigl(n e^{O(\beta)}\epsilon^{-2}\log(n/\delta)\bigr)$. 
The measurements act on individual copies and are supported on balls of radius $O\bigl(\beta^2+\beta\log(1/\epsilon)\bigr)$. Our sample complexity achieves the optimal exponential order in $\beta$, with optimal $\epsilon^{-2}$ dependence up to logarithmic factors.

Building on the stationarity-test framework \citep{Bergamaschi2026Stationarity}, we prove a unified Fisher-information coercivity bound showing that errors in the candidate Hamiltonian produce a detectable signal with only an $e^{-O(\beta)}$ loss in sensitivity. The key is a smooth frequency filter whose bandwidth remains independent of temperature, together with sharper estimates relating this signal to the Hamiltonian coefficients. For thermal metastable states, the guarantee holds under local drift $\zeta\le e^{-C\beta^{2D+1}}$, for a geometry-dependent constant $C>0$, and above an accuracy floor $\tO(e^{O(\beta)}\zeta)$. The argument applies to both Gibbs and thermal metastable states and, combined with local measurements and projected stochastic gradient descent (SGD), yields both learning guarantees.
\end{abstract}

\noindent\textbf{AI usage statement.} The proof was generated with OpenAI GPT 6 Astra Ultra. Gemini 4 Argon was used to assist with writing the manuscript. OpenAI Codex was used for manuscript revision, literature searches, mathematical exposition, proof discussions, and \LaTeX{} preparation. The author retains responsibility for the accuracy of the arguments, citations, and conclusions.

\clearpage
\tableofcontents
\clearpage

\section{Introduction}
\label{sec:intro}

Understanding the structure and dynamics of quantum many-body systems in thermal equilibrium is a central theme across condensed matter physics, statistical mechanics, and quantum information science \citep{Alhambra2023,KlieschEtAl2014,KuwaharaAlhambraAnshu2021AreaLaw}. At inverse temperature $\beta = 1/T$, a system with Hamiltonian $\bm{H}$ in thermal equilibrium is described by the Gibbs state
\begin{equation}\label{eq:gibbs-def}
 \bm{\rho}_{\bm{H}} = \frac{e^{-\beta \bm{H}}}{\Tr(e^{-\beta \bm{H}})}.
\end{equation}
Given copies of $\bm{\rho}_{\bm{H}}$—or more generally, of a locally thermalized \emph{metastable state} $\bm{\sigma}$ \citep{bergamaschi2025structuraltheoryquantummetastability}—the fundamental inverse problem of \emph{Hamiltonian learning} asks how to reconstruct the microscopic interaction coefficients and structure of $\bm{H}$ with optimal scaling in the system size $n$, target accuracy $\epsilon$, and inverse temperature $\beta$ \citep{AnshuEtAl2020Learning,HaahKothariTang2024Learning,BakshiEtAl2023Learning,Narayanan2024Learning,ChenAnshuNguyen2025Learning,DongLeng2026ScoreMatching,Bergamaschi2026Stationarity,WangYeChen2026MetastableLearning}.

In classical statistical mechanics, the Hammersley--Clifford theorem \citep{Hammersley1971MarkovFO} guarantees that a thermal Gibbs distribution is uniquely characterized by its local conditional marginals, enabling local learning of the underlying interactions. In the quantum setting, non-commutativity of the local terms $\bm{P}_j$ obstructs a naive conditional marginal factorization, motivating an extensive body of work on quantum belief propagation, approximate Markov networks, and the decay of correlations and conditional mutual information \citep{Araki1969,Hastings2007QBP,Kim2012PerturbativeTEE,LeiferPoulin2008,BrownPoulin2012,KlieschEtAl2014,KatoBrandao2019,BluhmCapelPerezHernandez2022,Kuwahara2025,Chen2025,Yang2026LocalMarkov}.

A powerful dynamical learning paradigm, recently formalized by Bergamaschi \citep{Bergamaschi2026Stationarity} as the \emph{stationarity test}, connects Hamiltonian learning directly to the rapidly developing theory of quantum Gibbs sampling and open-system thermalization 
\citep{KastoryanoTemme2013Mixing,ChenKastoryanoBrandaoGilyen2023,ChenKastoryanoGilyen2023,GilyenChenDoriguelloKastoryano2024,DingLiLin2025,ChenEtAl2025ThermalSimulation,RouzeFrancaAlhambra2026Thermalization}. The underlying principle is simple: $\bm{\rho}_{\bm{H}}$ is a stationary fixed point of any detailed-balanced quantum Markov chain (Lindbladian) $\cL_{\bm{H}}$ associated with $\bm{H}$. If a learner proposes a candidate Hamiltonian $\bm{K}$ and measures the local drift $\Tr[\bm{\sigma} \cL_{\bm{K}}^*[\bm{O}]]$ of local observables $\bm{O}$ on the input state $\bm{\sigma}$, stationarity yields both completeness and soundness:
\begin{itemize}
 \item \emph{Completeness:} When $\bm{K} = \bm{H}$ and $\bm{\sigma} = \bm{\rho}_{\bm{H}}$, the stationarity residual $\Tr[\bm{\rho}_{\bm{H}} \cL_{\bm{H}}^*[\bm{O}]]$ vanishes for all observables $\bm{O}$.
 \item \emph{Soundness (Coercivity):} When $\bm{K} \neq \bm{H}$, the stationarity residuals of local observables are bounded away from zero in proportion to the parameter error $\|\bm{K} - \bm{H}\|$, governed by the positivity of quantum entropy production \citep{Spohn,CM,Yang} and the Fisher information of the Gibbs sampler \citep{Bergamaschi2026Stationarity,bergamaschi2025structuraltheoryquantummetastability,WangYeChen2026MetastableLearning}.
\end{itemize}
Because the stationarity test only probes the quasi-local generator $\cL_{\bm{K}}$ at short times, it requires no assumptions on the global spectral gap, mixing time, or correlation decay of $\bm{\rho}_{\bm{H}}$, and extends naturally to \emph{thermal metastable states} satisfying $\max_i \|\cL_{i,\bm{H}}[\bm{\sigma}]\|_1 \le \zeta$ \citep{bergamaschi2025structuraltheoryquantummetastability,Bergamaschi2026Stationarity,WangYeChen2026MetastableLearning}—states that are approximately stationary under local thermal dynamics, even when far from global equilibrium. Examples include metastable states arising in slow-mixing low-temperature phases and self-correcting quantum memories \citep{Dennis_2002,AlickiEtAl2008ThermalStability,CastelnovoChamon2008ThermalTopology}.

Existing local and stationarity-based learning guarantees have sample complexity scaling as $e^{\poly(\beta)}$ in the inverse temperature $\beta$ \citep{ChenAnshuNguyen2025Learning,Bergamaschi2026Stationarity,WangYeChen2026MetastableLearning}. This raises the question:
\begin{center}
 \emph{Can we learn lattice Hamiltonians from Gibbs and thermal metastable states with sample complexity scaling as $e^{O(\beta)}$—linearly in $\beta$ in the exponent?}
\end{center}

\subsection{Main results}
\label{sec:main-results}

We resolve this question in the affirmative. Consider an $n$-qubit lattice $\Lambda \subset \mathbb{Z}^D$ in fixed spatial dimension $D$ and a finite-range Hamiltonian
\begin{equation}\label{eq:H-expansion}
 \bm{H}(\mathbf{h}) = \sum_{j=1}^m h_j \bm{P}_j, \qquad \mathbf{h} \in [-1,1]^m,
\end{equation}
where $\{\bm{P}_j\}_{j=1}^m$ is a known dictionary of distinct, non-identity Hermitian Pauli operators of weight at most $w$ and support diameter at most $R$, with at most $q$ terms acting on any site ($m \le qn$). Throughout, $(D, R, w, q)$ are fixed constants, and $\tO(\cdot)$ hides polynomial factors in $\beta$, $\log(1/\epsilon)$, and $\log\log(n/\delta)$.

\begin{theorem}[Learning from Gibbs states with linear-in-$\beta$ exponent]
\label{thm:main}
For any fixed geometry $(D,R,w,q)$, $\beta \ge 1$, and $0 < \epsilon, \delta < 1/2$, there is a single-copy learning algorithm that outputs $\widehat{\mathbf{h}} \in [-1,1]^m$ satisfying $\Pr\{\|\widehat{\mathbf{h}} - \mathbf{h}\|_\infty \le \epsilon\} \ge 1 - \delta$ using
\begin{equation}\label{thm:copies}
 N = \tO\!\left(e^{O(\beta)} \epsilon^{-2} \log(n/\delta)\right)
\end{equation}
copies of $\bm{\rho}_{\bm{H}}$ and total quantum gate and classical runtime
\begin{equation}\label{thm:actual-work}
 W = \tO\!\left(n e^{O(\beta)} \epsilon^{-2} \log(n/\delta)\right).
\end{equation}
Each single-copy measurement acts coherently within a spatial ball of radius
\begin{equation}\label{thm:radius}
 r_{\rm meas} = O\bigl(\beta^2 + \beta\log(1/\epsilon)\bigr).
\end{equation}
\end{theorem}

For high temperatures ($0 < \beta < 1$), rescaling $\bm{H}' = \beta \bm{H}$ and $\epsilon' = \beta\epsilon$ yields $N = \tO((\beta\epsilon)^{-2}\log(n/\delta))$.

Our second main result extends this $e^{O(\beta)}$ sample complexity to learning from copies of a \emph{thermal metastable state} $\bm{\sigma}$. Following \citep{bergamaschi2025structuraltheoryquantummetastability,Bergamaschi2026Stationarity}, $\bm{\sigma}$ is $\zeta$-locally metastable at inverse temperature $\beta$ if the site-summed completed Chen--Kastoryano--Gily\'en (CKG) Lindbladian $\cL_{i,\bm{H}}$ at every site $i \in \Lambda$ satisfies
\begin{equation}\label{meta:stationarity}
 \max_{i \in \Lambda} \|\cL_{i,\bm{H}}[\bm{\sigma}]\|_1 \le \zeta.
\end{equation}

\begin{theorem}[Learning from thermal metastable states with linear-in-$\beta$ exponent]
\label{meta:theorem}
For any fixed geometry $(D,R,w,q)$, there is a constant $C>0$ such that for all $\beta \ge 1$ and $0 < \epsilon, \delta < 1/2$, the following holds. Given independent copies of a state $\bm{\sigma}$ satisfying \eqref{meta:stationarity} with
\begin{equation}\label{meta:admission}
 0 \le \zeta \le \exp[-C\beta^{2D+1}]
\end{equation}
and target accuracy $\epsilon$ above the metastability error floor
\begin{equation}\label{meta:floor-explicit}
 \epsilon \ge \tO\!\left(e^{O(\beta)} \zeta\right),
\end{equation}
the single-copy local stochastic gradient descent algorithm outputs $\widehat{\mathbf{h}} \in [-1,1]^m$ satisfying $\Pr\{\|\widehat{\mathbf{h}} - \mathbf{h}\|_\infty \le \epsilon\} \ge 1 - \delta$ using
\begin{equation}\label{meta:copies}
 N = \tO\!\left(e^{O(\beta)} \epsilon^{-2} \log(n/\delta)\right), \qquad W = \tO\!\left(n e^{O(\beta)} \epsilon^{-2} \log(n/\delta)\right),
\end{equation}
with single-copy measurements localized to balls of radius \eqref{thm:radius}.
\end{theorem}

In Theorem~\ref{meta:theorem}, the required upper bound $\zeta \le e^{-C\beta^{2D+1}}$ is a state promise on $\bm{\sigma}$ that does not enter the sample complexity: once $\bm{\sigma}$ lies in the metastable regime, the sample complexity scales with a \emph{linear} exponent $e^{O(\beta)}$, and the error floor \eqref{meta:floor-explicit} scales nearly linearly in $\zeta$ with an $e^{O(\beta)}$ prefactor. As shown in Section~\ref{sec:metastable-proof}, an accuracy floor of order $\Omega(\zeta e^{c\beta}/\beta)$ is information-theoretically unavoidable.

Haah, Kothari, and Tang \citep[Theorem~5.3]{HaahKothariTang2024Learning} show that learning bounded-strength 2-local Hamiltonians to coefficientwise error $\epsilon$, with success probability at least $1-\delta$, requires $\Omega\bigl(e^{2\beta}\beta^{-2}\epsilon^{-2}\log(n/\delta)\bigr)$ copies in the worst case. Theorems~\ref{thm:main} and \ref{meta:theorem} achieve the optimal exponential order in $\beta$, with optimal $\epsilon^{-2}$ dependence up to logarithmic factors.

\paragraph{Sketch of the learning protocol.}
The same algorithm applies to both Gibbs and thermal metastable inputs $\bm{\tau} \in \{\bm{\rho}_{\bm{H}}, \bm{\sigma}\}$. Because $\bm{\tau}$ is stationary (or nearly stationary) under the CKG quantum Gibbs sampler at the true parameter $\mathbf{h}$, the short-time drift of each Pauli expectation $\Tr(\bm{P}_j\bm{\tau})$ under the candidate sampler provides a signed restoring signal toward $\mathbf{h}$. Starting from $\mathbf{x}_0 = \mathbf{0}$, at each iteration $t$ we perform two steps:
\begin{enumerate}
 \item \emph{Local short-time stationarity estimation:} For each dictionary term $\bm{P}_j$, let $\mathcal{G}_{j,\mathbf{x}_t}$ denote the candidate CKG generator restricted to a ball of radius $r_{\mathrm{meas}}$ around $\supp(\bm{P}_j)$ as in \eqref{thm:radius} (defined precisely in \eqref{oracle:finite-time-map}). For a short evolution time $t_{\mathrm{ev}} > 0$, we use fresh copies of $\bm{\tau}$ to estimate the finite-difference drift
 \begin{equation}\label{intro:protocol-residual}
  F_j(\mathbf{x}_t)
  = -\frac{\Tr\!\left(\bm{P}_j e^{t_{\mathrm{ev}}\mathcal{G}_{j,\mathbf{x}_t}}[\bm{\tau}]\right)
  - \Tr(\bm{P}_j\bm{\tau})}{t_{\mathrm{ev}}}
 \end{equation}
 by measuring $\bm{P}_j$ with and without applying the local channel $e^{t_{\mathrm{ev}}\mathcal{G}_{j,\mathbf{x}_t}}$, obtaining an empirical estimate $\widehat{F}_{t,j}$. Because $\mathcal{G}_{j,\mathbf{x}_t}$ acts strictly within radius $r_{\mathrm{meas}}$ of $\supp(\bm{P}_j)$, disjoint patches can be evolved and measured simultaneously on a single copy of $\bm{\tau}$.
 \item \emph{Projected stochastic gradient update:} Using step size $\alpha > 0$, we update all coordinates in parallel and clip them to $[-1, 1]$:
 \begin{equation}\label{intro:protocol-update}
  x_{t+1,j} = \min\!\left\{1, \max\!\left\{-1, x_{t,j} - \alpha\widehat{F}_{t,j}\right\}\right\},
  \qquad j = 1, \ldots, m.
 \end{equation}
\end{enumerate}
Repeating this trajectory $O(\log(n/\delta))$ times independently and taking the coordinatewise median boosts the success probability to $1-\delta$. Whereas Bergamaschi's parameter-learning protocol \citep[Algorithms~1--2]{Bergamaschi2026Stationarity} enumerates local candidate corrections via a discrete search, our projected stochastic gradient method updates the parameters directly along the estimated stationarity residuals; see Section~\ref{sec:gibbs-learning} for full details.

\subsection{Related works}
\label{sec:related-work}

\paragraph{Hamiltonian learning from thermal and metastable states.}
The problem of learning local Hamiltonians from thermal states has seen rapid progress over the past several years \citep{AnshuEtAl2020Learning,HaahKothariTang2024Learning,BakshiEtAl2023Learning,Narayanan2024Learning,ChenAnshuNguyen2025Learning,DongLeng2026ScoreMatching,Bergamaschi2026Stationarity,WangYeChen2026MetastableLearning,LewisTangWright2026Learning,ChenEtAl2026LearningHardness}.
Anshu, Arunachalam, Kuwahara, and Soleimanifar \citep{AnshuEtAl2020Learning} established the first polynomial sample complexity bounds for general local Hamiltonians via quantum belief propagation \citep{Hastings2007QBP,Kim2012PerturbativeTEE} and gradient optimization of the log-partition function. 
At high temperatures, Haah, Kothari, and Tang \citep{HaahKothariTang2024Learning} achieved optimal sample and time complexity via cluster expansions \citep{KP}. 
For general constant temperatures, Bakshi, Liu, Moitra, and Tang \citep{BakshiEtAl2023Learning} and Narayanan \citep{Narayanan2024Learning} developed polynomial-time algorithms via flat polynomial approximations and semidefinite programming, and Chen, Anshu, and Nguyen \citep{ChenAnshuNguyen2025Learning} subsequently achieved $\log(n)$ sample complexity and linear runtime in $n$ on $D$-dimensional lattices via local KMS conditions. 
More recently, Wang, Ye, and Chen \citep{WangYeChen2026MetastableLearning} developed a framework that extends Gibbs-state learning to metastable states of detailed-balanced Lindbladians. Bergamaschi \citep{Bergamaschi2026Stationarity} introduced the broader stationarity-test framework, with applications to structure learning from Gibbs states and parameter learning from thermal metastable states. Related developments include quantum score matching for thermal-state learning \citep{DongLeng2026ScoreMatching} and structure learning for open quantum systems \citep{LewisTangWright2026Learning}.
Table~\ref{tab:comparison} compares our guarantees with \citep{ChenAnshuNguyen2025Learning,Bergamaschi2026Stationarity,WangYeChen2026MetastableLearning}, detailing the sample complexity, measurement locality radius, allowance of the metastability drift $\zeta$, and runtime.

\paragraph{Quantum Gibbs sampling, rapid mixing, and metastability.}
Our work also builds on the theory of quantum Markov semigroups \citep{Lindblad76,Spohn,CM,KastoryanoTemme2013Mixing} and recent breakthroughs in constructing exact KMS-detailed-balanced quantum Gibbs samplers
\citep{ChenKastoryanoBrandaoGilyen2023,ChenKastoryanoGilyen2023,GilyenChenDoriguelloKastoryano2024,DingLiLin2025,ChenEtAl2025ThermalSimulation}. Mixing-time bounds have been established for commuting Hamiltonians \citep{BardetEtAl2023Thermalization,KochanowskiEtAl2025Thermalization} and for noncommuting lattice systems at high temperatures \citep{RouzeFrancaAlhambra2026Thermalization,BakshiLiuMoitraTang2025,ChenDingLu2026SizeIndependentLSI}. Further mixing-time guarantees cover one-dimensional spin chains at every finite temperature \citep{BergamaschiChen2025SpinChains,Kim2026RapidMixingChains,RouzeFranca2026DobrushinShlosman,ChenDing2026SpectralMixing,CapelPerezGarciaScandi2026MLSI,GaoGuoWang2026CMLSI,ChenDingLu2026SizeIndependentLSI}, and bounded-strength all-to-all systems at high temperatures \citep{Bergamaschi2026AllToAll,CaiZhanZlokapa2026SYK}.
Even in low-temperature regimes where global mixing is slow, the corresponding quantum Gibbs sampler prepares locally stationary \emph{metastable states} that satisfy approximate detailed balance, mutual information area law, and quantum Markov property \citep{bergamaschi2025structuraltheoryquantummetastability,chen2026notestrongquantummarkov,Bergamaschi2026Stationarity}.

\begin{table}[t]
\centering
\footnotesize
\setlength{\tabcolsep}{3pt}
\renewcommand{\arraystretch}{1.35}
\setlength{\comparisontablewidth}{1.12\textwidth}
\makebox[\textwidth][c]{%
\begin{tabular}{
 >{\raggedright\arraybackslash}p{\dimexpr0.18\comparisontablewidth-2\tabcolsep\relax} 
 >{\raggedright\arraybackslash}p{\dimexpr0.20\comparisontablewidth-2\tabcolsep\relax} 
 >{\raggedright\arraybackslash}p{\dimexpr0.22\comparisontablewidth-2\tabcolsep\relax} 
 >{\raggedright\arraybackslash}p{\dimexpr0.22\comparisontablewidth-2\tabcolsep\relax} 
 >{\raggedright\arraybackslash}p{\dimexpr0.18\comparisontablewidth-2\tabcolsep\relax}} 
\toprule
 & \textbf{Chen--Anshu--Nguyen} \citep{ChenAnshuNguyen2025Learning}
 & \textbf{Bergamaschi} \citep{Bergamaschi2026Stationarity}
 & \textbf{Wang--Ye--Chen} \citep{WangYeChen2026MetastableLearning}
 & \textbf{This work} \\
\midrule
\textbf{Gibbs sample complexity} ($\beta \ge 1$) &
$\tO\bigl(e^{\poly(\beta)}\epsilon^{-2}\log\frac{n}{\delta}\bigr)$ &
$\tO\bigl(e^{\poly(\beta)}\epsilon^{-2}\log\frac{n}{\delta}\bigr)$ &
$\tO\bigl(e^{\poly(\beta)}\epsilon^{-2}\log\frac{n}{\delta}\bigr)$ &
$\tO\bigl(e^{O(\beta)}\epsilon^{-2}\log\frac{n}{\delta}\bigr)$ \\
\textbf{Metastable sample complexity} ($\beta \ge 1$) &
— (Gibbs only) &
$\tO\bigl(e^{\poly(\beta)}\epsilon^{-2}\log\frac{n}{\delta}\bigr)$ &
$\tO\bigl(e^{\poly(\beta)}\epsilon^{-2}\log\frac{n}{\delta}\bigr)$ &
$\tO\bigl(e^{O(\beta)}\epsilon^{-2}\log\frac{n}{\delta}\bigr)$ \\
\textbf{Measurement radius} $r_{\rm meas}$ &
$O\bigl(\poly(\beta)\log\frac{1}{\epsilon}\bigr)$ &
$O\bigl(\poly(\beta)[1+\log\frac{1}{\epsilon}]\bigr)$  &
$O\bigl(\poly(\beta)\log\frac{1}{\epsilon}\bigr)$ &
$O\bigl(\beta^2 + \beta\log\frac{1}{\epsilon}\bigr)$ \\
\textbf{Allowance of $\zeta$} (for error $\epsilon$) &
$\zeta = 0$ &
$\zeta \le e^{-\poly(\beta)}\epsilon$ &
$\zeta \le \bigl(e^{-\poly(\beta)}\epsilon\bigr)^{\poly(\beta)}$ &
$\zeta \le \tO\bigl(e^{-O(\beta)}\epsilon\bigr)$, \newline $\zeta \le e^{-C\beta^{2D+1}}$ \\
\textbf{Metastability error floor} &
— &
$e^{\poly(\beta)} \zeta$ &
$e^{\poly(\beta)} \zeta^{1/\poly(\beta)}$ &
$\tO\bigl(e^{O(\beta)} \zeta\bigr)$ \\
\textbf{Total runtime} ($\beta \ge 1$) &
$\tO\bigl(n e^{\poly(\beta)}\epsilon^{-2}\log\frac{n}{\delta}\bigr)$ &
$\tO\bigl(n e^{\poly(\beta)}\epsilon^{-2}\log\frac{n}{\delta}\bigr)$ &
$\tO\bigl(n e^{\poly(\beta)}\epsilon^{-2}\log\frac{n}{\delta}\bigr)$ &
$\tO\bigl(n e^{O(\beta)}\epsilon^{-2}\log\frac{n}{\delta}\bigr)$ \\
\bottomrule
\end{tabular}%
}
\caption{Comparison of local and stationarity-based Hamiltonian learning guarantees for $n$-qubit finite-range lattice Hamiltonians on $\mathbb{Z}^D$ at inverse temperature $\beta \ge 1$, where $\epsilon$ is the $\ell_\infty$ coefficient error, $1-\delta$ is the success probability, $r_{\rm meas}$ is the spatial radius of each single-copy measurement, $\zeta$ is the local metastability drift \eqref{meta:stationarity}.}
\label{tab:comparison}
\end{table}

\subsection{Open questions}
\label{sec:open-questions}

Our results suggest several natural directions for future investigation:
\begin{enumerate}
 \item \emph{Relaxing the required upper bound on $\zeta$ to $e^{-O(\beta)}$.}
 In Theorem~\ref{meta:theorem}, once the supplied state $\bm{\sigma}$ satisfies the required upper bound $\zeta \le e^{-C\beta^{2D+1}}$, both the sample complexity $N = \tO(e^{O(\beta)}\epsilon^{-2}\log(n/\delta))$ and the achievable error floor $\epsilon \ge \tO(e^{O(\beta)}\zeta)$ scale with a linear exponent in $\beta$. The exponent $\beta^{2D+1}$ in this bound arises from reconstructing logarithmic commutators on a radius-$O(\beta^2)$ ball: its volume is $O(\beta^{2D})$, and the frequency reconstruction costs $e^{O(\beta)}$ per unit volume (Appendix~\ref{app:meta-comparison}). Can the required upper bound on $\zeta$ be relaxed to $\zeta \le e^{-O(\beta)}$, matching the single-qubit indistinguishability barrier $\zeta \lesssim \beta e^{-\beta}\epsilon$?

 \item \emph{Measurement locality.}
 Our algorithm uses measurements supported on balls of radius $r_{\rm meas}=O(\beta^2+\beta\log(1/\epsilon))$. Can this radius be reduced while retaining $\tO(e^{O(\beta)}\epsilon^{-2}\log(n/\delta))$ sample complexity, or is there an information-theoretic separation between Hamiltonian learning with different measurement radii?

 \item \emph{Sharp constants in the $e^{O(\beta)}$ exponent.}
 While the lower bound of \citep{HaahKothariTang2024Learning} shows that an $e^{\Omega(\beta)}$ sample complexity is unavoidable in the worst case due to thermal population suppression in commuting models, our upper bound $e^{C\beta}$ involves a constant $C$ governed by the local interaction degree and the frequency bandwidth ${\omega_b} = \Theta(1)$, chosen independently of temperature and system size, in the filtered Pauli twirl and modular conjugation (Lemma~\ref{lin:faithfulness}). 
Optimizing this exponent is highly relevant in learning low-temperature quantum systems.

 \item \emph{Structure learning from Gibbs and thermal metastable states.}
 Our guarantees use a supplied lattice geometry and a known family of candidate local interactions. Can one also learn the unknown interaction graph from Gibbs or thermal metastable samples while retaining an $e^{O(\beta)}$ sample complexity? 
\end{enumerate}

\section{Preliminaries and the Stationarity Test}
\label{sec:prelim}

\subsection{Lattice Hamiltonians, norms, and Lieb--Robinson locality}
\label{sec:setup-prelim}

Let $\Lambda \subset \mathbb{Z}^D$ be a finite lattice of $n$ qubit sites equipped with the standard lattice distance $d(\cdot,\cdot)$ in fixed spatial dimension $D \ge 1$. We consider Hamiltonians parameterized by a coefficient vector $\mathbf{h} = (h_1, \dots, h_m) \in [-1,1]^m$:
\begin{equation}\label{eq:display-2}
 \bm{H}(\mathbf{h}) = \sum_{j=1}^m h_j \bm{P}_j, \qquad \bm{\rho}_{\bm{H}} = \frac{e^{-\beta \bm{H}}}{\Tr(e^{-\beta \bm{H}})},
\end{equation}
where the interaction dictionary $\{\bm{P}_j\}_{j=1}^m$ consists of known, distinct, non-identity Hermitian Pauli operators on $\Lambda$. Each $\bm{P}_j$ has operator norm $\|\bm{P}_j\|_\infty = 1$, acts non-trivially on at most $w$ sites, and has support $\supp(\bm{P}_j)$ of diameter at most $R$. Each site $i \in \Lambda$ participates in at most $q$ dictionary terms, so that $m \le qn$. Throughout the paper, the geometric parameters $(D, R, w, q)$ are fixed constants, and $C, c > 0$ denote generic constants depending only on $(D, R, w, q)$.

\paragraph{Operator and state-weighted norms.}
For an operator $\bm{X}$ on $(\mathbb{C}^2)^{\otimes n}$, we write $\|\bm{X}\|_\infty$ (or $\|\bm{X}\|$) for the spectral operator norm, $\|\bm{X}\|_1 = \Tr\sqrt{\bm{X}^*\bm{X}}$ for the trace norm, and $\|\bm{X}\|_2 = (\Tr \bm{X}^*\bm{X})^{1/2}$ for the Hilbert--Schmidt norm \citep{Watrous}. For a state $\bm{\sigma} \ge 0$ and $v \in (0,1)$, the weighted Skew/KMS norms are defined by
\begin{equation}\label{eq:kms-norm}
 \|\bm{X}\|_{\bm{\sigma},v}^2 = \Tr\bigl(\bm{X}^* \bm{\sigma}^v \bm{X} \bm{\sigma}^{1-v}\bigr) = \|\bm{\sigma}^{v/2} \bm{X} \bm{\sigma}^{(1-v)/2}\|_2^2, \qquad \|\bm{X}\|_{\bm{\sigma},\mathrm{KMS}} := \|\bm{X}\|_{\bm{\sigma},1/2} = \|\bm{\sigma}^{1/4} \bm{X} \bm{\sigma}^{1/4}\|_2.
\end{equation}

\paragraph{Heisenberg dynamics and Lieb--Robinson bounds.}
For any Hamiltonian $\bm{K} = \bm{H}(\mathbf{k})$ with $\mathbf{k} \in [-1,1]^m$, let $\mathcal{U}_t^{\bm{K}}[\bm{X}] = e^{it\bm{K}} \bm{X} e^{-it\bm{K}}$ denote the Heisenberg time evolution. We use the same formula for complex times $t\in\mathbb C$; in particular, $\mathcal U_{-is}^{\bm K}[\bm X]=e^{s\bm K}\bm X e^{-s\bm K}$ for real $s$. 
By the Lieb--Robinson bound for finite-range lattice interactions \citep{LiebRobinson1972,NS,hastings2010localityquantumsystems,NachtergaeleSimsYoung2019,ChenLucasYin2023},
for any operator $\bm{A}$ supported on a local region $S$ and any ball $B$ of radius $r$ around $S$, the truncated Hamiltonian $\bm{K}_B$ (retaining only terms supported entirely within $B$) satisfies
\begin{equation}\label{eq:capped-lr}
 \norm{\mathcal{U}_t^{\bm{K}}[\bm{A}] - \mathcal{U}_t^{\bm{K}_B}[\bm{A}]}_\infty \le C |S| \norm{\bm{A}}_\infty \min\bigl\{1, e^{v_{\rm LR}|t| - c r}\bigr\},
\end{equation}
where $v_{\rm LR}, c, C > 0$ depend only on $(D, R, w, q)$.

\subsection{Quantum Gibbs samplers and the stationarity residual}
\label{sec:generator}

To test whether a candidate Hamiltonian $\bm{K} = \bm{H}(\mathbf{k})$ matches the unknown Hamiltonian $\bm{H} = \bm{H}(\mathbf{h})$, we employ the exact detailed-balanced quantum Gibbs sampler of Chen, Kastoryano, and Gily\'en \citep{ChenKastoryanoGilyen2023} (CKG). Let $\{\bm{A}_a\}$ range over all $3n$ single-site Pauli operators $\{\bm{X}_i, \bm{Y}_i, \bm{Z}_i\}_{i \in \Lambda}$. These are the system--bath coupling operators; the index $a$ specifies a lattice site and a Pauli choice. For inverse temperature $\beta > 0$ and any operator $\bm{A}$, define the Gaussian time filter and jump operators by
\begin{equation}\label{gen:jump}
 \bm{J}_{\bm{A},\bm{K}}(\omega) = \frac{1}{\sqrt{2\pi}}\int_{\mathbb{R}} f_\beta(u)e^{-i\omega u}e^{iu\bm{K}}\bm{A} e^{-iu\bm{K}}\,du, \qquad f_\beta(u) = (2/\pi)^{1/4}\beta^{-1/2}e^{-u^2/\beta^2}.
\end{equation}
For the single-site Pauli operators $\bm{A}_a$, set
\begin{equation}\label{gen:Q}
 \gamma(\omega) = e^{-\beta\max\{\omega+1/(2\beta),0\}}, \qquad
 \bm{Q}_{a,\bm{K}} = \int_{\mathbb{R}}\gamma(\omega) \bm{J}_{\bm{A}_a,\bm{K}}(\omega)^* \bm{J}_{\bm{A}_a,\bm{K}}(\omega)\,d\omega.
\end{equation}
The coherent compensation term $\bm{B}_{a,\bm{K}}$ and the completed Lindbladian generator $\cL_{\bm{K}}$ (in the Schr\"odinger picture) are given by
\begin{equation}\label{gen:generator}
\begin{aligned} 
 \bm{B}_{a,\bm{K}} &= \int_0^\infty \frac{\mathcal{U}_t^{\bm{K}}[\bm{Q}_{a,\bm{K}}] - \mathcal{U}_{-t}^{\bm{K}}[\bm{Q}_{a,\bm{K}}]}{\beta\sinh(2\pi t/\beta)}\,dt,\\
 \cL_{\bm A_a,\bm{K}}[\bm{X}] &= -i[\bm{B}_{a,\bm{K}}, \bm{X}] + \int_{\mathbb{R}}\gamma(\omega)\left(\bm{J}_{\bm{A}_a,\bm{K}}(\omega) \bm{X} \bm{J}_{\bm{A}_a,\bm{K}}(\omega)^* - \tfrac{1}{2}\bigl\{\bm{J}_{\bm{A}_a,\bm{K}}(\omega)^* \bm{J}_{\bm{A}_a,\bm{K}}(\omega), \bm{X}\bigr\}\right)d\omega,\\
 \cL_{\bm{K}} &= \sum_a \cL_{\bm A_a,\bm{K}}.
\end{aligned}
\end{equation}
Here $\cL_{\bm A_a,\bm K}$ is the generator associated with the single-site Pauli coupling operator $\bm A_a$. For a site $i$, define the sum over its three Pauli coupling operators by
\begin{equation}\label{gen:site-sum}
 \cL_{i,\bm H}:=\sum_{\bm A\in\{\bm X_i,\bm Y_i,\bm Z_i\}}\cL_{\bm A,\bm H},
 \qquad \cL_{\bm H}=\sum_{i\in\Lambda}\cL_{i,\bm H}.
\end{equation}
 The Heisenberg-picture adjoint $\cL_{\bm{K}}^*$ acts on observables $\bm{O}$ via
\begin{equation}\label{eq:display-3}
 \cL_{\bm A_a,\bm{K}}^*[\bm{O}] = i[\bm{B}_{a,\bm{K}}, \bm{O}] + \int_{\mathbb{R}}\gamma(\omega)\left(\bm{J}_{\bm{A}_a,\bm{K}}(\omega)^* \bm{O} \bm{J}_{\bm{A}_a,\bm{K}}(\omega) - \tfrac{1}{2}\bigl\{\bm{J}_{\bm{A}_a,\bm{K}}(\omega)^* \bm{J}_{\bm{A}_a,\bm{K}}(\omega), \bm{O}\bigr\}\right)d\omega.
\end{equation}

\begin{lemma}[Stationarity and normalization of the CKG generator \citep{ChenKastoryanoGilyen2023}]
\label{gen:stationary}
For every single-site Pauli operator $\bm{A}_a$ and any candidate Hamiltonian $\bm{K}$, $0 \le \bm{Q}_{a,\bm{K}} \le \bm{I}$, the integral in \eqref{gen:generator} converges absolutely in operator norm, and the Gibbs state $\bm{\rho}_{\bm{K}} = e^{-\beta \bm{K}}/\Tr(e^{-\beta \bm{K}})$ is an exact stationary state of each generator: $\cL_{\bm A_a,\bm{K}}[\bm{\rho}_{\bm{K}}] = 0$.
\end{lemma}

\begin{lemma}[Imaginary-time conjugation of Gaussian-filtered operators {\citep[Lemma IX.2]{Chen2025}}]
\label{lem:gaussian-conjugation}
Our filter \eqref{gen:jump} corresponds to the uncertainty parameter $\sigma=\beta^{-1}$ in \citep[Lemma IX.2]{Chen2025}. Taking the imaginary-time parameter in that lemma to be $\pm\beta/2$, for any Hermitian $\bm K$ and any operator $\bm A$ we obtain
\begin{equation}\label{gen:gaussian-conjugation}
\begin{aligned}
 e^{\pm\beta\bm K/2}\bm J_{\bm A,\bm K}(\omega)e^{\mp\beta\bm K/2}
 &=e^{\pm\beta\omega/2+1/4}\bm J_{\bm A,\bm K}(\omega\pm1/\beta),\\
 \norm{e^{\pm\beta\bm K/2}\bm J_{\bm A,\bm K}(\omega)e^{\mp\beta\bm K/2}}_\infty
 &\le C\sqrt{\beta}\,e^{\beta|\omega|/2}\norm{\bm A}_\infty.
\end{aligned}
\end{equation}
The constant $C$ is independent of the system size and $\bm K$.
\end{lemma}

\begin{lemma}[Quasi-locality of the Gibbs sampler {\citep[Lemma A.4 and Corollary A.6]{Bergamaschi2026Stationarity}}]\label{lem:bath-locality}
For $\beta\ge1$, let $\bm K_{a,r}$ retain the candidate interactions supported entirely within the radius-$r$ ball around the site supporting $\bm A_a$. Then
\begin{equation}\label{oracle:bath-locality}
 \norm{\cL_{\bm A_a,\bm K}^*-\cL_{\bm A_a,\bm K_{a,r}}^*}_{\infty\to\infty}
 \le \poly(\beta)e^{-cr/\beta},
\end{equation}
where the polynomial prefactor and $c>0$ depend only on the fixed locality parameters.
\end{lemma}

\paragraph{The stationarity residual, stationarity tests, and the Fisher quadratic form.}
Let $\bm{\tau}$ denote the supplied input state, which may be either the Gibbs state $\bm{\rho}_{\bm{H}}$ or a $\zeta$-locally metastable state $\bm{\sigma}$. Following the stationarity-test paradigm of Bergamaschi \citep[Definition 3.1]{Bergamaschi2026Stationarity}, for any candidate Hamiltonian $\bm{K} = \bm{H}(\mathbf{k})$ we define the stationarity residual map $\mathbf{R}_{\bm{\tau}}: [-1,1]^m \to \mathbb{R}^m$ and the local stationarity tests $T_a(\bm{U}, \bm{K}; \bm{\tau})$ by
\begin{equation}\label{eq:R}
 \mathbf{R}_{\bm{\tau}}(\mathbf{k}) := -\Bigl(\Tr\bigl[\bm{\tau} \cL_{\bm{K}}^*[\bm{P}_j]\bigr]\Bigr)_{j=1}^m, \qquad T_a(\bm{U}, \bm{K}; \bm{\tau}) := \Tr\bigl[\bm{\tau} \cL_{\bm A_a,\bm{K}}^*[\bm{U}]\bigr].
\end{equation}
When $\bm{\tau} = \bm{\rho}_{\bm{H}}$, we also write $\mathbf{R}(\mathbf{k}) = \mathbf{R}_{\bm{\rho}_{\bm{H}}}(\mathbf{k})$ for brevity. Denoting the coefficient error by $\mathbf{e} = \mathbf{k} - \mathbf{h} \in \mathbb{R}^m$ and the Hamiltonian error by $\bm{E} = \bm{K} - \bm{H} = \sum_{j=1}^m e_j \bm{P}_j$, the key analytical quantity governing both learning problems is the \emph{centered radial residual}:
\begin{equation}\label{eq:centered-radial}
 \bigl\langle \mathbf{e}, \mathbf{R}_{\bm{\tau}}(\mathbf{k}) - \mathbf{R}_{\bm{\tau}}(\mathbf{h}) \bigr\rangle = -\sum_a \Bigl(T_a(\bm{E}, \bm{K}; \bm{\tau}) - T_a(\bm{E}, \bm{H}; \bm{\tau})\Bigr).
\end{equation}

Writing $\bm{J}_{\bm{A}_a,\bm{K}}(\omega, t) := \mathcal{U}_t^{\bm{K}}[\bm{J}_{\bm{A}_a,\bm{K}}(\omega)]$, the positive Fisher quadratic form associated with the CKG generator indexed by $a$ and an auxiliary state $\bm{\tau}_a$ is given by \citep{bergamaschi2025structuraltheoryquantummetastability,Bergamaschi2026Stationarity}
\begin{equation}\label{meta:Q-definition}
 \mathcal{Q}_a(\bm{U}, \bm{K}; \bm{\tau}_a) := \int_0^1 dv \int_{\mathbb{R}^2} dt\,d\omega\, h_{v-1/2}(\omega) g_{v-1/2}(t) \norm{\bigl[\bm{J}_{\bm{A}_a,\bm{K}}(\omega, t), \bm{U}\bigr]}_{\bm{\tau}_a,v}^2,
\end{equation}
where $\|\bm{X}\|_{\bm{\tau}_a,v} = \|\bm{\tau}_a^{v/2} \bm{X} \bm{\tau}_a^{(1-v)/2}\|_2$ and the positive kernels $h_z(\omega), g_z(t)$ for $z = v - 1/2 \in (-1/2, 1/2)$ are
\begin{equation}\label{eq:fisher-kernels}
 h_z(\omega) = \exp\!\left(z\beta\omega - \frac{z^2}{2} - \frac{1}{8} - \frac{\beta}{2}\left|\omega - \frac{z}{\beta}\right|\right), \qquad g_z(t) = \frac{2\cos(\pi z)\cosh(2\pi t/\beta)}{\beta\bigl[\cosh(4\pi t/\beta) + \cos(2\pi z)\bigr]}.
\end{equation}
Notice that \eqref{meta:Q-definition} is well-defined and nonnegative for any density operator $\bm{\tau}_a$. When $\bm{\tau} = \bm{\rho}_{\bm{H}}$, exact stationarity $\cL_{\bm{H}}[\bm{\rho}_{\bm{H}}] = 0$ gives $\mathbf{R}_{\bm{\rho}_{\bm{H}}}(\mathbf{h}) = \mathbf{0}$, while $\cL_{\bm{K}}[\bm{\rho}_{\bm{K}}] = 0$ and $\log \bm{\rho}_{\bm{H}} - \log \bm{\rho}_{\bm{K}} = \beta \bm{E} + c\bm{I}$ yield the exact identity relating the stationarity test to Fisher information and entropy production \citep{Spohn,CM,ChenKastoryanoGilyen2023,bergamaschi2025structuraltheoryquantummetastability,Bergamaschi2026Stationarity}:
\begin{equation}\label{eq:radial}
 \bigl\langle \mathbf{e}, \mathbf{R}_{\bm{\rho}_{\bm{H}}}(\mathbf{k}) \bigr\rangle = -\Tr\bigl[\cL_{\bm{K}}[\bm{\rho}_{\bm{H}}] \bm{E}\bigr] = \frac{1}{\beta}\EP_{\cL_{\bm{K}}}(\bm{\rho}_{\bm{H}}\|\bm{\rho}_{\bm{K}}) = \beta\sum_a \mathcal{Q}_a(\bm{E}, \bm{K}; \bm{\rho}_{\bm{H}}),
\end{equation}
where $\EP_{\cL}(\bm{\rho}\|\bm{\sigma}) := -\Tr[\cL[\bm{\rho}](\log \bm{\rho} - \log \bm{\sigma})]$ is the entropy production.

\paragraph{Approximate detailed balance.}
For each $a$, let $\bm H_a$ denote a reference Hamiltonian and $\bm\tau_a$ its associated reference state. In the Gibbs application we take $\bm H_a=\bm H$ and $\bm\tau_a=\bm\rho_{\bm H}$. In the metastable application, $\bm H_a$ is the true Hamiltonian truncated to a region $A_a$ around the site supporting $\bm A_a$:
\begin{equation}\label{meta:reference-hamiltonian}
 \bm H_a:=\sum_{j:\,\supp(\bm P_j)\subseteq A_a}h_j\bm P_j.
\end{equation}
The region $A_a$ and reference state $\bm\tau_a$ are specified in the metastable construction below. 
For any such pair $(\bm H_a,\bm\tau_a)$ with Gibbs state $\bm{\rho}_{\bm{H}_a} = e^{-\beta \bm{H}_a}/\Tr(e^{-\beta \bm{H}_a})$, define $\mathcal{D}_{\bm{\tau}_a}^{\bm{H}_a}$ by
\begin{equation}\label{meta:adb-defect}
 \mathcal{D}_{\bm{\tau}_a}^{\bm{H}_a}[\bm{Z}] := \bm{Z}\sqrt{\bm{\tau}_a} - \sqrt{\bm{\tau}_a}\, \bm{\rho}_{\bm{H}_a}^{-1/2}\bm{Z}\bm{\rho}_{\bm{H}_a}^{1/2}.
\end{equation}
The approximate detailed-balance (ADB) error of $\bm{\tau}_a$ with respect to $(\bm{H}_a, \bm{A})$ is defined as \citep{bergamaschi2025structuraltheoryquantummetastability}
\begin{equation}\label{meta:ADB}
 \mathsf{ADB}_{\bm{H}_a,\bm{A}}(\bm{\tau}_a) := \int_{\mathbb{R}^2} \gamma(\omega) g(t) \norm{\mathcal{D}_{\bm{\tau}_a}^{\bm{H}_a}\bigl[\bm{J}_{\bm{A},\bm{H}_a}(\omega, t)\bigr]}_2^2\,d\omega\,dt, \qquad g(t) = \frac{1}{\beta\cosh(2\pi t/\beta)}.
\end{equation}
For an exact Gibbs state, $\mathcal{D}_{\bm{\rho}_{\bm{H}}}^{\bm{H}}[\bm{Z}] = 0$ identically for every operator $\bm{Z}$, so $\mathsf{ADB}_{\bm{H},\bm{A}}(\bm{\rho}_{\bm{H}}) = 0$.
For $\bm\tau_a>0$ and Hermitian $\bm A$, \citep[Eqs.~(43)--(46)]{WangYeChen2026MetastableLearning} give
\begin{equation}\label{app:adb-gradient}
 \mathsf{ADB}_{\bm H_a,\bm A}(\bm\tau_a)
 =\int_{-1/2}^{1/2}\iint h_z(\omega)g_z^{\rm ADB}(t)
 \norm{[\bm J_{\bm A,\bm H_a}(\omega,t),\log\bm\tau_a+\beta\bm H_a]}_{\bm\tau_a,1/2+z}^2\,dt\,d\omega\,dz,
\end{equation}
where
\begin{equation}\label{app:adb-time-kernel}
 g_z^{\rm ADB}(t)=\frac1{2\pi\beta}
 \log\frac{\cosh(2\pi t/\beta)+\cos(\pi z)}{\cosh(2\pi t/\beta)-\cos(\pi z)}.
\end{equation}
Replacing $g_z^{\rm ADB}$ by $g_z$ gives entropy production. The kernel comparison in \citep[Theorem B.1]{WangYeChen2026MetastableLearning} with Gaussian width $\beta^{-1}$ yields
\begin{equation}\label{app:absolute-ep-adb}
 \mathsf{ADB}_{\bm H_a,\bm A}(\bm\tau_a)\le C\EP_{\cL_{\bm A,\bm H_a}}(\bm\tau_a\|\bm\rho_{\bm H_a}).
\end{equation}

\subsection{Technical overview}
\label{sec:tech-overview}

Our algorithm treats the stationarity residual $\mathbf{R}_{\bm{\tau}}(\mathbf{k})$ from \eqref{eq:R} as a vector field on the parameter cube $[-1, 1]^m$ that vanishes at $\mathbf{k} = \mathbf{h}$ when $\bm{\tau} = \bm{\rho}_{\bm{H}}$ (and has $\ell_\infty$ norm at most $\tO(\zeta)$ when $\bm{\tau} = \bm{\sigma}$). To recover $\mathbf{h}$ by iteratively stepping along $-\mathbf{R}_{\bm{\tau}}(\mathbf{k})$ with $e^{O(\beta)}$ sample complexity, we establish two main properties: (i) a radial coercivity bound with a \emph{linear} exponent in $\beta$, showing that $-\mathbf{R}_{\bm{\tau}}(\mathbf{k})$ points toward $\mathbf{h}$ up to the baseline drift,
\begin{equation}\label{overview:radial-coercivity}
 \langle\mathbf k-\mathbf h,\mathbf R_{\bm\tau}(\mathbf k)\rangle
 \ge ce^{-C\beta}\|\mathbf k-\mathbf h\|_2^2
 -\|\mathbf R_{\bm\tau}(\mathbf h)\|_\infty\|\mathbf k-\mathbf h\|_1,
\end{equation}
and (ii) a local measurement and stochastic gradient scheme that estimates $\mathbf{R}_{\bm{\tau}}(\mathbf{k})$ and updates $\mathbf{k}$ without accumulating error across the system size $n$.

\paragraph{1. Fisher coercivity via smooth frequency filtering (Section~\ref{sec:coercivity}).}
For Gibbs states, the entropy-production identity \eqref{eq:radial} expresses the radial residual as the positive Fisher sum $\beta\sum_a \mathcal{Q}_a(\bm{E}, \bm{K}; \bm{\rho}_{\bm{H}})$. To lower-bound this sum by the squared coefficient error, we start from the temperature-independent Hilbert--Schmidt lower bound on local Pauli commutators:
\begin{equation}\label{overview:pauli-signal}
 2^{-n}\sum_a\Tr([\bm A_a,\bm E]^\dagger[\bm A_a,\bm E])
 \ge 8\norm{\mathbf k-\mathbf h}_2^2.
\end{equation}
We first show that a smooth filter with temperature-independent bandwidth preserves the local commutator signal in operator norm, and then transfer this filtered signal to the KMS norm. The resulting thermal-weight loss is only $e^{O(\beta)}$. Writing $\bm{F}_{\bm A_a,\bm K}$ for the filtered Pauli operator obtained by smoothly averaging $\bm{J}_{\bm{A}_a,\bm{K}}(\omega)$ over $[-\omega_b, \omega_b]$, Lieb--Robinson locality gives, uniformly for $\beta\ge1$,
\begin{equation}\label{overview:fixed-band-error}
 \left(\sum_a\norm{[\bm F_{\bm A_a,\bm K},\bm E]-[\bm A_a,\bm E]}_\infty^2\right)^{1/2}
 \le\frac{C}{\omega_b}\norm{\mathbf k-\mathbf h}_2.
\end{equation}
Choosing a sufficiently large, temperature-independent bandwidth $\omega_b=\Theta(1)$ makes the filtering error at most half the original commutator signal in \eqref{overview:pauli-signal}, preserving its lower bound. Passing to the KMS norm on $[-\omega_b,\omega_b]$ then costs only $e^{O(\beta\omega_b)}=e^{O(\beta)}$ in the thermal weights, yielding
\begin{equation}\label{eq:central-message}
 \bigl\langle \mathbf{k} - \mathbf{h}, \mathbf{R}_{\bm{\rho}_{\bm{H}}}(\mathbf{k}) \bigr\rangle
 = \beta \sum_a \mathcal{Q}_a(\bm{E}, \bm{K}; \bm{\rho}_{\bm{H}})
 \ge c e^{-C\beta}\norm{\mathbf{k} - \mathbf{h}}_2^2.
\end{equation}

\paragraph{2. Extension to thermal metastable states (Section~\ref{sec:metastable-proof}).}
When the input is a $\zeta$-locally metastable state $\bm{\sigma}$, $\log\bm{\sigma}$ is no longer globally equal to $-\beta\bm{H} + c\bm{I}$, so the exact identity \eqref{eq:radial} acquires a non-equilibrium remainder governed by commutators with $\log\bm{\sigma} + \beta\bm{H}$. Using generator quasi-locality \eqref{oracle:bath-locality}, we localize each stationarity test $T_a$ to a patch around site $a$ of radius $O(\beta^2)$ (volume $V = O(\beta^{2D})$) and regularize the reduced patch state $\bm{\sigma}_a^{(\lambda)}$. On each patch, approximate detailed balance \eqref{meta:ADB} controls both the KMS norm comparison and the remainder involving commutators with $\log\bm{\sigma}_a^{(\lambda)}+\beta\bm H_a$ up to an error scaling as $e^{O(\beta V)}\sqrt{\zeta} = e^{O(\beta^{2D+1})}\sqrt{\zeta}\,\|\mathbf{k}-\mathbf{h}\|_2^2$. Provided $\zeta\le e^{-C\beta^{2D+1}}$, this remainder is absorbed into the $e^{-O(\beta)}$ Fisher lower bound, yielding
\begin{equation}\label{overview:metastable-coercivity}
 \bigl\langle \mathbf k-\mathbf h,
 \mathbf R_{\bm\sigma}(\mathbf k)-\mathbf R_{\bm\sigma}(\mathbf h)\bigr\rangle
 \ge ce^{-C\beta}\norm{\mathbf k-\mathbf h}_2^2,
\end{equation}
while the remaining baseline drift $\mathbf R_{\bm\sigma}(\mathbf h)$ determines the attainable learning accuracy.

\paragraph{3. Short-time local evolution estimator and local SGD (Section~\ref{sec:gibbs-learning}).}
Each coordinate $R_{\bm\tau,j}(\mathbf k)$ is the negative time derivative of $\Tr(\bm P_j e^{t\cL_{\bm K}}[\bm\tau])$ at $t=0$. By generator quasi-locality \eqref{oracle:bath-locality}, we restrict $\cL_{\bm{K}}$ to a local patch around $\supp(\bm{P}_j)$ and approximate this derivative by a short-time finite difference (evolving and measuring disjoint patches in parallel), which retains the $e^{-O(\beta)}$ radial coercivity of $\mathbf{R}_{\bm{\tau}}$.
To see why coercivity alone drives convergence, consider an ideal residual map $\mathbf F$ with $\mathbf F(\mathbf h)=\mathbf0$ satisfying
\begin{equation}\label{overview:descent-assumptions}
 \langle\mathbf x-\mathbf h,\mathbf F(\mathbf x)\rangle\ge\mu\|\mathbf x-\mathbf h\|_2^2,
 \qquad
 \|\mathbf F(\mathbf x)-\mathbf F(\mathbf y)\|_2\le L\|\mathbf x-\mathbf y\|_2.
\end{equation}
Let $\mathsf P$ be Euclidean projection onto $[-1,1]^m$. For the update $\mathbf x_{t+1}=\mathsf P(\mathbf x_t-\alpha\mathbf F(\mathbf x_t))$, projection cannot increase the distance to $\mathbf h$, so
\begin{equation}\label{overview:descent-contraction}
\begin{aligned}
 \|\mathbf x_{t+1}-\mathbf h\|_2^2
 &\le\|\mathbf x_t-\mathbf h\|_2^2
 -2\alpha\langle\mathbf x_t-\mathbf h,\mathbf F(\mathbf x_t)\rangle
 +\alpha^2\|\mathbf F(\mathbf x_t)\|_2^2\\
 &\le(1-2\alpha\mu+\alpha^2L^2)\|\mathbf x_t-\mathbf h\|_2^2.
\end{aligned}
\end{equation}
Taking $\alpha=\mu/L^2$ gives contraction by $1-\mu^2/L^2$, yielding geometric convergence to $\mathbf{h}$.

In the actual algorithm, the updates use noisy finite-time estimates. Applying \eqref{overview:descent-contraction} globally to all $m = \Theta(n)$ coordinates would accumulate $O(n)$ sampling variance. However, because each single-step update at coordinate $j$ depends only on parameters within its local patch, after $T = \tO(L^2/\mu^2) = \tO(e^{O(\beta)})$ steps the iterate $x_{T,j}$ depends only on an $n$-independent light-cone of $\tO(e^{O(\beta)})$ coordinates. Applying the contraction analysis to this localized region bounds the coordinatewise mean-square error up to the truncation, sampling, and metastable-drift floor, yielding $\tO(e^{O(\beta)}\epsilon^{-2}\log(n/\delta))$ sample complexity.

\section{Fisher-Information Coercivity}
\label{sec:coercivity}

We establish a Fisher-information coercivity bound that applies to both Gibbs and thermal metastable states. 
The common argument combines a lower bound on the local Fisher quadratic forms with a bound on their difference from the centered radial residual. 
We first state the resulting radial coercivity bound.

Recall from \eqref{eq:R}--\eqref{eq:centered-radial} that for an input state $\bm{\tau}$, candidate $\bm{K} = \bm{H}(\mathbf{k})$, parameter error $\mathbf{e} = \mathbf{k} - \mathbf{h}$, and Hamiltonian error $\bm{E} = \bm{K} - \bm{H}$, the centered radial residual is $\langle \mathbf{e}, \mathbf{R}_{\bm{\tau}}(\mathbf{k}) - \mathbf{R}_{\bm{\tau}}(\mathbf{h})\rangle$.

\begin{theorem}[Radial coercivity for Gibbs and thermal metastable states]\label{thm:unified-radial-coercivity}
There exist constants $c,C>0$ depending only on $(D,R,w,q)$ such that, for all $\beta\ge1$ and $\mathbf k,\mathbf h\in[-1,1]^m$,
\begin{equation}\label{eq:unified-radial-coercivity}
 \bigl\langle\mathbf k-\mathbf h,
 \mathbf R_{\bm\tau}(\mathbf k)-\mathbf R_{\bm\tau}(\mathbf h)\bigr\rangle
 \ge ce^{-C\beta}\norm{\mathbf k-\mathbf h}_2^2.
\end{equation}
This holds for $\bm\tau=\bm\rho_{\bm H}$ and for every $\zeta$-locally metastable state $\bm\tau=\bm\sigma$ satisfying $\zeta\le\zeta_*(\beta)$, with $\zeta_*(\beta)=\exp[-C\beta^{2D+1}]$ chosen in \eqref{meta:scale-values}. In the Gibbs case, $\mathbf R_{\bm\rho_{\bm H}}(\mathbf h)=\mathbf0$.
\end{theorem}

The Gibbs and metastable cases are proved in Sections~\ref{sec:gibbs-coercivity} and~\ref{meta:localization}, respectively. The proof proceeds in three steps:
\begin{enumerate}
 \item We construct a family of filtered commutators $[\mathcal{U}_t^{\bm{K}_a}[\bm{F}_{\bm A_a,\bm K_a}], \bm{U}_a]$ at a constant bandwidth ${\omega_b} = \Theta(1)$ independent of $\beta$ and $n$, and bound their operator-norm distance to the bare Pauli commutators $[\bm{A}_a, \bm{E}]$ (Section~\ref{lin:section}).
 \item We prove a lower bound on the KMS norm of approximate local commutators under approximate detailed balance (Lemma~\ref{lin:faithfulness}), with the Gibbs-state bound as the $\eta = 0$ case (Section~\ref{meta:faithfulness-section}).
 \item We combine a paired-weight AM--GM inequality for the Fisher-information kernels with Cauchy--Schwarz applied to the frequency integral to prove the universal Fisher coercivity bound $\beta\sum_a \mathcal{Q}_a(\bm{U}_a, \bm{K}_a; \bm{\tau}_a) \ge \mu_\beta \|\mathbf{e}\|_2^2$ with $\mu_\beta = c e^{-C\beta}$ (Section~\ref{boundary:coercivity-section}).
\end{enumerate}

\subsection{Filtered commutator approximation}
\label{lin:section}

We construct filtered commutators for both Gibbs and thermal metastable states. We first specify the common local setup and the filtered operators, then prove the approximation using real-time commutator continuity. For each single-site Pauli operator $\bm{A}_a$, consider the bare local commutator $[\bm A_a,\bm E]$.
The commutator $[\bm{A}_a,\bm{E}]$ is supported on a local interaction neighborhood $S_a$ around site $a$ of dimension $d_a = 2^{|S_a|} \le d_* = O(1)$. The Pauli-orthogonality calculation in Appendix~\ref{app:pauli-signal} gives
\begin{equation}\label{lin:paulisignal}
 2^{-n}\sum_a \Tr([\bm{A}_a,\bm{E}]^\dagger [\bm{A}_a,\bm{E}]) = 8 \sum_{j=1}^m |\supp(\bm{P}_j)| e_j^2 \ge 8\norm{\mathbf{e}}_2^2.
\end{equation}

Throughout this subsection, for each $a$, let $\bm{K}_a$ be any Hamiltonian on the dictionary $\{\bm{P}_j\}$ with coefficients bounded in $[-2,2]$ (such as $\bm{K}_a = \bm{K}$ for Gibbs states, or $\bm{K}_a = \bm{H}_a + \bm{U}_a$ for metastable states), and let $\bm{U}_a = \sum_{j \in I_a} e_j \bm{P}_j$ be any sub-sum of $\bm{E} = \sum_{j=1}^m e_j \bm{P}_j$ that retains all terms within distance $\ell \ge R$ of $a$ (such as $\bm{U}_a = \bm{E}$ or the radius-$\ell$ truncation of $\bm{E}$). Because every dictionary term $\bm{P}_j$ non-commuting with $\bm{A}_a$ has support within distance $R \le \ell$ of $a$, we have the exact identity
\begin{equation}\label{eq:Ua-commutator-identity}
 [\bm{A}_a, \bm{U}_a] = [\bm{A}_a, \bm{E}].
\end{equation}

To construct a filtered operator, choose a smooth, real, even cutoff $\chi \in C_c^\infty((-1,1))$ with $\chi=1$ on $[-1/2,1/2]$. For bandwidth ${\omega_b}\ge1$, define
\begin{equation}\label{eq:display-9}
 g(u):=\frac{1}{2\pi}\int_{\mathbb R}\chi(\omega)e^{-i\omega u}\,d\omega,
 \qquad g_{\omega_b}(u):={\omega_b} g({\omega_b} u).
\end{equation}
The kernel $g$ is real and even, with $\int_{\mathbb R}g(u)\,du=1$, and smoothness and compact support of $\chi$ imply that $g$ is a Schwartz function satisfying, for all integers $N,m\ge0$,
\begin{equation}\label{lin:schwartz-moments}
 \sup_{u\in\mathbb R}(1+|u|)^N|g^{(m)}(u)|<\infty,
 \qquad \int_{\mathbb R}|g(u)|(1+|u|)^N\,du<\infty.
\end{equation}
For any Hermitian operator $\bm A$ and Hamiltonian $\bm G$, define the Hermitian filtered operator
\begin{equation}\label{lin:B}
 \bm F_{\bm A,\bm G} := c_\beta \int_{\mathbb R}\chi(\omega/\omega_b)\bm J_{\bm A,\bm G}(\omega)\,d\omega
 =\int_{\mathbb R}g_{\omega_b}(u)e^{-u^2/\beta^2}\mathcal U_u^{\bm G}[\bm A]\,du,
\end{equation}
where $c_\beta=(\sqrt{2\pi}f_\beta(0))^{-1}$. For the auxiliary candidate $\bm K_a$ and coupling operator $\bm A_a$, we use $\bm F_{\bm A_a,\bm K_a}$.
Unlike a sharp indicator $\mathbf{1}_{[-{\omega_b},{\omega_b}]}$ whose sinc kernel decays as $1/|u|$, the smooth cutoff has finite polynomial moments independent of $\beta$:
\begin{equation}\label{lin:smooth-cutoff-moment}
 \int_{\mathbb R}|g_{\omega_b}(u)|\,|u|(1+|u|)^D\,du
 =\frac1{\omega_b}\int_{\mathbb R}|g(v)|\,|v|(1+|v|/{\omega_b})^D\,dv
 \le\frac C{\omega_b}.
\end{equation}

The approximation below holds for every bandwidth $\omega_b\ge1$; we will subsequently choose a sufficiently large constant bandwidth independent of $\beta$ and $n$. To control the filtered commutators, we first bound the change in the bare commutators under real-time evolution.

\begin{lemma}[Real-time commutator continuity]\label{lin:continuity}
There exists a constant $C > 0$ depending only on $(D, R, w, q)$ such that for all $\mathbf{k}, \mathbf{h} \in [-1,1]^m$, $\mathbf{e} = \mathbf{k} - \mathbf{h}$, any auxiliary family $(\bm{K}_a, \bm{U}_a)$ as above, and all $u \in \mathbb{R}$,
\begin{equation}\label{eq:display-4}
 \left(\sum_a \norm{\bigl[\mathcal{U}_u^{\bm{K}_a}[\bm{A}_a] - \bm{A}_a, \bm{U}_a\bigr]}_\infty^2\right)^{1/2} \le C |u|(1 + |u|)^D \norm{\mathbf{e}}_2, \qquad \left(\sum_a \norm{[\bm{A}_a,\bm{E}]}_\infty^2\right)^{1/2} \le C \norm{\mathbf{e}}_2.
\end{equation}
\end{lemma}
\begin{proof}
Since each $\bm{P}_j$ has support of size at most $w$ and each site meets at most $q$ terms, bounded overlap gives $\left(\sum_a \norm{[\bm{A}_a,\bm{E}]}_\infty^2\right)^{1/2} \le C\norm{\mathbf{e}}_2$. For the time-evolved difference, write $\mathcal{U}_u^{\bm{K}_a}[\bm{A}_a] - \bm{A}_a = i\int_0^u \mathcal{U}_s^{\bm{K}_a}[[\bm{K}_a, \bm{A}_a]]\,ds$. The local commutator $[\bm{K}_a, \bm{A}_a]$ has $O(1)$ norm and support diameter uniformly in $a$. By the capped Lieb--Robinson bound \eqref{eq:capped-lr} applied to $\bm{K}_a$, the nonnegative influence matrix
\begin{equation}\label{eq:display-5}
 M_{aj}(s) := \norm{\bigl[\mathcal{U}_s^{\bm{K}_a}[[\bm{K}_a, \bm{A}_a]], \bm{P}_j\bigr]}_\infty
\end{equation}
satisfies $M_{aj}(s) \le C\min\{1,e^{-c(d(a,\supp(\bm{P}_j))-v_{\rm LR}|s|)}\}$, where we absorb the bounded support radius of $[\bm{K}_a,\bm{A}_a]$ into $C$.
For fixed $a$, at most $C(1+r)^D$ interaction terms meet the radius-$r$ ball since each site belongs to at most $q$ terms. For fixed $j$, at most $C(1+r)^D$ coupling operators are supported within distance $r$ of its support, since $|\supp(\bm{P}_j)|\le w$ and there are three Pauli coupling operators per site. Splitting either sum at radius $v_{\rm LR}|s|+O(1)$ and summing the exponentially decaying outer shells therefore gives
\begin{equation}\label{eq:display-6}
 \sup_a\sum_j M_{aj}(s) \le C(1+|s|)^D,
 \qquad
 \sup_j\sum_a M_{aj}(s) \le C(1+|s|)^D.
\end{equation}
Since $\bm{U}_a = \sum_{j \in I_a} e_j\bm{P}_j$ with $|e_j|\mathbf{1}_{j \in I_a} \le |e_j|$, the triangle inequality and Schur's test (via Cauchy--Schwarz) yield
\begin{equation}\label{eq:display-7}
 \begin{aligned}
 \left(\sum_a\norm{\bigl[\mathcal{U}_s^{\bm{K}_a}[[\bm{K}_a,\bm{A}_a]],\bm{U}_a\bigr]}_\infty^2\right)^{1/2}
 &\le \left(\sum_a\Bigl(\sum_j M_{aj}(s)|e_j|\Bigr)^2\right)^{1/2}\\
 &\le \left[\Bigl(\sup_a\sum_j M_{aj}(s)\Bigr)\Bigl(\sup_j\sum_a M_{aj}(s)\Bigr)\right]^{1/2}\norm{\mathbf{e}}_2\\
 &\le C(1+|s|)^D\norm{\mathbf{e}}_2.
 \end{aligned}
\end{equation}
Integrating over $s \in [0, |u|]$ with the triangle inequality gives
\begin{equation}\label{eq:display-8}
 \left(\sum_a\norm{\bigl[\mathcal{U}_u^{\bm{K}_a}[\bm{A}_a]-\bm{A}_a,\bm{U}_a\bigr]}_\infty^2\right)^{1/2}
 \le C\norm{\mathbf{e}}_2\int_0^{|u|}(1+s)^D\,ds
 \le C|u|(1+|u|)^D\norm{\mathbf{e}}_2.
\end{equation}
\end{proof}

Integrating the real-time continuity estimate against the smooth filter gives the desired approximation uniformly for $\beta\ge1$.

\begin{lemma}[Uniform filtered commutator approximation]\label{lin:band}
For all $\beta \ge 1$, ${\omega_b} \ge 1$, and $|t| \le 1$, the filtered commutators $[\mathcal{U}_t^{\bm{K}_a}[\bm{F}_{\bm A_a,\bm K_a}], \bm{U}_a]$ satisfy
\begin{equation}\label{lin:banderror}
 \left(\sum_a \norm{[\mathcal{U}_t^{\bm{K}_a}[\bm{F}_{\bm A_a,\bm K_a}],\bm{U}_a] - [\bm{A}_a,\bm{E}]}_\infty^2\right)^{1/2} \le C({\omega_b}^{-1} + |t|)\norm{\mathbf{e}}_2.
\end{equation}
\end{lemma}
\begin{proof}
Since $\int_{\mathbb{R}} g_{\omega_b}(u)\,du = 1$ and $1 - e^{-u^2/\beta^2} \le u^2/\beta^2$, the Gaussian weight defect satisfies $\bigl|\int_{\mathbb{R}} g_{\omega_b}(u)(e^{-u^2/\beta^2} - 1)\,du\bigr| \le \beta^{-2}\int_{\mathbb{R}} |g_{\omega_b}(u)|u^2\,du = O(\beta^{-2}{\omega_b}^{-2})$. Applying $\mathcal{U}_t^{\bm{K}_a}$ to \eqref{lin:B} and using $[\bm{A}_a, \bm{U}_a] = [\bm{A}_a,\bm{E}]$ gives
\begin{equation}\label{eq:display-10}
\begin{aligned}
 [\mathcal{U}_t^{\bm{K}_a}[\bm{F}_{\bm A_a,\bm K_a}],\bm{U}_a] - [\bm{A}_a,\bm{E}]=\int_{\mathbb R}g_{\omega_b}(u)e^{-u^2/\beta^2}
 \bigl[\mathcal{U}_{u+t}^{\bm{K}_a}[\bm{A}_a]-\bm{A}_a,\bm{U}_a\bigr]\,du\\
 +\left(\int_{\mathbb R}g_{\omega_b}(u)(e^{-u^2/\beta^2}-1)\,du\right)[\bm{A}_a,\bm{E}].
\end{aligned}
\end{equation}
The triangle inequality and Lemma~\ref{lin:continuity}, together with $e^{-u^2/\beta^2}\le1$ and $\beta\ge1$, therefore yield
\begin{equation}\label{eq:display-11}
 \left(\sum_a\norm{[\mathcal{U}_t^{\bm{K}_a}[\bm{F}_{\bm A_a,\bm K_a}],\bm{U}_a] - [\bm{A}_a,\bm{E}]}_\infty^2\right)^{1/2}
 \le C\norm{\mathbf{e}}_2 \int_{\mathbb{R}} |g_{\omega_b}(u)|\,|u + t|(1 + |u + t|)^D\,du + O({\omega_b}^{-2})\norm{\mathbf{e}}_2.
\end{equation}
Substituting $v={\omega_b} u$ for ${\omega_b}\ge1$ and $|t|\le1$ gives
\begin{equation}\label{eq:display-12}
\begin{aligned}
 \int_{\mathbb R}|g_{\omega_b}(u)|\,|u+t|(1+|u+t|)^D\,du
 &=\int_{\mathbb R}|g(v)|\,|v/{\omega_b}+t|(1+|v/{\omega_b}+t|)^D\,dv\\
 &\le C({\omega_b}^{-1}+|t|)\int_{\mathbb R}|g(v)|(1+|v|)^{D+1}\,dv
 \le C({\omega_b}^{-1}+|t|).
\end{aligned}
\end{equation}
Absorbing ${\omega_b}^{-2}\le{\omega_b}^{-1}$ proves \eqref{lin:banderror}.
\end{proof}

\subsection{KMS lower bound under approximate detailed balance}
\label{meta:faithfulness-section}

We next prove a lemma that converts the Hilbert--Schmidt signal \eqref{lin:paulisignal} of the local commutators $[\bm{A}_a,\bm{E}]$ into a lower bound on $\sum_a \|\bm{X}_a\|_{\bm{\tau}_a,\mathrm{KMS}}^2$ under approximate detailed balance, with the Gibbs-state bound as the $\eta = 0$ special case. We use the time-zero ADB bound of \citep[Lemma C.1]{bergamaschi2025structuraltheoryquantummetastability} (see also \citep[Appendix B.1]{WangYeChen2026MetastableLearning}):

\begin{lemma}[Time-zero ADB reduction]\label{meta:timezero}
There is an absolute constant $C > 0$ such that for any reference pair $(\bm{H}_a, \bm{\tau}_a)$ and operator $\bm{A}$,
\begin{equation}\label{meta:timezero-bound}
 \int_{\mathbb{R}} \gamma(\omega)\norm{\mathcal{D}_{\bm{\tau}_a}^{\bm{H}_a}\bigl[\bm{J}_{\bm{A},\bm{H}_a}(\omega)\bigr]}_2^2\,d\omega \le C\,\mathsf{ADB}_{\bm{H}_a,\bm{A}}(\bm{\tau}_a).
\end{equation}
\end{lemma}

\paragraph{Filtered local rotations and intertwining bounds.}
Let $\chi_{\omega_b}(\omega) = \chi(\omega/{\omega_b})$ be our smooth cutoff supported on $[-{\omega_b}, {\omega_b}]$ at a constant bandwidth ${\omega_b} = \Theta(1)$. For any single-site Pauli operator $\bm A$, use $\bm F_{\bm A,\bm H_a}$ from \eqref{lin:B} with the reference Hamiltonian $\bm H_a$.
Since $\|[\bm H_a,\bm A]\|_\infty=O(1)$, Duhamel's formula gives $\|\mathcal U_u^{\bm H_a}[\bm A]-\bm A\|_\infty\le C|u|$. Using $\int g_{\omega_b}(u)\,du=1$, $1-e^{-u^2/\beta^2}\le u^2/\beta^2$, and the Schwartz moments of $g$, we obtain, for $\beta,\omega_b\ge1$,
\begin{equation}\label{meta:rotation-approx}
 \begin{aligned}
 \norm{\bm F_{\bm A,\bm H_a}-\bm A}_\infty
 &\le\int_{\mathbb R}|g_{\omega_b}(u)|\left(C|u|+\frac{u^2}{\beta^2}\right)\,du\\
 &=O(\omega_b^{-1}+\beta^{-2}\omega_b^{-2})=O(\omega_b^{-1}).
 \end{aligned}
\end{equation}
Moreover, if $\mathsf{ADB}_{\bm{H}_a,\bm{A}}(\bm{\tau}_a) \le \eta$, Lemma~\ref{meta:timezero} and Cauchy--Schwarz over $\omega \in [-{\omega_b}, {\omega_b}]$ give
\begin{equation}\label{meta:filtered-defect}
 \norm{\mathcal{D}_{\bm{\tau}_a}^{\bm{H}_a}[\bm{F}_{\bm A,\bm H_a}]}_2^2 \le c_\beta^2 \left(\int_{-{\omega_b}}^{\omega_b} \frac{|\chi_{\omega_b}(\omega)|^2}{\gamma(\omega)}\,d\omega\right)\left(\int_{\mathbb{R}} \gamma(\omega)\norm{\mathcal{D}_{\bm{\tau}_a}^{\bm{H}_a}\bigl[\bm{J}_{\bm{A},\bm{H}_a}(\omega)\bigr]}_2^2\,d\omega\right) \le C \beta {\omega_b} e^{\beta{\omega_b}} \eta.
\end{equation}

Applying Lemma~\ref{lem:gaussian-conjugation} inside \eqref{lin:B}, using $c_\beta=O(\sqrt\beta)$ and $\supp\chi_{\omega_b}\subset[-\omega_b,\omega_b]$, bounds the modular conjugation by $C\beta\omega_b e^{\beta\omega_b/2}$. Together with \eqref{meta:filtered-defect}, this gives, for fixed $\omega_b=\Theta(1)$ and $\beta\ge1$,
\begin{equation}\label{meta:Fbeta}
 \norm{e^{\pm\beta\bm H_a/2}\bm F_{\bm A,\bm H_a}e^{\mp\beta\bm H_a/2}}_\infty \le e^{O(\beta)}, \qquad \norm{\mathcal{D}_{\bm{\tau}_a}^{\bm{H}_a}[\bm{F}_{\bm A,\bm H_a}]}_2 \le e^{O(\beta)} \sqrt{\eta}.
\end{equation}

\paragraph{Multi-site Pauli products and KMS trace transfer.}
Now let $\bm{P}$ be a Pauli operator supported on a local region $S_a$ of dimension $2^{|S_a|} \le d_* = O(1)$. Write $\bm P=\bm A_1\cdots\bm A_\ell$, where $\bm A_1,\ldots,\bm A_\ell$ are its nonidentity single-site Pauli factors in a fixed site order. Define its filtered version by
\begin{equation}\label{meta:filtered-product}
 \widetilde{\bm P}:=\bm F_{\bm A_1,\bm H_a}\cdots\bm F_{\bm A_\ell,\bm H_a},
\end{equation}
All factors use the same reference Hamiltonian $\bm H_a$ and the filter in \eqref{lin:B}. The order is fixed because the filtered factors need not commute; in particular, $\widetilde{\bm P}$ need not be Hermitian or unitary. 
For $\bm P=\bm I$, the empty product gives $\widetilde{\bm P}=\bm I$. Using the identity \citep[Eq.~(152)]{Bergamaschi2026Stationarity}
\begin{equation}\label{meta:defect-product}
 \mathcal{D}_{\bm{\tau}_a}^{\bm{H}_a}[\bm{U}\bm{V}] = \bm{U} \mathcal{D}_{\bm{\tau}_a}^{\bm{H}_a}[\bm{V}] + \mathcal{D}_{\bm{\tau}_a}^{\bm{H}_a}[\bm{U}] e^{\beta\bm H_a/2}\bm{V}e^{-\beta\bm H_a/2}
\end{equation}
and the multiplicativity of conjugation by $e^{\beta\bm H_a/2}$, we immediately obtain from \eqref{meta:rotation-approx} and \eqref{meta:Fbeta} that
\begin{equation}\label{meta:product-bounds}
 \norm{\widetilde{\bm{P}} - \bm{P}}_\infty \le O({\omega_b}^{-1}), \qquad \norm{e^{\beta\bm H_a/2}\widetilde{\bm{P}}e^{-\beta\bm H_a/2}}_\infty \le e^{O(\beta)}, \qquad \norm{\mathcal{D}_{\bm{\tau}_a}^{\bm{H}_a}[\widetilde{\bm{P}}]}_2 \le e^{O(\beta)} \sqrt{\eta}.
\end{equation}
\begin{lemma}[Refined KMS trace transfer]\label{lem:trace-transfer}
Suppose $\mathsf{ADB}_{\bm H_a,\bm A}(\bm\tau_a)\le\eta$ for every single-site Pauli $\bm A$ on $S_a$. For any two filtered Pauli products $\widetilde{\bm P},\widetilde{\bm Q}$ supported before filtering on $S_a$, and any operator $\bm X$,
\begin{equation}\label{meta:trace-comparison}
 \bigl|\Tr(\widetilde{\bm{P}}\bm{\tau}_a \widetilde{\bm{Q}} \bm{X})\bigr| \le e^{O(\beta)}\bigl(\norm{\bm{X}}_{\bm{\tau}_a,\mathrm{KMS}} + \sqrt{\eta}\norm{\bm{X}}_\infty\bigr).
\end{equation}
\end{lemma}
\begin{proof}
This refines \citep[Lemma 6.9]{Bergamaschi2026Stationarity} using the filtered-product bounds \eqref{meta:product-bounds}. By the definition \eqref{meta:adb-defect},
\begin{equation}\label{meta:trace-transfer-defect-identities}
\begin{aligned}
 \widetilde{\bm P}\sqrt{\bm\tau_a}
 &=\sqrt{\bm\tau_a}\,e^{\beta\bm H_a/2}\widetilde{\bm P}e^{-\beta\bm H_a/2}
   +\mathcal D_{\bm\tau_a}^{\bm H_a}[\widetilde{\bm P}],\\
 \sqrt{\bm\tau_a}\widetilde{\bm Q}
 &=e^{-\beta\bm H_a/2}\widetilde{\bm Q}e^{\beta\bm H_a/2}\sqrt{\bm\tau_a}
   +\mathcal D_{\bm\tau_a}^{\bm H_a}[\widetilde{\bm Q}^*]^*.
\end{aligned}
\end{equation}
Substituting the first identity into $\widetilde{\bm P}\bm\tau_a\widetilde{\bm Q}=(\widetilde{\bm P}\sqrt{\bm\tau_a})(\sqrt{\bm\tau_a}\widetilde{\bm Q})$, and then using the second only in the term without $\mathcal D_{\bm\tau_a}^{\bm H_a}[\widetilde{\bm P}]$, gives
\begin{equation}\label{meta:trace-transfer-expansion}
\begin{aligned}
 \widetilde{\bm P}\bm\tau_a\widetilde{\bm Q}
 ={}&\sqrt{\bm\tau_a}\,e^{\beta\bm H_a/2}\widetilde{\bm P}e^{-\beta\bm H_a}
 \widetilde{\bm Q}e^{\beta\bm H_a/2}\sqrt{\bm\tau_a}+\mathcal D_{\bm\tau_a}^{\bm H_a}[\widetilde{\bm P}]\sqrt{\bm\tau_a}\widetilde{\bm Q}+\sqrt{\bm\tau_a}\,e^{\beta\bm H_a/2}\widetilde{\bm P}e^{-\beta\bm H_a/2}
 \mathcal D_{\bm\tau_a}^{\bm H_a}[\widetilde{\bm Q}^*]^*.
\end{aligned}
\end{equation}
For any operator $\bm B$, Cauchy--Schwarz and Schatten H\"older give
\begin{equation}\label{meta:trace-transfer-kms}
\begin{aligned}
 |\Tr(\sqrt{\bm\tau_a}\bm B\sqrt{\bm\tau_a}\bm X)|
 &\le\norm{\bm\tau_a^{1/4}\bm B\bm\tau_a^{1/4}}_2\norm{\bm X}_{\bm\tau_a,\mathrm{KMS}}\\
 &\le\norm{\bm B}_\infty\norm{\bm X}_{\bm\tau_a,\mathrm{KMS}},
\end{aligned}
\end{equation}
since $\Tr\bm\tau_a=1$. The product bounds apply also to $\widetilde{\bm Q}^*$, whose Hermitian filtered factors occur in reverse order; taking adjoints therefore also bounds the negative modular conjugation of $\widetilde{\bm Q}$. Thus the first term of \eqref{meta:trace-transfer-expansion}, paired with $\bm X$ inside the trace, is bounded by $e^{O(\beta)}\norm{\bm X}_{\bm\tau_a,\mathrm{KMS}}$. The other two terms contribute at most
\begin{equation}\label{meta:trace-transfer-error}
\begin{aligned}
 &\Bigl(\norm{\mathcal D_{\bm\tau_a}^{\bm H_a}[\widetilde{\bm P}]}_2\norm{\widetilde{\bm Q}}_\infty
 +\norm{e^{\beta\bm H_a/2}\widetilde{\bm P}e^{-\beta\bm H_a/2}}_\infty
 \norm{\mathcal D_{\bm\tau_a}^{\bm H_a}[\widetilde{\bm Q}^*]}_2\Bigr)\norm{\bm X}_\infty\le e^{O(\beta)}\sqrt\eta\,\norm{\bm X}_\infty,
\end{aligned}
\end{equation}
by Cauchy--Schwarz, $\norm{\sqrt{\bm\tau_a}}_2=1$, and \eqref{meta:product-bounds}. Combining these bounds proves \eqref{meta:trace-comparison}.
\end{proof}

We now combine the Pauli-twirl estimate proved in Appendix~\ref{app:pauli-twirl} with \eqref{meta:trace-comparison} to establish the KMS lower bound for approximate local commutators.

\begin{lemma}[KMS lower bound for approximate local commutators]\label{lin:faithfulness}\label{meta:heterogeneous}
For each $a$, let $(\bm{H}_a, \bm{\tau}_a)$ be a reference Hamiltonian and state satisfying our uniform bounded-interaction assumptions, and let $[\bm{A}_a, \bm{E}]$ be supported on $S_a$ with $d_a = 2^{|S_a|} \le d_* = O(1)$. Suppose that for every single-site Pauli $\bm{A}$ on $S_a$, $\mathsf{ADB}_{\bm{H}_a, \bm{A}}(\bm{\tau}_a) \le \eta$. If a family of (possibly nonlocal) operators $(\bm{X}_a)_a$ satisfies
\begin{equation}\label{eq:display-13}
 \left(\sum_a \norm{\bm{X}_a - [\bm{A}_a,\bm{E}]}_\infty^2\right)^{1/2} \le \frac{\sqrt{2}}{d_*} \norm{\mathbf{e}}_2, \qquad \left(\sum_a \norm{\bm{X}_a}_\infty^2\right)^{1/2} \le C_0 \norm{\mathbf{e}}_2,
\end{equation}
then for all $\beta \ge 1$,
\begin{equation}\label{eq:unified-frame-preabsorb}
 \left(\sum_a \norm{\bm{X}_a}_{\bm{\tau}_a,\mathrm{KMS}}^2\right)^{1/2} \ge \left(c e^{-C\beta} - C'\sqrt{\eta}\right)\norm{\mathbf{e}}_2,
\end{equation}
where $c, C, C' > 0$ depend only on $(D, R, w, q)$ and $C_0$. In particular, whenever $\eta \le (c / (2C'))^2 e^{-2C\beta}$ (which holds with $\eta = 0$ for Gibbs states $\bm{\tau}_a = \bm{\rho}_{\bm{H}}$, $\bm{H}_a = \bm{H}$),
\begin{equation}\label{lin:faithfulnessresult}
 \sum_a \norm{\bm{X}_a}_{\bm{\tau}_a,\mathrm{KMS}}^2 \ge \kappa_\beta \norm{\mathbf{e}}_2^2, \qquad \kappa_\beta = \frac{c^2}{4} e^{-2C\beta}.
\end{equation}
\end{lemma}
\begin{proof}
By Appendix~\ref{app:pauli-twirl}, for each $a$ there exist Pauli operators $\bm{P}_a,\bm{Q}_a$ supported on $S_a$ such that
\begin{equation}\label{eq:display-27}
 \bigl|\Tr(\bm{P}_a\bm{\tau}_a\bm{Q}_a[\bm{A}_a,\bm{E}])\bigr| \ge d_*^{-1}\bigl(2^{-n}\Tr([\bm{A}_a,\bm{E}]^\dagger[\bm{A}_a,\bm{E}])\bigr)^{1/2}.
\end{equation}
By \eqref{lin:paulisignal}, the vector of these traces has Euclidean norm at least $2\sqrt{2} d_*^{-1}\|\mathbf{e}\|_2$.

Using $\|\bm{P}_a \bm{\tau}_a \bm{Q}_a\|_1 = 1$, replacing $[\bm{A}_a,\bm{E}]$ by $\bm{X}_a$ changes the Euclidean norm of this trace vector by at most $\left(\sum_a \norm{\bm{X}_a - [\bm{A}_a,\bm{E}]}_\infty^2\right)^{1/2} \le \frac{\sqrt{2}}{d_*} \|\mathbf{e}\|_2$. 
Next, replacing $\bm{P}_a, \bm{Q}_a$ by their filtered products $\widetilde{\bm{P}}_a, \widetilde{\bm{Q}}_a$ (formed with respect to $\bm{H}_a$ at a sufficiently large constant bandwidth ${\omega_b} = \Theta(1)$) changes the trace vector norm by at most $O({\omega_b}^{-1}) C_0 \|\mathbf{e}\|_2 \le \frac{1}{\sqrt{2}d_*}\|\mathbf{e}\|_2$, leaving
\begin{equation}\label{eq:display-28}
 \left(\sum_a \bigl|\Tr(\widetilde{\bm{P}}_a \bm{\tau}_a \widetilde{\bm{Q}}_a \bm{X}_a)\bigr|^2\right)^{1/2} \ge \frac{1}{\sqrt{2}d_*} \norm{\mathbf{e}}_2.
\end{equation}
Applying the trace-transfer bound \eqref{meta:trace-comparison} coordinatewise to each triple $(\bm{H}_a, \bm{\tau}_a, \bm{X}_a)$ and using the Euclidean triangle inequality along with $(\sum_a \|\bm{X}_a\|_\infty^2)^{1/2} \le C_0\|\mathbf{e}\|_2$ gives
\begin{equation}
 \frac{1}{\sqrt{2}d_*}\norm{\mathbf{e}}_2 \le e^{C\beta}\left[\left(\sum_a \norm{\bm{X}_a}_{\bm{\tau}_a,\mathrm{KMS}}^2\right)^{1/2} + C_0\sqrt{\eta}\norm{\mathbf{e}}_2\right].
\end{equation}
Dividing by $e^{C\beta}$ yields \eqref{eq:unified-frame-preabsorb} with $c = 1/(\sqrt{2}d_*)$ and $C' = C_0$, and \eqref{lin:faithfulnessresult} follows immediately whenever $C'\sqrt{\eta} \le \frac{1}{2}c e^{-C\beta}$.
\end{proof}

\subsection{Universal Fisher coercivity}
\label{boundary:coercivity-section}

We now combine the filtered commutator approximation (Lemma~\ref{lin:band}) and the KMS lower bound for approximate local commutators (Lemma~\ref{lin:faithfulness}) with the Fisher quadratic form \eqref{meta:Q-definition} to prove our main positive-coercivity theorem for both Gibbs and metastable states.

\begin{theorem}[Universal Fisher coercivity]\label{thm:fisher-coercivity}\label{cor:heterogeneous-fisher}
For each $a$, let $(\bm{H}_a, \bm{\tau}_a)$ be a reference pair, let $\bm{U}_a = \sum_{j \in I_a} e_j \bm{P}_j$ retain all terms of $\bm{E} = \bm{K} - \bm{H}$ within distance $\ell \ge R$ of $a$, and let $\bm{K}_a$ have coefficients bounded in $[-2,2]$. There exist constants $c, C, c', C' > 0$ depending only on $(D, R, w, q)$ such that if $\mathsf{ADB}_{\bm{H}_a, \bm{A}}(\bm{\tau}_a) \le \eta \le c' e^{-C'\beta}$ for every single-site Pauli $\bm{A}$ on $S_a$, then for all $\beta \ge 1$,
\begin{equation}\label{meta:heterogeneous-frame}
 \boxed{\;\beta \sum_a \mathcal{Q}_a(\bm{U}_a, \bm{K}_a; \bm{\tau}_a) \ge \mu_\beta \norm{\mathbf{e}}_2^2, \qquad \mu_\beta = c e^{-C\beta}.\;}
\end{equation}
\end{theorem}
\begin{proof}
Fix a sufficiently large constant bandwidth ${\omega_b} = \Theta(1)$ and a sufficiently small constant time window $t_0 > 0$ so that the right-hand side of Lemma~\ref{lin:band} satisfies $C({\omega_b}^{-1} + t_0) \le \frac{\sqrt{2}}{d_*}$. Then for all $|t| \le t_0$, Lemma~\ref{lin:faithfulness} applied to $[\mathcal{U}_t^{\bm{K}_a}[\bm{F}_{\bm A_a,\bm K_a}], \bm{U}_a]$ gives the uniform KMS lower bound
\begin{equation}\label{eq:display-17}
 \sum_a \norm{[\mathcal{U}_t^{\bm{K}_a}[\bm{F}_{\bm A_a,\bm K_a}],\bm{U}_a]}_{\bm{\tau}_a,\mathrm{KMS}}^2 \ge \kappa_\beta \norm{\mathbf{e}}_2^2, \qquad |t| \le t_0, \quad \kappa_\beta = c_1 e^{-C_1\beta}.
\end{equation}

Next, consider the Fisher quadratic form \eqref{meta:Q-definition} with $z = v - 1/2 \in (-1/2, 1/2)$. For any density matrix $\bm{\tau} = \sum_i p_i |i\rangle\langle i|$ and any operator $\bm{Z}$, we have $\|\bm{Z}\|_{\bm{\tau},1/2+z}^2 = \sum_{i,j} p_i^{1/2+z}p_j^{1/2-z}|\langle i|\bm{Z}|j\rangle|^2$. The weighted AM--GM inequality gives, entrywise,
\begin{equation}\label{eq:paired-weight-entrywise}
 h_z(\omega)p_i^{1/2+z}p_j^{1/2-z}
 +h_{-z}(\omega)p_i^{1/2-z}p_j^{1/2+z}
 \ge 2\sqrt{h_z(\omega)h_{-z}(\omega)}\sqrt{p_i p_j}.
\end{equation}
Multiplying by $|\langle i|\bm Z|j\rangle|^2$ and summing proves, for arbitrary $\bm Z$,
\begin{equation}\label{eq:paired-weight}
 h_z(\omega)\norm{\bm{Z}}_{\bm{\tau},1/2+z}^2 + h_{-z}(\omega)\norm{\bm{Z}}_{\bm{\tau},1/2-z}^2 \ge 2\sqrt{h_z(\omega)h_{-z}(\omega)}\norm{\bm{Z}}_{\bm{\tau},\mathrm{KMS}}^2.
\end{equation}
From \eqref{eq:fisher-kernels}, the geometric mean of $h_z(\omega)$ and $h_{-z}(\omega)$ simplifies to
\begin{equation}\label{eq:geom-mean-h}
 \sqrt{h_z(\omega)h_{-z}(\omega)} = \exp\!\left[-\frac{z^2}{2} - \frac{1}{8} - \frac{\beta}{2}\max\!\left\{|\omega|, \frac{|z|}{\beta}\right\}\right].
\end{equation}
For $|z| \le 1/4$, $|\omega| \le {\omega_b}$, and $|t| \le t_0$, we have $\sqrt{h_z(\omega)h_{-z}(\omega)} \ge c_2 e^{-\beta{\omega_b}/2}$ and $g_z(t) = g_{-z}(t) \ge c_3/\beta$. Restricting the integral in \eqref{meta:Q-definition} to $|z| \le 1/4$, $|\omega| \le {\omega_b}$, and $|t| \le t_0$ therefore gives
\begin{equation}\label{eq:Q-restricted}
 \mathcal{Q}_a(\bm{U}_a, \bm{K}_a; \bm{\tau}_a) \ge \frac{c_4 e^{-\beta{\omega_b}/2}}{\beta} \int_{-t_0}^{t_0} dt \int_{-{\omega_b}}^{\omega_b} d\omega\, \norm{\bigl[\bm{J}_{\bm{A}_a,\bm{K}_a}(\omega, t), \bm{U}_a\bigr]}_{\bm{\tau}_a,\mathrm{KMS}}^2.
\end{equation}

Finally, by \eqref{lin:B} we have $[\mathcal{U}_t^{\bm{K}_a}[\bm{F}_{\bm A_a,\bm K_a}],\bm{U}_a] = c_\beta \int_{-{\omega_b}}^{\omega_b} \chi(\omega/{\omega_b})[\bm{J}_{\bm{A}_a,\bm{K}_a}(\omega, t), \bm{U}_a]\,d\omega$, where $c_\beta^2 = \Theta(\beta)$ and $\int_{-{\omega_b}}^{\omega_b} |\chi(\omega/{\omega_b})|^2\,d\omega = \Theta({\omega_b}) = \Theta(1)$. Cauchy--Schwarz in $\omega$ therefore gives
\begin{equation}\label{eq:display-18}
 \norm{[\mathcal{U}_t^{\bm{K}_a}[\bm{F}_{\bm A_a,\bm K_a}],\bm{U}_a]}_{\bm{\tau}_a,\mathrm{KMS}}^2 \le C \beta \int_{-{\omega_b}}^{\omega_b} \norm{\bigl[\bm{J}_{\bm{A}_a,\bm{K}_a}(\omega, t), \bm{U}_a\bigr]}_{\bm{\tau}_a,\mathrm{KMS}}^2\,d\omega.
\end{equation}
Summing \eqref{eq:Q-restricted} over $a$ and substituting \eqref{eq:display-18} and the KMS lower bound \eqref{eq:display-17} yields
\begin{equation}\label{eq:display-19}
 \sum_a \mathcal{Q}_a(\bm{U}_a, \bm{K}_a; \bm{\tau}_a) \ge \frac{c_5 e^{-\beta{\omega_b}/2}}{\beta^2}\int_{-t_0}^{t_0} \sum_a \norm{[\mathcal{U}_t^{\bm{K}_a}[\bm{F}_{\bm A_a,\bm K_a}],\bm{U}_a]}_{\bm{\tau}_a,\mathrm{KMS}}^2\,dt \ge \frac{2 c_5 t_0 \kappa_\beta}{\beta^2} e^{-\beta{\omega_b}/2}\norm{\mathbf{e}}_2^2.
\end{equation}
Multiplying by $\beta$ gives \eqref{meta:heterogeneous-frame} with $\mu_\beta = \frac{2 c_5 t_0 \kappa_\beta}{\beta} e^{-\beta{\omega_b}/2} \ge c e^{-C\beta}$.
\end{proof}

\section{Stationarity-Test Identities and Fisher Information}
\label{sec:metastable-proof}

With the universal Fisher coercivity bound (Theorem~\ref{thm:fisher-coercivity}) in hand, the proofs of radial coercivity for Gibbs states and thermal metastable states diverge only in how accurately the centered stationarity residual $\langle \mathbf{e}, \mathbf{R}_{\bm{\tau}}(\mathbf{k})-\mathbf{R}_{\bm{\tau}}(\mathbf{h})\rangle$ agrees with the positive Fisher sum $\beta\sum_a \mathcal{Q}_a(\bm{U}_a, \bm{K}_a; \bm{\tau}_a)$. Together, Corollaries~\ref{thm:coercivity} and~\ref{cor:meta-coercivity} below establish Theorem~\ref{thm:unified-radial-coercivity}.

\subsection{Exact identity for Gibbs states}
\label{sec:gibbs-coercivity}

When the supplied state is the global Gibbs state $\bm{\tau} = \bm{\rho}_{\bm{H}}$, we choose for every $a$ the global auxiliary data
\begin{equation}\label{eq:gibbs-choice}
 \bm{\tau}_a = \bm{\rho}_{\bm{H}}, \qquad \bm{H}_a = \bm{H}, \qquad \bm{U}_a = \bm{E} = \bm{K} - \bm{H}, \qquad \bm{K}_a = \bm{K}.
\end{equation}
Here $\eta=0$ and $\mathbf R_{\bm\rho_{\bm H}}(\mathbf h)=\mathbf0$. By \eqref{eq:radial}, the radial residual equals $\beta\sum_a \mathcal{Q}_a(\bm{E}, \bm{K}; \bm{\rho}_{\bm{H}})$ without remainder, so Theorem~\ref{thm:fisher-coercivity} immediately yields the following corollary.

\begin{corollary}[Linear-in-$\beta$ coercivity for Gibbs states]\label{thm:coercivity}
There exist constants $c, C > 0$ depending only on $(D, R, w, q)$ such that for all $\beta \ge 1$ and all $\mathbf{k}, \mathbf{h} \in [-1,1]^m$,
\begin{equation}\label{coercivity:mu}
 \boxed{\;\bigl\langle \mathbf{e}, \mathbf{R}_{\bm{\rho}_{\bm{H}}}(\mathbf{k}) \bigr\rangle = \frac{1}{\beta}\EP_{\cL_{\bm{K}}}(\bm{\rho}_{\bm{H}}\|\bm{\rho}_{\bm{K}}) \ge \mu_\beta \norm{\mathbf{e}}_2^2, \qquad \mu_\beta = c e^{-C\beta}.\;}
\end{equation}
\end{corollary}

\subsection{Approximate identity for metastable states}
\label{meta:localization}

Now suppose the supplied state is a $\zeta$-locally metastable state $\bm{\tau} = \bm{\sigma}$ satisfying $\max_{i \in \Lambda} \|\cL_{i,\bm{H}}[\bm{\sigma}]\|_1 \le \zeta$. Write $B(a,r):=\{i\in\Lambda:d(i,\supp(\bm A_a))\le r\}$ for the radius-$r$ ball centered at the site supporting $\bm A_a$. Because $\log \bm{\sigma}$ is no longer globally equal to $-\beta \bm{H} + c\bm{I}$, we introduce for each $a$ the localized auxiliary data on balls of radii $\ell < b < L$:
\begin{equation}\label{meta:regularized-state}
 \begin{aligned}
 A_a &= B(a, L), \\
 \bm{\sigma}_a^{(\lambda)} &= (1 - \lambda)\Tr_{A_a^c}(\bm{\sigma}) + \lambda\bm{\rho}_{\bm{H}_a}, \qquad \lambda \in (0, 1/2), \\
 \bm{U}_a &= \sum_{d(a, \supp(\bm{P}_j)) \le \ell} e_j \bm{P}_j, \\
 \bm{K}_a &= \bm{H}_a + \bm{U}_a.
 \end{aligned}
\end{equation}
\begin{proposition}[Comparison with Fisher information for metastable states]\label{prop:meta-comparison}
There is a constant $C>0$, depending only on $(D,R,w,q)$, such that the following holds. Let $\beta\ge1$, $\mathbf k,\mathbf h\in[-1,1]^m$, and suppose $\bm\sigma$ satisfies \eqref{meta:stationarity} with $\zeta\le e^{-C\beta^{2D+1}}$. The parameters in \eqref{meta:regularized-state} can be chosen, depending only on $\beta$ and the fixed locality parameters, with
\[
 \ell,b=O(\beta^2),\qquad L=O(\beta^{2D+2}),\qquad
 \log(1/\lambda)=O(\beta^{2D+1}),
\]
so that
\begin{equation}\label{eq:meta-comparison-box} 
 \left|\langle \mathbf{e}, \mathbf{R}_{\bm{\sigma}}(\mathbf{k})-\mathbf{R}_{\bm{\sigma}}(\mathbf{h})\rangle - \beta\sum_a \mathcal{Q}_a(\bm{U}_a, \bm{K}_a; \bm{\sigma}_a^{(\lambda)})\right| \le \Delta_{\mathrm{meta}}\norm{\mathbf{e}}_2^2
\end{equation}
with $0\le\Delta_{\rm meta}\le\mu_\beta/2$, where $\mu_\beta=c_\mu e^{-C_\mu\beta}$ is the Fisher lower-bound constant from Theorem~\ref{thm:fisher-coercivity}.
\end{proposition}

We first state three lemmas proved in Appendices~\ref{app:joint-derivative}, \ref{app:regularized-adb}, and~\ref{app:continuity-proof}, respectively. Throughout these lemmas and their proofs, $\beta\ge1$, and all constants and polynomial degrees depend only on the fixed locality parameters.

\begin{lemma}[Localization and regularization of stationarity tests]\label{lem:meta-localization}
For the patch data \eqref{meta:regularized-state} with $L>\ell+R$, the coefficient derivatives satisfy, uniformly for $\mathbf k\in[-2,2]^m$,
\begin{equation}\label{meta:joint-derivative}
 \norm{\partial_{k_\ell}\cL_{\bm A_a,\bm{K}}^*[\bm{P}_j]}_\infty \le \poly(\beta)e^{-c[d(a,j)+d(a,\ell)]/\beta}.
\end{equation}
The global and regularized patch test differences satisfy
\begin{equation}\label{meta:localized-regularized-tests}
\begin{aligned}
 &\left|\sum_a\bigl[T_a(\bm E,\bm K;\bm\sigma)-T_a(\bm E,\bm H;\bm\sigma)
 -T_a(\bm U_a,\bm K_a;\bm\sigma_a^{(\lambda)})
 +T_a(\bm U_a,\bm H_a;\bm\sigma_a^{(\lambda)})\bigr]\right|\\
 &\qquad\le\poly(\beta,\ell)
 \bigl[e^{-c\ell/\beta}+e^{-cL/\beta}+\lambda\bigr]\norm{\mathbf e}_2^2.
\end{aligned}
\end{equation}
\end{lemma}

\begin{lemma}[Logarithm, ADB, and commutators of regularized patches]\label{lem:meta-regularized-adb}
For the patch data \eqref{meta:regularized-state},
\begin{equation}\label{meta:log-bound}
 \norm{\log\bm\sigma_a^{(\lambda)}+\beta\bm H_a}_\infty\le M:=C\{\log(1/\lambda)+\beta L^D\}.
\end{equation}
Let $\mathcal P_1(S)$ denote the single-site Pauli operators on $S$. The ADB error on $B(a,b)$ satisfies
\begin{equation}\label{meta:adb-local}
 \eta_a:=\max_{\bm A\in\mathcal P_1(B(a,b))}\mathsf{ADB}_{\bm H_a,\bm A}(\bm\sigma_a^{(\lambda)})
 \le CM\{\zeta+\poly(\beta)e^{-c(L-b)/\beta}\}.
\end{equation}
For every Pauli string $\bm P$ supported on $s$ sites in $B(a,b)$ and every $u\ge1+\log(\beta+1)$,
\begin{equation}\label{app:ordered-cutoff-bound}
 \left(\int_0^1\norm{[\bm P,\log\bm\sigma_a^{(\lambda)}+\beta\bm H_a]}_{\bm\sigma_a^{(\lambda)},v}^2\,dv\right)^{1/2}
 \le e^{C\beta(s+u)}\sqrt{\eta_a}+CM e^{-u}.
\end{equation}
\end{lemma}

Applying the preceding lemma to Pauli strings supported in $B(a,b)$ refines the stationarity-test comparison of \citep[Lemma 6.14]{Bergamaschi2026Stationarity} as follows.
\begin{lemma}[Quantitative stationarity-test comparison]\label{lem:meta-continuity}
For the patch data \eqref{meta:regularized-state}, let $M$ and $\eta_a$ be as in Lemma~\ref{lem:meta-regularized-adb}, and suppose $b\ge2(\ell+R)$. Set $V=C_0(1+b)^D$, with $C_0$ large enough that $|B(a,b)|\le V$. For every $u\ge1+\log(\beta+1)$,
\begin{equation}\label{meta:continuity-interface}
\begin{aligned}
 &\bigl|T_a(\bm U_a,\bm K_a;\bm\sigma_a^{(\lambda)})-T_a(\bm U_a,\bm H_a;\bm\sigma_a^{(\lambda)})
       +\beta\mathcal Q_a(\bm U_a,\bm K_a;\bm\sigma_a^{(\lambda)})\bigr|\\
 &\quad\le\poly(\beta,\ell,b)\norm{\bm U_a}_\infty^2
 \left\{M e^{-c(b-\ell)/\beta}
       +e^{CV+C\beta(V+u)}\sqrt{\eta_a}+M e^{CV-u}\right\}.
\end{aligned}
\end{equation}
\end{lemma}

\begin{proof}[Proof of Proposition~\ref{prop:meta-comparison}]
\emph{Combining the estimates.}
By \eqref{eq:centered-radial} and Lemmas~\ref{lem:meta-localization}--\ref{lem:meta-continuity},
\begin{equation}\label{meta:comparison-error-sum}
\begin{aligned}
 &\left|\langle\mathbf e,\mathbf R_{\bm\sigma}(\mathbf k)-\mathbf R_{\bm\sigma}(\mathbf h)\rangle
 -\beta\sum_a\mathcal Q_a(\bm U_a,\bm K_a;\bm\sigma_a^{(\lambda)})\right|\\
 &\le\left|\sum_a\bigl[T_a(\bm E,\bm K;\bm\sigma)-T_a(\bm E,\bm H;\bm\sigma)
 -T_a(\bm U_a,\bm K_a;\bm\sigma_a^{(\lambda)})
 +T_a(\bm U_a,\bm H_a;\bm\sigma_a^{(\lambda)})\bigr]\right|\\
 &\quad+\sum_a\left|T_a(\bm U_a,\bm K_a;\bm\sigma_a^{(\lambda)})
 -T_a(\bm U_a,\bm H_a;\bm\sigma_a^{(\lambda)})
 +\beta\mathcal Q_a(\bm U_a,\bm K_a;\bm\sigma_a^{(\lambda)})\right|\\
 &\le\poly(\beta,\ell)\bigl[e^{-c\ell/\beta}+e^{-cL/\beta}+\lambda\bigr]\norm{\mathbf e}_2^2\\
 &\quad+\poly(\beta,\ell,b)\sum_a\norm{\bm U_a}_\infty^2
 \left[M e^{-c(b-\ell)/\beta}+e^{CV+C\beta(V+u)}\sqrt{\eta_a}+M e^{CV-u}\right].
\end{aligned}
\end{equation}
To sum these weights, put $I_a:=\{j:d(a,\supp(\bm P_j))\le\ell\}$. There are at most $C(1+\ell)^D$ indices in $I_a$, and each fixed $j$ belongs to at most $C(1+\ell)^D$ such sets. Hence Cauchy--Schwarz gives
\begin{equation}\label{meta:patch-overlap}
 \sum_a\norm{\bm U_a}_\infty^2
 \le\sum_a|I_a|\sum_{j\in I_a}|e_j|^2
 \le C(1+\ell)^{2D}\norm{\mathbf e}_2^2.
\end{equation}
The triangle inequality and $\eta:=\max_a\eta_a$ therefore give \eqref{eq:meta-comparison-box} with
\begin{equation}\label{eq:Delta-meta-explicit}
\begin{aligned}
 \Delta_{\rm meta}\le\poly(\beta,\ell,b)\bigl[&e^{-c\ell/\beta}+e^{-cL/\beta}+\lambda
       +M e^{-c(b-\ell)/\beta}\\
       &+e^{CV+C\beta(V+u)}\sqrt\eta+M e^{CV-u}\bigr],
 \qquad \eta:=\max_a\eta_a,
\end{aligned}
\end{equation}
where the polynomial prefactor includes the overlap factor $(1+\ell)^{2D}$.

\emph{Choosing the scales.}
To preserve the Fisher lower bound $\mu_\beta=c_\mu e^{-C_\mu\beta}$, choose
\begin{equation}\label{meta:scale-values}
\begin{aligned}
 \ell&=c_\ell\beta^2,& b&=c_b\beta^2,& V&=C_0(1+b)^D,\\
 L&=b+c_L\beta^2V,& \lambda&=e^{-c_\lambda\beta V},&
 \zeta&\le\zeta_*(\beta):=e^{-C\beta^{2D+1}},
\end{aligned}
\end{equation}
with sufficiently large constants and $c_b>2(c_\ell+R)$. These give $L=O(\beta^{2D+2})$ and $M=\poly(\beta)$. The scale constants make each of the first four errors in \eqref{eq:Delta-meta-explicit}, including the polynomial prefactor, at most $\mu_\beta/12$.
Next, choose
\[
 u=c_u\{V+\beta+\log(M+1)+\log(\poly(\beta,\ell,b)+1)\}=O(V)
\]
with a sufficiently large fixed constant $c_u$ so that
\[
 \poly(\beta,\ell,b)M e^{CV-u}\le\mu_\beta/12.
\]
Finally, to ensure
\[
 \poly(\beta,\ell,b)e^{CV+C\beta(V+u)}\sqrt\eta\le\mu_\beta/12,
\]
it suffices to require $\eta\le e^{-K\beta V}$ for a sufficiently large constant $K$, since $u=O(V)$. By \eqref{meta:adb-local}, this is ensured by making each of $CM\zeta$ and $CM\poly(\beta)e^{-c(L-b)/\beta}$ at most $\tfrac12e^{-K\beta V}$. The latter is controlled by choosing $c_L$ sufficiently large; since $V=O(\beta^{2D})$ and $M=\poly(\beta)$, the former is ensured by $\zeta\le e^{-C\beta^{2D+1}}$ with sufficiently large $C$. This gives the required upper bound on $\zeta$ in \eqref{meta:scale-values} and also ensures the ADB hypothesis of Theorem~\ref{thm:fisher-coercivity}. Summing the six errors gives $\Delta_{\rm meta}\le\mu_\beta/2$, completing \eqref{eq:meta-comparison-box}.
\end{proof}

Proposition~\ref{prop:meta-comparison} and Theorem~\ref{thm:fisher-coercivity} give the following consequence.

\begin{corollary}[Linear-in-$\beta$ coercivity and baseline drift for metastable states]\label{cor:meta-coercivity}
Under the required upper bound $\zeta \le \zeta_*(\beta) = \exp[-C\beta^{2D+1}]$ in \eqref{meta:scale-values}, for all $\beta \ge 1$ and $\mathbf{k}, \mathbf{h} \in [-1,1]^m$,
\begin{equation}\label{meta:radial-result}
 \bigl\langle \mathbf{e},\, \mathbf{R}_{\bm{\sigma}}(\mathbf{k}) - \mathbf{R}_{\bm{\sigma}}(\mathbf{h})\bigr\rangle \ge \frac{\mu_\beta}{2} \norm{\mathbf{e}}_2^2, \qquad \mu_\beta = c e^{-C\beta},
\end{equation}
and the baseline residual at the true Hamiltonian $\bm{H} = \bm{H}(\mathbf{h})$ satisfies
\begin{equation}\label{meta:baseline-statement}
 \norm{\mathbf{R}_{\bm{\sigma}}(\mathbf{h})}_\infty \le b_0 := \zeta\,\poly\bigl(\beta, \log(1/\zeta)\bigr) = \tO(\zeta).
\end{equation}
\end{corollary}
\begin{proof}
Inequality \eqref{meta:radial-result} follows by subtracting the comparison error in Proposition~\ref{prop:meta-comparison} from the Fisher lower bound \eqref{meta:heterogeneous-frame}, using $\Delta_{\mathrm{meta}}\le\mu_\beta/2$. 

For the baseline bound \eqref{meta:baseline-statement}, grouping the generators associated with $\bm X_i,\bm Y_i,\bm Z_i$ at each site $i\in\Lambda$, the local metastability condition $\|\cL_{i,\bm{H}}[\bm{\sigma}]\|_1 \le \zeta$ and the generator locality bound \eqref{oracle:bath-locality} imply
\begin{equation}\label{eq:display-29}
 \bigl|\Tr\bigl[\bm{\sigma} \cL_{i,\bm{H}}^*[\bm{P}_j]\bigr]\bigr| \le \min\bigl\{C\zeta,\; \poly(\beta)e^{-c d(i,j)/\beta}\bigr\}.
\end{equation}
For $\zeta>0$, sum over sites within radius $O(\beta[\log(\beta+1)+\log(1/\zeta)])$ of $\supp(\bm P_j)$ and over the exponentially decaying outer shells to obtain \eqref{meta:baseline-statement}. 
\end{proof}

\paragraph{Optimality of the metastability accuracy floor.}
Following the indistinguishability argument of Wang, Ye, and Chen \citep[Appendix E.1]{WangYeChen2026MetastableLearning}, we use a low-temperature single-qubit example to exhibit an information-theoretic accuracy floor of order $\Omega(\zeta e^\beta/\beta)$ for learning from $\zeta$-locally metastable states.
Consider a single qubit with Hamiltonian $\bm H(h)=-h\bm Z$, reference parameter $h_0=1/2$, and common input state $\bm\sigma=\bm\rho_{\bm H(h_0)}$. Write $p_+$ and $p_-$ for its populations in the $\bm Z$ eigenstates $|0\rangle$ and $|1\rangle$, respectively; these have energies $-h$ and $+h$, and $p_-=p_+e^{-2\beta h_0}$.
For either generator associated with the single-site Pauli operators $\bm X$ and $\bm Y$, let $r_{\rm up}(h)$ and $r_{\rm down}(h)$ denote the transition rates $|0\rangle\to|1\rangle$ and $|1\rangle\to|0\rangle$, respectively. The two generators have identical rates. Each contributes $r_{\rm up}(h)p_+-r_{\rm down}(h)p_-$ to the derivative of the excited-state population; the generator associated with $\bm Z$ and the coherent term vanish on this diagonal state. Detailed balance gives $r_{\rm up}(h)=e^{-2\beta h}r_{\rm down}(h)$, and Lemma~\ref{gen:stationary} gives $r_{\rm down}(h)\le1$.
For the two competing parameters $h_\pm=h_0\pm d$ with $\beta d\le1/4$, summing the two generators and taking the trace norm of the diagonal drift gives the factor $4$ in
\begin{equation}\label{meta:one-qubit-drift}
 \norm{\cL_{\bm{H}(h_\pm)}[\bm{\sigma}]}_1 \le 4 r_{\rm down}(h_\pm)\bigl|p_- - p_+ e^{-2\beta h_\pm}\bigr| \le 8\beta d\, e^{-2\beta h_0 + 2\beta d}.
\end{equation}
Choosing $d = c \zeta e^\beta / \beta$ for a sufficiently small constant $c > 0$ ensures that the \emph{exact same} state $\bm{\sigma} = \bm{\rho}_{\bm{H}(h_0)}$ is $\zeta$-locally metastable for both $\bm{H}(h_+)$ and $\bm{H}(h_-)$. Because the input states are identical for both Hamiltonians, no algorithm—regardless of sample size or measurement power—can achieve coefficient error $\epsilon < d = \Omega(\zeta e^\beta / \beta)$.

\section{Local Stochastic Gradient Descent and Proofs of Main Theorems}
\label{sec:gibbs-learning}

Having established linear-in-$\beta$ coercivity for Gibbs states (Corollary~\ref{thm:coercivity}) and thermal metastable states (Corollary~\ref{cor:meta-coercivity}), we now measure a finite-time stationarity test by local Lindblad evolution. Its coercivity and locality yield both learning guarantees through the same projected stochastic gradient descent algorithm.

\subsection{Measuring stationarity by short-time evolution}
\label{oracle:section}

We measure the change of a local Pauli expectation under the candidate Gibbs sampler, following \citep[Lemma 4.3 and Section 6.3]{Bergamaschi2026Stationarity}. The evolution time depends on the target accuracy only through the logarithmic dependence of the patch radius. The additional estimate below shows that the finite-time test retains our $e^{-O(\beta)}$ radial coercivity, so it can be used directly in the stochastic gradient descent algorithm.

For each coordinate $j$, let $B_j(r)$ be the radius-$r$ neighborhood of $\supp(\bm P_j)$, and let $\bm K_{B_j(r)}$ retain only candidate interactions supported in this region. Writing $\cL_{i,\bm K}$ for the sum of the generators associated with $\bm X_i,\bm Y_i,\bm Z_i$, define the truncated instantaneous residual by
\begin{equation}\label{oracle:patch-generator}
 R^{(r)}_{\bm\tau,j}(\mathbf k):=-\Tr\!\left(\bm P_j\sum_{i\in B_j(r/2)}\cL_{i,\bm K_{B_j(r)}}[\bm\tau]\right).
\end{equation}
The summed generator acts only on $B_j(r)$. The quasi-locality bound \eqref{oracle:bath-locality}, summed over distance shells, gives
\begin{equation}\label{cutoff:bias}
 \sup_{\mathbf k\in[-1,1]^m}\|\mathbf R^{(r)}_{\bm\tau}(\mathbf k)-\mathbf R_{\bm\tau}(\mathbf k)\|_\infty\le\nu/4,
 \qquad r=C\beta\log(C\beta/\nu),
\end{equation}
for sufficiently large $C$; the spatial estimate is detailed in Appendix~\ref{app:residual-estimation}. For an evolution time $t_{\rm ev}>0$, the measured expectation map is
\begin{equation}\label{oracle:finite-time-map}
 F_{\nu,j}(\mathbf k):=
 -\frac{\Tr\!\left(\bm P_j e^{t_{\rm ev}\sum_{i\in B_j(r/2)}\cL_{i,\bm K_{B_j(r)}}}[\bm\tau]\right)-\Tr(\bm P_j\bm\tau)}{t_{\rm ev}}.
\end{equation}
Thus we only need to measure $\bm P_j$ on fresh copies, with or without the local evolution.

\begin{lemma}[Local short-time estimator and preservation of coercivity]\label{lem:oracle}
Suppose $\max_i\|\cL_{i,\bm H}[\bm\tau]\|_1\le\zeta$ and
$\langle\mathbf k-\mathbf h,\mathbf R_{\bm\tau}(\mathbf k)-\mathbf R_{\bm\tau}(\mathbf h)\rangle\ge2\mu\|\mathbf k-\mathbf h\|_2^2$, where $0<\mu\le1$. Take $0<\nu<1/2$, the radius in \eqref{cutoff:bias}, and
\begin{equation}\label{oracle:evolution-time}
 t_{\rm ev}=\frac{c\mu}{\beta^2r^{3D}},
\end{equation}
with $c>0$ sufficiently small. Then the map \eqref{oracle:finite-time-map} depends only on coefficients in $B_j(r)$ at coordinate $j$ and satisfies
\begin{equation}\label{cutoff:polynomial-oracle}
 \|\mathbf F_\nu(\mathbf x)-\mathbf F_\nu(\mathbf y)\|_2\le C\beta r^{2D}\|\mathbf x-\mathbf y\|_2,
 \qquad \mathbf x,\mathbf y\in[-1,1]^m.
\end{equation}
Moreover,
\begin{equation}\label{oracle:finite-time-coercivity}
\begin{aligned}
 \langle\mathbf k-\mathbf h,\mathbf F_\nu(\mathbf k)\rangle
 &\ge\mu\|\mathbf k-\mathbf h\|_2^2-(Cr^D\zeta+\nu)\|\mathbf k-\mathbf h\|_1,\\
 \|\mathbf F_\nu(\mathbf h)\|_\infty
 &\le Cr^D\zeta+\nu/4.
\end{aligned}
\end{equation}
Each coordinate has a single-copy unbiased estimator bounded by $2/t_{\rm ev}$. All coordinates can be measured in parallel using $O(r^D)$ colors of disjoint patches.
\end{lemma}

\begin{proof}[Proof sketch; details in Appendix~\ref{app:short-time-coercivity}]
The local generator has norm $O(\beta r^D)$, and its coefficient dependence is controlled by \eqref{app:patch-generator-bounds}. Duhamel's formula and channel contractivity give \eqref{cutoff:polynomial-oracle} and the stronger bound on the finite-time correction
\begin{equation}\label{oracle:finite-time-correction}
 \left\|(\mathbf F_\nu-\mathbf R^{(r)}_{\bm\tau})(\mathbf x)
       -(\mathbf F_\nu-\mathbf R^{(r)}_{\bm\tau})(\mathbf y)\right\|_2
 \le Ct_{\rm ev}\beta^2r^{3D}\|\mathbf x-\mathbf y\|_2.
\end{equation}
The choice \eqref{oracle:evolution-time} therefore loses at most half the original coercivity. At the true Hamiltonian, contractivity and local stationarity give $\|\mathbf F_\nu(\mathbf h)\|_\infty\le Cr^D\zeta+\nu/4$. Combining this with the spatial error \eqref{cutoff:bias} proves \eqref{oracle:finite-time-coercivity}.

For the estimator, choose a fair bit $b\in\{0,1\}$. If $b=1$, apply $e^{t_{\rm ev}\sum_{i\in B_j(r/2)}\cL_{i,\bm K_{B_j(r)}}}$; if $b=0$, leave the copy unchanged. Measure $\bm P_j$, obtaining $x\in\{-1,1\}$, and output
\begin{equation}\label{oracle:unbiased-outcome}
 Z_j=\frac{2(-1)^b x}{t_{\rm ev}},\qquad
 \mathbb E Z_j=F_{\nu,j}(\mathbf k),\qquad |Z_j|\le 2/t_{\rm ev}.
\end{equation}
Disjoint patches can be processed on the same copy. 
The CKG simulation algorithm implements the local channel to diamond-norm error $O(\nu t_{\rm ev})$ using $\poly(\beta,r,\log(1/(\nu t_{\rm ev})))$ gates, accounting for the $O(r^D)$ local generators in the patch \citep[Theorem I.2]{ChenKastoryanoGilyen2023}.
This contributes at most $\nu$ bias to a coordinate estimate. 
\end{proof}

\subsection{Local projected stochastic gradient descent under baseline drift}
\label{sgd:section}

We now use the finite-time map from Lemma~\ref{lem:oracle}. Set $L=C\beta r^{2D}$. The map is local and $L$-Lipschitz, has outcomes bounded by $2/t_{\rm ev}$, and satisfies \eqref{oracle:finite-time-coercivity}. 
Here $\zeta=0$ for Gibbs inputs. We use the measured residual as the direction in projected stochastic gradient descent.

\paragraph{The projected stochastic gradient update.}
Let $\mathsf P:\mathbb R^m\to[-1,1]^m$ be the Euclidean projection onto the coefficient cube, given coordinatewise by $[\mathsf P(\mathbf x)]_j=\min\{1,\max\{-1,x_j\}\}$.
Fix a total iteration count $T\ge1$, initialize $\mathbf{x}_0=\mathbf0$, and take $L\ge\mu$. At each iteration $t=0,\ldots,T-1$, use one batch of $S$ fresh copies of $\bm\tau$ per patch color to obtain the estimator
\begin{equation}\label{sgd:estimator-decomposition}
 \widehat{\mathbf F}(\mathbf x_t)
 =\mathbf F_\nu(\mathbf x_t)+\mathbf b_t+\boldsymbol\xi_t.
\end{equation}
Let $\mathcal F_t$ denote the history before this batch. Here $\mathbf b_t=\mathbb E[\widehat{\mathbf F}(\mathbf x_t)\mid\mathcal F_t]-\mathbf F_\nu(\mathbf x_t)$ is the bias due to imperfect implementation, with $\|\mathbf b_t\|_\infty\le\nu$, and $\boldsymbol\xi_t$ is the mean-zero sampling noise. For every set of coordinates $Q$, they satisfy
\begin{equation}\label{sgd:variance}
 \|\mathbf b_t\|_\infty\le\nu,
 \qquad \mathbb E[\boldsymbol\xi_{t,Q}\mid\mathcal F_t]=\mathbf0,
 \qquad
 \mathbb E[\|\boldsymbol\xi_{t,Q}\|_2^2\mid\mathcal F_t]
 \le\frac{4|Q|}{S t_{\rm ev}^2}.
\end{equation}
The last bound follows from the coordinatewise conditional variance bound $4/(S t_{\rm ev}^2)$ and linearity of expectation; it does not require independence across coordinates measured on the same copy. We then update
\begin{equation}\label{sgd:update}
 \mathbf x_{t+1}=\mathsf P\bigl(\mathbf x_t-\alpha\widehat{\mathbf F}(\mathbf x_t)\bigr),
 \qquad \alpha=\frac{\mu}{4L^2}.
\end{equation}
Each iteration consumes $O(r^D S)$ copies.

\paragraph{Local updates.}
Applying \eqref{sgd:variance} to all $m$ coefficients gives a total variance bound $4m/(S t_{\rm ev}^2)$, introducing a factor proportional to the system size. To avoid this factor, fix a target coefficient $h_j$, associated with the Pauli term $\bm P_j$, and define
\begin{equation}\label{sgd:coefficient-region}
 Q=\bigl\{\ell:\operatorname{dist}(\supp(\bm P_\ell),\supp(\bm P_j))\le(T+1)(r+R)\bigr\}.
\end{equation}
The cardinality of $Q$ satisfies
\begin{equation}\label{sgd:volume}
 |Q|\le V:=C[1+(T+1)(r+R)]^D.
\end{equation}
Define a full-vector comparison process $\mathbf x_t^{(Q)}\in[-1,1]^m$, initialized to zero on $Q$ and fixed to the true coefficient vector $\mathbf h$ outside $Q$ at every iteration. Update only its coordinates in $Q$ using \eqref{sgd:update} with residual $\mathbf F_\nu(\mathbf x_t^{(Q)})_Q$ and the same realized bias and noise $(\mathbf b_{t,Q},\boldsymbol\xi_{t,Q})$ as the actual algorithm. This process is only used in the analysis and is adapted to $\mathcal F_t$.

Each update at coordinate $\ell$ depends only on coefficients whose supports lie in $B_\ell(r)$. Since $\operatorname{diam}(\supp(\bm P_\ell))\le R$, every such coefficient index $q$ satisfies
\begin{equation}\label{sgd:dependency-distance}
 \operatorname{dist}(\supp(\bm P_q),\supp(\bm P_j))
 \le r+R+\operatorname{dist}(\supp(\bm P_\ell),\supp(\bm P_j)).
\end{equation}
Thus the agreement radius shrinks by at most $r+R$ per iteration. Induction shows that the two processes agree at time $t$ on all coefficient indices $\ell$ satisfying $\operatorname{dist}(\supp(\bm P_\ell),\supp(\bm P_j))\le(T-t+1)(r+R)$, and hence
\begin{equation}\label{eq:display-21}
 x_{T,j}=x_{T,j}^{(Q)}.
\end{equation}

\begin{lemma}[Single-run mean-square error for local stochastic gradient descent]\label{lem:local-sgd-mse}
Suppose $\mathbf F_\nu$ is radius-$r$ local, $L$-Lipschitz, and satisfies \eqref{oracle:finite-time-coercivity}, with $0<\mu\le L$. For an iteration count $T\ge1$ and $S\ge1$ fresh copies per patch color per iteration, set $V=C[1+(T+1)(r+R)]^D$ as in \eqref{sgd:volume}. The output of \eqref{sgd:update} satisfies, for every coordinate $j$,
\begin{equation}\label{sgd:mse}
 \mathbb E|x_{T,j}-h_j|^2
 \le V e^{-\alpha\mu T}
 +\frac{16V(Cr^D\zeta+\nu)^2}{\mu^2}
 +\frac{4\alpha V}{\mu S t_{\rm ev}^2}.
\end{equation}
\end{lemma}
\begin{proof}
Fix $Q$. Since $\mathbf x_t^{(Q)}-\mathbf h$ is supported on $Q$, \eqref{oracle:finite-time-coercivity}, Lipschitz continuity, and $\|\mathbf b_t\|_\infty\le\nu$ give
\begin{equation}\label{sgd:direction-bounds}
\begin{aligned}
 \langle\mathbf x_{t,Q}^{(Q)}-\mathbf h_Q,\mathbf F_\nu(\mathbf x_t^{(Q)})_Q+\mathbf b_{t,Q}\rangle
 &\ge\mu\|\mathbf x_{t,Q}^{(Q)}-\mathbf h_Q\|_2^2
 -(Cr^D\zeta+2\nu)\sqrt{|Q|}\|\mathbf x_{t,Q}^{(Q)}-\mathbf h_Q\|_2,\\
 \|\mathbf F_\nu(\mathbf x_t^{(Q)})_Q+\mathbf b_{t,Q}\|_2^2
 &\le2L^2\|\mathbf x_{t,Q}^{(Q)}-\mathbf h_Q\|_2^2+2|Q|(Cr^D\zeta+2\nu)^2.
\end{aligned}
\end{equation}
Using the update \eqref{sgd:update}, projection is nonexpansive and fixes $\mathbf h_Q$. Conditional on $\mathcal F_t$, both $\mathbf x_t^{(Q)}$ and $\mathbf F_\nu(\mathbf x_t^{(Q)})_Q+\mathbf b_{t,Q}$ are fixed and the cross term with $\boldsymbol\xi_{t,Q}$ has zero expectation by \eqref{sgd:variance}. Expanding the squared distance gives
\begin{equation}\label{sgd:projection-expansion}
\begin{aligned}
 \mathbb E[\|\mathbf x_{t+1,Q}^{(Q)}-\mathbf h_Q\|_2^2\mid\mathcal F_t]
 &\le\|\mathbf x_{t,Q}^{(Q)}-\mathbf h_Q\|_2^2
 -2\alpha\langle\mathbf x_{t,Q}^{(Q)}-\mathbf h_Q,\mathbf F_\nu(\mathbf x_t^{(Q)})_Q+\mathbf b_{t,Q}\rangle\\
 &\quad+\alpha^2\|\mathbf F_\nu(\mathbf x_t^{(Q)})_Q+\mathbf b_{t,Q}\|_2^2
 +\frac{4\alpha^2|Q|}{S t_{\rm ev}^2}.
\end{aligned}
\end{equation}
Substituting the two bounds in \eqref{sgd:direction-bounds} yields
\begin{equation}\label{sgd:projection}
\begin{aligned}
 \mathbb E[\|\mathbf x_{t+1,Q}^{(Q)}-\mathbf h_Q\|_2^2\mid\mathcal F_t]
 &\le(1-2\alpha\mu+2\alpha^2L^2)\|\mathbf x_{t,Q}^{(Q)}-\mathbf h_Q\|_2^2\\
 &\quad+2\alpha(Cr^D\zeta+2\nu)\sqrt{|Q|}\|\mathbf x_{t,Q}^{(Q)}-\mathbf h_Q\|_2
 +2\alpha^2|Q|(Cr^D\zeta+2\nu)^2+\frac{4\alpha^2|Q|}{S t_{\rm ev}^2}.
\end{aligned}
\end{equation}
Using $2ab\le(\mu/2)a^2+(2/\mu)b^2$, $\alpha=\mu/(4L^2)$, and $\alpha\mu\le1/4$ gives the one-step contraction
\begin{equation}\label{sgd:recurrence}
 \mathbb E[\|\mathbf x_{t+1,Q}^{(Q)}-\mathbf h_Q\|_2^2\mid\mathcal F_t]
 \le(1-\alpha\mu)\|\mathbf x_{t,Q}^{(Q)}-\mathbf h_Q\|_2^2
 +\frac{16\alpha|Q|(Cr^D\zeta+\nu)^2}{\mu}
 +\frac{4\alpha^2|Q|}{S t_{\rm ev}^2}.
\end{equation}
Iterating from $\|\mathbf x_{0,Q}^{(Q)}-\mathbf h_Q\|_2^2\le|Q|\le V$ and using \eqref{eq:display-21} proves \eqref{sgd:mse}.
\end{proof}

\subsection{Proofs of Theorems~\ref{thm:main} and \ref{meta:theorem}}
\label{completion:section}

Both main theorems now follow from Lemma~\ref{lem:local-sgd-mse} by using the corresponding coercivity constant $\mu$ and local metastability bound $\zeta$.

\begin{proof}[Proof of Theorems~\ref{thm:main} and \ref{meta:theorem}]
Corollaries~\ref{thm:coercivity} and \ref{cor:meta-coercivity}, followed by Lemma~\ref{lem:oracle}, give \eqref{oracle:finite-time-coercivity} with
\begin{equation}\label{sgd:state-parameters}
 \mu=\begin{cases}
  \mu_\beta/2,&\bm\tau=\bm\rho_{\bm H},\\
  \mu_\beta/4,&\bm\tau=\bm\sigma,
 \end{cases}
 \qquad \mu_\beta=ce^{-C\beta}.
\end{equation}
In Lemma~\ref{lem:local-sgd-mse}, we choose the oracle truncation tolerance $\nu$, iteration count $T$, and per-color batch size $S$ as
\begin{equation}\label{sgd:counts}
 \nu = \Theta\!\left(\frac{\mu \epsilon}{\sqrt{V}}\right), \qquad T = \Theta\!\left(\frac{L^2}{\mu^2}\log\frac{V}{\epsilon}\right), \qquad S = \left\lceil C\frac{4\alpha V}{\mu \epsilon^2 t_{\rm ev}^2}\right\rceil.
\end{equation}
Since $r,L=\poly(\beta,\log(1/\nu))$ and $t_{\rm ev}^{-1}=\mu^{-1}\poly(\beta,\log(1/\nu))$ by Lemma~\ref{lem:oracle}, with $\mu=\Theta(e^{-C\beta})$, the bound on the size of the local region in \eqref{sgd:volume} satisfies $V = O((T(r+R))^D) = O((T r)^D) = \tO(\mu^{-2D}) = \tO(e^{O(\beta)})$, and hence $\log(1/\nu) = O(\beta + \log(1/\epsilon))$. For Gibbs states ($\zeta = 0$) and for metastable states under the accuracy floor $\epsilon \ge \tO(e^{O(\beta)}\zeta)$ in \eqref{meta:floor-explicit} (which ensures $Cr^D\zeta \le \frac{\mu\epsilon}{64\sqrt{V}}$), each of the three terms on the right-hand side of \eqref{sgd:mse} is at most $\epsilon^2/48$. Thus $\mathbb{E}|x_{T,j} - h_j|^2 \le \epsilon^2/16$, and Markov's inequality yields $\Pr(|x_{T,j} - h_j| > \epsilon) \le 1/16$ for each coordinate $j \in \{1, \dots, m\}$.

To boost the success probability to $1 - \delta$ across all $m \le qn$ coordinates, we run $K_{\rm rep} = \Theta(\log(n/\delta))$ independent trajectories and output the coordinatewise median $\widehat{h}_j = \operatorname{median}(x_{T,j}^{(1)}, \dots, x_{T,j}^{(K_{\rm rep})})$. By Hoeffding's inequality and a union bound over $j \in \{1, \dots, m\}$, $\Pr\{\|\widehat{\mathbf{h}} - \mathbf{h}\|_\infty \le \epsilon\} \ge 1 - \delta$.
The total number of consumed state copies across all $K_{\rm rep}$ runs, $T$ iterations, and $O(r^D)$ patch colors is
\begin{equation}\label{sgd:copies}
 N = O\bigl(K_{\rm rep} T r^D S\bigr) = \tO\!\left(\mu^{-(2D+4)} \epsilon^{-2} \log(n/\delta)\right) = \tO\!\left(e^{O(\beta)} \epsilon^{-2} \log(n/\delta)\right).
\end{equation}
Inserting $\log(1/\nu) = O(\beta + \log(1/\epsilon))$ into \eqref{cutoff:bias} gives the coherent measurement radius $r_{\rm meas} = O(\beta^2 + \beta\log(1/\epsilon))$ in \eqref{thm:radius}, and multiplying $N$ by the $O(n)\poly(\beta,\log(1/\epsilon))$ per-copy cost of local channel simulation and classical processing gives \eqref{thm:actual-work} and \eqref{meta:copies}.
\end{proof}

\paragraph{Unified measurement oracle and learning algorithm.}
Algorithms~\ref{alg:residual} and \ref{alg:gibbs} summarize the single-copy measurement subroutine and the projected stochastic gradient learner for both Gibbs and thermal metastable inputs $\bm{\tau} \in \{\bm{\rho}_{\bm{H}}, \bm{\sigma}\}$.

\begin{algorithm}[H]
\caption{Stationarity measurement by short-time local evolution}\label{alg:residual}
\begin{algorithmic}[1]
\Require Candidate $\mathbf k$, fresh copies of $\bm\tau$, tolerance $\nu$, evolution time $t_{\rm ev}$ from \eqref{oracle:evolution-time}, batch size $S$.
\Ensure Empirical residual vector $\widehat{\mathbf{F}}(\mathbf{k})$.
\State Construct the radius-$r$ patches and their $O(r^D)$ disjoint-patch colors using Lemma~\ref{lem:oracle}.
\For{each patch color}
 \For{$s=1,\ldots,S$}
  \State Take a fresh copy of $\bm{\tau}$.
  \State For each coordinate $j$ in this color, draw an independent fair bit $b_j$.
  \State If $b_j=1$, apply $e^{t_{\rm ev}\sum_{i\in B_j(r/2)}\cL_{i,\bm K_{B_j(r)}}}$ on its patch; otherwise leave that patch unchanged.
  \State Measure $\bm P_j$, obtaining $x_j\in\{-1,1\}$, and record $2(-1)^{b_j}x_j/t_{\rm ev}$.
 \EndFor
 \State Average the $S$ weighted outcomes for each coordinate in this color.
\EndFor
\State \Return the vector of coordinate averages $\widehat{\mathbf{F}}(\mathbf{k})$.
\end{algorithmic}
\end{algorithm}

\begin{algorithm}[H]
\caption{Unified Hamiltonian learning from Gibbs or thermal metastable samples}\label{alg:gibbs}\label{alg:metastable}
\begin{algorithmic}[1]
\Require Known Pauli dictionary and lattice, inverse temperature $\beta$, fresh copies of $\bm{\tau} \in \{\bm{\rho}_{\bm H}, \bm{\sigma}\}$, accuracy $\epsilon$ (satisfying \eqref{meta:floor-explicit} when $\bm{\tau} = \bm{\sigma}$), failure probability $\delta$.
\State Set $\mu=\mu_\beta/2$ for Gibbs inputs or $\mu=\mu_\beta/4$ for metastable inputs. Choose $\nu,T,S$ by \eqref{sgd:counts}, $r,t_{\rm ev}$ by \eqref{cutoff:bias} and \eqref{oracle:evolution-time}, $L=C\beta r^{2D}$, $\alpha=\mu/(4L^2)$, and an odd $K_{\rm rep}=\Theta(\log(n/\delta))$.
\For{$q=1,\ldots,K_{\rm rep}$}
 \State Initialize $\mathbf{x}_0=\mathbf{0}$.
 \For{$t=0,\ldots,T-1$}
  \State Call Algorithm~\ref{alg:residual} at $\mathbf{x}_t$ using fresh copies of $\bm{\tau}$.
  \State Set $\mathbf{x}_{t+1}=\mathsf{P}(\mathbf{x}_t-\alpha\widehat{\mathbf{F}}(\mathbf{x}_t))$.
 \EndFor
 \State Store $\mathbf{x}_T^{(q)}=\mathbf{x}_T$.
\EndFor
\State \Return $\widehat h_j=\operatorname{median}_{q}x_{T,j}^{(q)}$ for every $j$.
\end{algorithmic}
\end{algorithm}

\appendix
\section{Pauli Estimates and Locality of Stationarity Tests}\label{app:pauli-identities-section}

\subsection{A lower bound from Pauli commutators}\label{app:pauli-signal}
For completeness, we prove \eqref{lin:paulisignal}, the summed qubit version of the identity stated in \citep[Eq.~(60)]{Bergamaschi2026Stationarity}. Fix a single-site Pauli operator $\bm A_a$. Each dictionary Pauli $\bm P_j$ either commutes or anticommutes with $\bm A_a$; in the latter case $[\bm A_a,\bm P_j]=2\bm A_a\bm P_j$. Left multiplication by $\bm A_a$ preserves Hilbert--Schmidt inner products, so distinct nonzero commutators are orthogonal:
\begin{equation}\label{app:commutator-orthogonality}
 2^{-n}\Tr\bigl((\bm A_a\bm P_j)^*\bm A_a\bm P_k\bigr)
 =2^{-n}\Tr(\bm P_j\bm P_k)=\delta_{jk}.
\end{equation}
Expanding $\bm E=\sum_j e_j\bm P_j$ therefore gives
\begin{equation}\label{app:single-seed-signal}
 2^{-n}\Tr([\bm A_a,\bm E]^*[\bm A_a,\bm E])
 =4\sum_{j:\,\{\bm A_a,\bm P_j\}=0}e_j^2.
\end{equation}
At each site in $\supp(\bm P_j)$, exactly two of the three single-site Paulis anticommute with $\bm P_j$; outside its support all three commute. Summing over all indices $a$ yields
\begin{equation}\label{app:summed-pauli-signal}
 2^{-n}\sum_a\Tr([\bm A_a,\bm E]^*[\bm A_a,\bm E])
 =8\sum_j|\supp(\bm P_j)|e_j^2
 \ge8\norm{\mathbf e}_2^2,
\end{equation}
where the last inequality uses that every dictionary Pauli is nonidentity.

\subsection{The local Pauli-twirl estimate}\label{app:pauli-twirl}
We prove \eqref{eq:display-27} directly. Let $\bm\tau_{a,S_a}=\Tr_{S_a^c}\bm\tau_a$ and $d_a=2^{|S_a|}$. All operators inside $\Tr_{S_a}$ below are restricted to $S_a$. Let $\mathcal P(S_a)$ consist of the $d_a^2$ Pauli strings without additional phases. Pauli orthogonality and the Pauli-twirl identity give, for any operator $\bm Z$ on $S_a$,
\begin{equation}\label{app:pauli-identities}
 \sum_{\bm P\in\mathcal P(S_a)}|\Tr_{S_a}(\bm P\bm Z)|^2
 =d_a\Tr_{S_a}(\bm Z^*\bm Z),\qquad
 \sum_{\bm Q\in\mathcal P(S_a)}\bm Q\bm Z\bm Q
 =d_a\Tr_{S_a}(\bm Z)\bm I_{S_a}.
\end{equation}
Applying the first identity with $\bm Z=\bm\tau_{a,S_a}\bm Q[\bm A_a,\bm E]$ and then the second to $\bm\tau_{a,S_a}^2$ yields
\begin{equation}\label{app:pauli-twirl-square}
\begin{aligned}
 \sum_{\bm P,\bm Q\in\mathcal P(S_a)}
 |\Tr_{S_a}(\bm P\bm\tau_{a,S_a}\bm Q[\bm A_a,\bm E])|^2&=d_a\sum_{\bm Q\in\mathcal P(S_a)}
 \Tr_{S_a}([\bm A_a,\bm E]^*\bm Q\bm\tau_{a,S_a}^2\bm Q[\bm A_a,\bm E])\\
 &=d_a^2\Tr_{S_a}(\bm\tau_{a,S_a}^2)
 \Tr_{S_a}([\bm A_a,\bm E]^*[\bm A_a,\bm E]).
\end{aligned}
\end{equation}
There are $d_a^4$ pairs and $\Tr_{S_a}(\bm\tau_{a,S_a}^2)\ge d_a^{-1}$. Hence some pair $\bm P_a,\bm Q_a$ satisfies
\begin{equation}\label{app:pauli-pair}
 |\Tr_{S_a}(\bm P_a\bm\tau_{a,S_a}\bm Q_a[\bm A_a,\bm E])|
 \ge\frac1{d_a}\left(\frac1{d_a}\Tr_{S_a}([\bm A_a,\bm E]^*[\bm A_a,\bm E])\right)^{1/2}.
\end{equation}
Since $[\bm A_a,\bm E]$, $\bm P_a$, and $\bm Q_a$ are supported on $S_a$, partial trace and normalization give
\begin{equation}\label{app:pauli-local-global}
\begin{aligned}
 \Tr_{S_a}(\bm P_a\bm\tau_{a,S_a}\bm Q_a[\bm A_a,\bm E])
 &=\Tr(\bm P_a\bm\tau_a\bm Q_a[\bm A_a,\bm E]),\\
 d_a^{-1}\Tr_{S_a}([\bm A_a,\bm E]^*[\bm A_a,\bm E])
 &=2^{-n}\Tr([\bm A_a,\bm E]^*[\bm A_a,\bm E]).
\end{aligned}
\end{equation}
Here the full trace is on the $n$-qubit system, with local operators extended by the identity; on a smaller reference patch the same identity uses its number of qubits. Combining \eqref{app:pauli-pair}--\eqref{app:pauli-local-global} with $d_a\le d_*$ proves \eqref{eq:display-27}.

\subsection{Localization and regularization of stationarity tests}\label{app:joint-derivative}

\begin{lemma}[Uniform second-derivative bound]\label{lem:uniform-second-derivative}
For $\beta\ge1$ and $\bm K=\sum_j k_j\bm P_j$ with $\mathbf k\in[-2,2]^m$, the single-bath generator satisfies
\begin{equation}\label{app:uniform-second-derivative}
 \bigl\|\partial_{k_\ell}^2\cL_{\bm A_a,\bm K}^*[\bm P_j]\bigr\|_\infty
 \le M_\beta=\poly(\beta),
\end{equation}
uniformly in $a,j,\ell$ and the system size, with constants depending only on $(D,R,w,q)$.
\end{lemma}
\begin{proof}
For a coefficient-independent operator $\bm X$, Duhamel's formula and $\|\bm P_\ell\|_\infty=1$ give
\begin{equation}\label{app:duhamel-second-derivative}
\begin{aligned}
 \partial_{k_\ell}\mathcal U_u^{\bm K}[\bm X]
 &=i\int_0^u\mathcal U_{u-v}^{\bm K}
       [[\bm P_\ell,\mathcal U_v^{\bm K}[\bm X]]]\,dv,\\
 \|\partial_{k_\ell}^s\mathcal U_u^{\bm K}[\bm X]\|_\infty
 &\le(2|u|)^s\|\bm X\|_\infty,\qquad s=0,1,2.
\end{aligned}
\end{equation}
Here the second derivative is bounded by iterating the first identity. Since $f_\beta$ is independent of $k_\ell$, differentiating \eqref{gen:jump} gives
\begin{equation}\label{app:differentiated-jump-filter}
 \partial_{k_\ell}^s\bm J_{\bm A_a,\bm K}(\omega)
 =\frac{1}{\sqrt{2\pi}}\int_{\mathbb R}f_\beta(u)e^{-i\omega u}
 \partial_{k_\ell}^s\mathcal U_u^{\bm K}[\bm A_a]\,du.
\end{equation}
Plancherel and $0\le\gamma\le1$ therefore imply
\begin{equation}\label{app:jump-derivative-moments}
\begin{aligned}
 \left\|\int_{\mathbb R}\gamma(\omega)
 (\partial_{k_\ell}^s\bm J_{\bm A_a,\bm K}(\omega))^*(\partial_{k_\ell}^s\bm J_{\bm A_a,\bm K}(\omega))\,d\omega\right\|_\infty
 &\le\left\|\int_{\mathbb R}
 (\partial_{k_\ell}^s\bm J_{\bm A_a,\bm K}(\omega))^*(\partial_{k_\ell}^s\bm J_{\bm A_a,\bm K}(\omega))\,d\omega\right\|_\infty\\
 &=\left\|\int_{\mathbb R}|f_\beta(u)|^2
 (\partial_{k_\ell}^s\mathcal U_u^{\bm K}[\bm A_a])^*
 (\partial_{k_\ell}^s\mathcal U_u^{\bm K}[\bm A_a])\,du\right\|_\infty\\
 &\le\int_{\mathbb R}|f_\beta(u)|^2
 \|\partial_{k_\ell}^s\mathcal U_u^{\bm K}[\bm A_a]\|_\infty^2\,du
\le C_s\beta^{2s},\qquad s=0,1,2.
\end{aligned}
\end{equation}
Consequently, operator Cauchy--Schwarz gives, for $p,q\le2$ and any observable $\bm O$,
\begin{equation}\label{app:jump-cross-derivative-bound}
 \left\|\int_{\mathbb R}\gamma(\omega)
 (\partial_{k_\ell}^p\bm J_{\bm A_a,\bm K}(\omega))^*\bm O(\partial_{k_\ell}^q\bm J_{\bm A_a,\bm K}(\omega))\,d\omega\right\|_\infty
 \le C\beta^{p+q}\|\bm O\|_\infty.
\end{equation}
Let $\mathcal D_{a,\bm K}^*$ denote the dissipative part of \eqref{eq:display-3}. The product rule, with $p+q=2$ in \eqref{app:jump-cross-derivative-bound}, yields
\begin{equation}\label{app:dissipative-second-derivative}
 \|\partial_{k_\ell}^2\mathcal D_{a,\bm K}^*[\bm O]\|_\infty
 \le C\beta^2\|\bm O\|_\infty.
\end{equation}
For the coherent term, use $[\bm K,\bm J_{\bm A_a,\bm K}(\omega)]=\bm J_{[\bm K,\bm A_a],\bm K}(\omega)$. Since $[\bm K,\bm A_a]$ and its coefficient derivatives are uniformly bounded, the same Gaussian moment estimates apply to $\bm J_{[\bm K,\bm A_a],\bm K}(\omega)$. 
Expanding the commutator and then applying operator Cauchy--Schwarz gives
\begin{equation}\label{app:commutator-derivative-bound}
\begin{aligned}
 [\bm K,\bm Q_{a,\bm K}]
 &=\int_{\mathbb R}\gamma(\omega)\bigl(
 \bm J_{\bm A_a,\bm K}(\omega)^*\bm J_{[\bm K,\bm A_a],\bm K}(\omega)
 -\bm J_{[\bm K,\bm A_a],\bm K}(\omega)^*\bm J_{\bm A_a,\bm K}(\omega)\bigr)\,d\omega,\\
 \|\partial_{k_\ell}^s[\bm K,\bm Q_{a,\bm K}]\|_\infty
 &\le C'\beta^s,\qquad s=0,1,2.
\end{aligned}
\end{equation}
To handle the kernel at $t=0$, write the difference of evolutions as an integral of a commutator. Duhamel's formula and the product rule then give
\begin{equation}\label{app:coherent-derivative-cancellation}
\begin{aligned}
 \mathcal U_t^{\bm K}[\bm Q_{a,\bm K}]-\mathcal U_{-t}^{\bm K}[\bm Q_{a,\bm K}]
 &=i\int_{-t}^t\mathcal U_v^{\bm K}[[\bm K,\bm Q_{a,\bm K}]]\,dv,\\
 \left\|\partial_{k_\ell}^2\bigl(\mathcal U_t^{\bm K}[\bm Q_{a,\bm K}]
       -\mathcal U_{-t}^{\bm K}[\bm Q_{a,\bm K}]\bigr)\right\|_\infty
 &\le C\int_{-t}^t(\beta+|v|)^2\,dv
 \le Ct(\beta+t)^2.
\end{aligned}
\end{equation}
This factor of $t$ cancels the kernel's singularity. The same integrable bound holds for the lower derivatives, justifying differentiation under the integral, and rescaling $t=\beta x$ yields
\begin{equation}\label{app:coherent-second-derivative-bound}
 \|\partial_{k_\ell}^2\bm B_{a,\bm K}\|_\infty
 \le C\int_0^\infty\frac{t(\beta+t)^2}{\beta\sinh(2\pi t/\beta)}\,dt
 =C\beta^3\int_0^\infty\frac{x(1+x)^2}{\sinh(2\pi x)}\,dx
 \le C'\beta^3.
\end{equation}
Combining \eqref{app:dissipative-second-derivative} and \eqref{app:coherent-second-derivative-bound} with $\|[\bm X,\bm P_j]\|_\infty\le2\|\bm X\|_\infty$ proves \eqref{app:uniform-second-derivative}.
\end{proof}

\begin{proof}[Proof of Lemma~\ref{lem:meta-localization}]
We first prove the joint derivative decay \eqref{meta:joint-derivative}, and then telescope the test difference in \eqref{meta:localized-regularized-tests} into three steps: localizing $\bm{E} \to \bm{U}_a$, truncating $\bm{H} \to \bm{H}_a$, and regularizing $\bm{\sigma} \to \bm{\sigma}_a^{(\lambda)}$.

\paragraph{1. Joint derivative decay \eqref{meta:joint-derivative}.}
To convert the zeroth-order spatial decay \eqref{oracle:bath-locality} into a bound on $\partial_{k_\ell}\cL_{\bm{A}_a,\bm{K}}^*[\bm{P}_j]$, we interpolate with the second derivative $\partial_{k_\ell}^2\cL_{\bm{A}_a,\bm{K}}^*[\bm{P}_j]$, which is uniformly bounded by $M_\beta = \poly(\beta)$ for $\mathbf{k}\in[-2,2]^m$ by Lemma~\ref{lem:uniform-second-derivative}.

We now interpolate. For any twice differentiable function $f:[0,1]\to\mathcal{B}(\mathcal{H})$ with $\|f(u)\|_\infty\le\varepsilon$ and $\|f''(u)\|_\infty\le M$, Taylor's theorem gives
\begin{equation}\label{app:taylor-derivative-bound}
 f(h)-f(0)=hf'(0)+\int_0^h(h-u)f''(u)\,du,
 \qquad
 \|f'(0)\|_\infty\le\frac{2\varepsilon}{h}+\frac{Mh}{2}
\end{equation}
for $0<h\le1$. Optimizing with $h=\min\{1,2\sqrt{\varepsilon/M}\}$ yields
\begin{equation}\label{app:derivative-interpolation}
 \|f'(0)\|_\infty\le C\bigl(\sqrt{M\varepsilon}+\varepsilon\bigr).
\end{equation}
For any $\mathbf{k}\in[-2,2]^m$, choose a sign $\theta\in\{-1,1\}$ so that $\mathbf{k}+\theta u\mathbf{e}_\ell\in[-2,2]^m$ for $u\in[0,1]$. We apply \eqref{app:derivative-interpolation} with $M=M_\beta$ to two functions along this segment:
\begin{itemize}
 \item \emph{Observable distance $d(a,j)$:} Take $r_1=d(a,j)/2$. For $d(a,j)>2R$, the truncated generator $\cL_{\bm{A}_a,(\bm{K}+\theta u\bm{P}_\ell)_{a,r_1}}^*$ is supported outside $\supp(\bm{P}_j)$ and annihilates $\bm{P}_j$, so \eqref{oracle:bath-locality} gives
 \begin{equation}\label{app:truncated-observable-bound}
  \|f_1(u)\|_\infty := \|\cL_{\bm A_a,\bm K+\theta u\bm P_\ell}^*[\bm P_j]\|_\infty
  \le\poly(\beta)e^{-c d(a,j)/\beta},\qquad u\in[0,1].
 \end{equation}
 Applying \eqref{app:derivative-interpolation} to $f_1(u)$ gives $\|\partial_{k_\ell}\cL_{\bm A_a,\bm K}^*[\bm P_j]\|_\infty \le \poly(\beta)e^{-c_1 d(a,j)/\beta}$.
 \item \emph{Perturbation distance $d(a,\ell)$:} Take $r_2=d(a,\ell)/2$. For $d(a,\ell)>2R$, $\bm{K}_{a,r_2}$ excludes $\bm{P}_\ell$ and is independent of $k_\ell$. Thus
 \begin{equation}\label{app:coefficient-locality-difference}
  f_2(u):=\cL_{\bm A_a,\bm K+\theta u\bm P_\ell}^*[\bm P_j]
        -\cL_{\bm A_a,\bm K_{a,r_2}}^*[\bm P_j]
 \end{equation}
 satisfies $f_2'(0)=\theta\partial_{k_\ell}\cL_{\bm A_a,\bm K}^*[\bm P_j]$, $\|f_2''(u)\|_\infty\le M_\beta$, and $\|f_2(u)\|_\infty\le\poly(\beta)e^{-c d(a,\ell)/\beta}$ by \eqref{oracle:bath-locality}. Applying \eqref{app:derivative-interpolation} gives $\|\partial_{k_\ell}\cL_{\bm A_a,\bm K}^*[\bm P_j]\|_\infty \le \poly(\beta)e^{-c_2 d(a,\ell)/\beta}$.
\end{itemize}
Taking the geometric mean of the two bounds proves \eqref{meta:joint-derivative}.

\paragraph{2. Localization and regularization of tests \eqref{meta:localized-regularized-tests}.}
For any perturbation $\bm{U}$, background Hamiltonian $\bm{G}$, and state $\bm{\tau}$, introduce the shorthand
\begin{equation}\label{app:test-increment}
 \Delta_a(\bm{U}, \bm{G}; \bm{\tau}) := T_a(\bm{U}, \bm{G}+\bm{U}; \bm{\tau}) - T_a(\bm{U}, \bm{G}; \bm{\tau}) = \Tr\bigl[\bm{\tau}\bigl(\cL_{\bm{A}_a,\bm{G}+\bm{U}}^* - \cL_{\bm{A}_a,\bm{G}}^*\bigr)[\bm{U}]\bigr].
\end{equation}
For $\bm U=\sum_j u_j\bm P_j$, the fundamental theorem of calculus gives the common quadratic representation
\begin{equation}\label{app:test-increment-quadratic}
 \Delta_a(\bm U,\bm G;\bm\tau)
 =\sum_{j,p}u_ju_p\int_0^1
 \Tr\bigl[\bm\tau\,\partial_{k_p}\cL_{\bm A_a,\bm G+t\bm U}^*[\bm P_j]\bigr]dt.
\end{equation}
Since $\bm{K} = \bm{H} + \bm{E}$ and $\bm{K}_a = \bm{H}_a + \bm{U}_a$, the summand in \eqref{meta:localized-regularized-tests} is $\Delta_a(\bm{E}, \bm{H}; \bm{\sigma}) - \Delta_a(\bm{U}_a, \bm{H}_a; \bm{\sigma}_a^{(\lambda)})$, which we telescope into three steps:
\begin{equation}\label{app:test-three-step-decomposition}
\begin{aligned}
 \Delta_a(\bm{E}, \bm{H}; \bm{\sigma}) - \Delta_a(\bm{U}_a, \bm{H}_a; \bm{\sigma}_a^{(\lambda)})
 &= \bigl[\Delta_a(\bm{E}, \bm{H}; \bm{\sigma}) - \Delta_a(\bm{U}_a, \bm{H}; \bm{\sigma})\bigr]\\
 &\quad + \bigl[\Delta_a(\bm{U}_a, \bm{H}; \bm{\sigma}) - \Delta_a(\bm{U}_a, \bm{H}_a; \bm{\sigma})\bigr]\\
 &\quad + \bigl[\Delta_a(\bm{U}_a, \bm{H}_a; \bm{\sigma}) - \Delta_a(\bm{U}_a, \bm{H}_a; \bm{\sigma}_a^{(\lambda)})\bigr].
\end{aligned}
\end{equation}
For Steps (ii) and (iii), we will use the same coefficient estimate. Since $|I_a|\le C(1+\ell)^D$ and each index $j$ belongs to at most $C(1+\ell)^D$ sets $I_a$, Cauchy--Schwarz gives
\begin{equation}\label{app:coefficient-overlap}
 \sum_a\left(\sum_{j\in I_a}|e_j|\right)^2
 \le \left(\sup_a|I_a|\right)
      \left(\sup_j\#\{a:j\in I_a\}\right)\norm{\mathbf e}_2^2
 \le C(1+\ell)^{2D}\norm{\mathbf e}_2^2.
\end{equation}
We bound the three terms by spatial decay in $\ell$, spatial decay in $L$, and the trace-norm regularization error, respectively:
\begin{itemize}
 \item \emph{Step (i): Localizing $\bm{E}\to\bm{U}_a$.} Adding and subtracting $\Tr[\bm{\sigma}(\cL_{\bm{A}_a,\bm{K}}^* - \cL_{\bm{A}_a,\bm{H}}^*)[\bm{U}_a]]$ splits the first difference into two parts, which we expand by integrating $\partial_{k_p}\cL_{\bm{A}_a,\cdot}^*[\bm{P}_j]$ along the segments $\bm{H}(s) = \bm{H}+s\bm{E}$ and $\widetilde{\bm{H}}(s) = \bm{H}+\bm{U}_a+s(\bm{E}-\bm{U}_a)$ for $s\in[0,1]$:
 \begin{equation}\label{app:error-localization-expansion}
 \begin{aligned}
  &\Delta_a(\bm{E}, \bm{H}; \bm{\sigma}) - \Delta_a(\bm{U}_a, \bm{H}; \bm{\sigma})\\
  &\quad = \Tr\bigl[\bm{\sigma}(\cL_{\bm{A}_a,\bm{K}}^*-\cL_{\bm{A}_a,\bm{H}}^*)[\bm{E}-\bm{U}_a]\bigr]
   + \Tr\bigl[\bm{\sigma}(\cL_{\bm{A}_a,\bm{K}}^*-\cL_{\bm{A}_a,\bm{H}+\bm{U}_a}^*)[\bm{U}_a]\bigr]\\
  &\quad = \sum_{j\notin I_a}\sum_{p=1}^m e_j e_p \int_0^1 \Tr\bigl[\bm{\sigma}\,\partial_{k_p}\cL_{\bm{A}_a,\bm{H}(s)}^*[\bm{P}_j]\bigr]\,ds
   + \sum_{j\in I_a}\sum_{p\notin I_a} e_j e_p \int_0^1 \Tr\bigl[\bm{\sigma}\,\partial_{k_p}\cL_{\bm{A}_a,\widetilde{\bm{H}}(s)}^*[\bm{P}_j]\bigr]\,ds.
 \end{aligned}
 \end{equation}
 In both sums, every pair $(j,p)$ has at least one index outside $I_a=\{j:d(a,j)\le\ell\}$, so $d(a,j)+d(a,p)\ge\ell$. By \eqref{meta:joint-derivative}, each derivative has norm at most $\poly(\beta)e^{-c\ell/(2\beta)}e^{-c[d(a,j)+d(a,p)]/(2\beta)}$. 
 Set $w_{aj}=e^{-c d(a,j)/(2\beta)}$. Weighted Cauchy--Schwarz and polynomial volume growth give
 \begin{equation}\label{app:weighted-coefficient-sum}
  \sum_a\left(\sum_jw_{aj}|e_j|\right)^2
  \le\left(\sup_a\sum_jw_{aj}\right)
      \left(\sup_j\sum_aw_{aj}\right)\norm{\mathbf e}_2^2
  \le\poly(\beta)\norm{\mathbf e}_2^2.
 \end{equation}
 Multiplying by the extracted factor $\poly(\beta)e^{-c\ell/(2\beta)}$ and reducing $c$ proves
 \begin{equation}\label{meta:signed-schur}
  \left|\sum_a\bigl[\Delta_a(\bm E,\bm H;\bm\sigma)-\Delta_a(\bm U_a,\bm H;\bm\sigma)\bigr]\right|
  \le\poly(\beta)e^{-c\ell/\beta}\norm{\mathbf e}_2^2.
 \end{equation}
 \item \emph{Step (ii): Truncating $\bm{H}\to\bm{H}_a$.} Set $\mathcal D_{a,t}^*:=\cL_{\bm A_a,\bm H+t\bm U_a}^*-\cL_{\bm A_a,\bm H_a+t\bm U_a}^*$ for $t\in[0,1]$. By the fundamental theorem of calculus in $t$,
 \begin{equation}\label{app:background-truncation-expansion}
  \Delta_a(\bm{U}_a, \bm{H}; \bm{\sigma}) - \Delta_a(\bm{U}_a, \bm{H}_a; \bm{\sigma})
  = \Tr\bigl[\bm{\sigma}(\mathcal{D}_{a,1}^* - \mathcal{D}_{a,0}^*)[\bm{U}_a]\bigr]
  = \sum_{j,p\in I_a} e_j e_p \int_0^1 \Tr\bigl[\bm{\sigma}\,\partial_{k_p}\mathcal{D}_{a,t}^*[\bm{P}_j]\bigr]\,dt.
 \end{equation}
 Because $\bm{H}+t\bm{U}_a$ and $\bm{H}_a+t\bm{U}_a$ agree on $A_a=B(a,L)$, \eqref{oracle:bath-locality} bounds $\|\mathcal{D}_{a,t}^*[\bm{P}_j]\|_\infty \le \poly(\beta)e^{-cL/\beta}$ uniformly under bounded variations of $k_p$, and \eqref{app:derivative-interpolation} gives $\|\partial_{k_p}\mathcal{D}_{a,t}^*[\bm{P}_j]\|_\infty \le \poly(\beta)e^{-c'L/\beta}$ for all $j,p\in I_a$. Hence
 \begin{equation}\label{meta:quadratic-truncation}
  \bigl|\Delta_a(\bm U_a,\bm H;\bm\sigma)-\Delta_a(\bm U_a,\bm H_a;\bm\sigma)\bigr|
  \le\poly(\beta)e^{-c'L/\beta}
  \left(\sum_{j\in I_a}|e_j|\right)^2.
 \end{equation}
 Summing over $a$ and applying \eqref{app:coefficient-overlap} gives $\poly(\beta,\ell)e^{-c'L/\beta}\norm{\mathbf e}_2^2$.
 \item \emph{Step (iii): Regularizing $\bm{\sigma}\to\bm{\sigma}_a^{(\lambda)}$.} Write the third difference as $\Tr_{A_a}[(\Tr_{A_a^c}\bm{\sigma}-\bm{\sigma}_a^{(\lambda)})\bm{O}_a]$, where $\bm{O}_a := (\cL_{\bm{A}_a,\bm{K}_a}^*-\cL_{\bm{A}_a,\bm{H}_a}^*)[\bm{U}_a]$. Because $\bm{A}_a$, $\bm{U}_a$, $\bm{H}_a$, and $\bm{K}_a$ are supported inside $A_a$, $\bm{O}_a$ acts only on $A_a$, so $\Tr[\bm{\sigma}\bm{O}_a] = \Tr_{A_a}[\Tr_{A_a^c}(\bm{\sigma})\bm{O}_a]$. By \eqref{meta:regularized-state}, $\|\Tr_{A_a^c}(\bm{\sigma}) - \bm{\sigma}_a^{(\lambda)}\|_1 \le 2\lambda$. Meanwhile, expanding $\bm{O}_a = \sum_{j,p\in I_a} e_j e_p \int_0^1 \partial_{k_p}\cL_{\bm{A}_a,\bm{H}_a+t\bm{U}_a}^*[\bm{P}_j]\,dt$ and using $\|\partial_{k_p}\cL_{\bm{A}_a,\cdot}^*[\bm{P}_j]\|_\infty \le \poly(\beta)$ gives $\|\bm{O}_a\|_\infty \le \poly(\beta)\left(\sum_{j\in I_a}|e_j|\right)^2$, and hence
 \begin{equation}\label{app:test-regularization-bound}
  \bigl|\Delta_a(\bm U_a,\bm H_a;\bm\sigma)-\Delta_a(\bm U_a,\bm H_a;\bm\sigma_a^{(\lambda)})\bigr|
  \le 2\lambda\|\bm{O}_a\|_\infty \le \lambda\,\poly(\beta)\left(\sum_{j\in I_a}|e_j|\right)^2.
 \end{equation}
 Summing over $a$ and applying \eqref{app:coefficient-overlap} gives $\lambda\,\poly(\beta,\ell)\norm{\mathbf e}_2^2$.
\end{itemize}
Adding the three bounds proves \eqref{meta:localized-regularized-tests}.
\end{proof}

\subsection{Spatial truncation of the stationarity test}\label{app:residual-estimation}
Let $B_j(r)$ be the patch used in \eqref{oracle:patch-generator}. The difference between the full and truncated Heisenberg observables satisfies
\begin{equation}\label{app:spatial-error-split}
\begin{aligned}
 \|\cL_{\bm K}^*[\bm P_j]-\sum_{i\in B_j(r/2)}\cL_{i,\bm K_{B_j(r)}}^*[\bm P_j]\|_\infty
 &\le\sum_{i\notin B_j(r/2)}\|\cL_{i,\bm K}^*[\bm P_j]\|_\infty+\sum_{i\in B_j(r/2)}
 \|(\cL_{i,\bm K}^*-\cL_{i,\bm K_{B_j(r)}}^*)[\bm P_j]\|_\infty\\
 &\le\poly(\beta)e^{-cr/\beta}.
\end{aligned}
\end{equation}
For the first term, the adjoint of a generator truncated to half the distance from $i$ to $\supp(\bm P_j)$ annihilates $\bm P_j$. Applying \eqref{oracle:bath-locality} and summing the exponentially decaying distance shells gives the last line. For the second term, if $i\in B_j(r/2)$, then $B(i,r/2)\subseteq B_j(r)$. Hence truncating either $\bm K$ or $\bm K_{B_j(r)}$ to $B(i,r/2)$ gives the same Hamiltonian $\bm K_{B(i,r/2)}$. Inserting its generator, using $\norm{\bm P_j}_\infty=1$, and applying \eqref{oracle:bath-locality} twice gives
\begin{equation}\label{app:retained-site-truncation}
\begin{aligned}
 \norm{(\cL_{i,\bm K}^*-\cL_{i,\bm K_{B_j(r)}}^*)[\bm P_j]}_\infty&\le\norm{\cL_{i,\bm K}^*-\cL_{i,\bm K_{B(i,r/2)}}^*}_{\infty\to\infty}
 +\norm{\cL_{i,\bm K_{B_j(r)}}^*-\cL_{i,\bm K_{B(i,r/2)}}^*}_{\infty\to\infty}\\
 &\quad\le 2\poly(\beta)e^{-cr/(2\beta)}.
\end{aligned}
\end{equation}
Summing over the $O((1+r)^D)$ retained sites yields
\begin{equation}\label{app:retained-sites-sum}
 \sum_{i\in B_j(r/2)}\norm{(\cL_{i,\bm K}^*-\cL_{i,\bm K_{B_j(r)}}^*)[\bm P_j]}_\infty
 \le\poly(\beta)(1+r)^D e^{-cr/(2\beta)}
 \le\poly(\beta)e^{-c'r/\beta},
\end{equation}
where the last inequality absorbs the polynomial factor by reducing the exponential decay constant. Taking $r=C\beta\log(C\beta/\nu)$ proves \eqref{cutoff:bias}. The same choice, with a larger constant if necessary, also ensures
\begin{equation}\label{app:patch-stationarity-error}
 \sum_{i\in B_j(r/2)}
 \|\cL_{i,\bm H_{B_j(r)}}-\cL_{i,\bm H}\|_{1\to1}\le\nu/4.
\end{equation}

By duality of induced norms \citep[Proposition~3.51, Eq.~(3.152)]{LinWiebe2026QASC}, any linear map $\Phi$ and its trace adjoint $\Phi^*$ satisfy
\begin{equation}\label{app:induced-norm-duality}
 \|\Phi\|_{1\to1}=\|\Phi^*\|_{\infty\to\infty}.
\end{equation}

\section{Stationarity-Test Comparison for Metastable States}\label{app:meta-comparison}
We prove Lemmas~\ref{lem:meta-regularized-adb} and~\ref{lem:meta-continuity}.
\subsection{Logarithm, ADB, and commutators of regularized patches}\label{app:regularized-adb}
\begin{proof}[Proof of Lemma~\ref{lem:meta-regularized-adb}]
\textit{Logarithm and single-site ADB bounds.}
We justify \eqref{meta:log-bound} and \eqref{meta:adb-local} for the regularized state in \eqref{meta:regularized-state}. Since $\bm\sigma_a^{(\lambda)}\ge\lambda\bm\rho_{\bm H_a}$ and $\Tr(e^{-\beta\bm H_a})\le2^{|A_a|}e^{\beta\norm{\bm H_a}_\infty}$,
\begin{equation}\label{app:regularized-spectrum}
 \lambda_{\min}(\bm\sigma_a^{(\lambda)})\ge\lambda\,2^{-|A_a|}e^{-2\beta\norm{\bm H_a}_\infty},\qquad
 \norm{\log\bm\sigma_a^{(\lambda)}}_\infty\le\log(1/\lambda)+|A_a|\log2+2\beta\norm{\bm H_a}_\infty.
\end{equation}
The triangle inequality, $|A_a|=O(L^D)$, $\norm{\bm H_a}_\infty=O(L^D)$, and $\beta\ge1$ prove \eqref{meta:log-bound}.

For a site $i\in B(a,b)$, let $\cL_{i,\bm H_a}$ be the sum of the generators associated with $\bm X_i,\bm Y_i,\bm Z_i$, as in the metastability promise. The generator locality estimate \eqref{oracle:bath-locality}, duality of the induced norms, and contractivity of partial trace give
\begin{equation}\label{app:patch-stationarity}
\begin{aligned}
 \norm{\cL_{i,\bm H_a}[\bm\sigma_a^{(\lambda)}]}_1
 &=(1-\lambda)\norm{\cL_{i,\bm H_a}[\Tr_{A_a^c}\bm\sigma]}_1\\
 &\le\norm{\cL_{i,\bm H}[\bm\sigma]}_1
   +\poly(\beta)e^{-c(L-b)/\beta}\\
 &\le\zeta+\poly(\beta)e^{-c(L-b)/\beta}.
\end{aligned}
\end{equation}

Each individual generator has nonnegative entropy production relative to $\bm\rho_{\bm H_a}$. Applying \eqref{app:absolute-ep-adb} to each Pauli $\bm A$ on site $i$ gives
\begin{equation}\label{app:site-entropy-adb}
\begin{aligned}
 \mathsf{ADB}_{\bm H_a,\bm A}(\bm\sigma_a^{(\lambda)})
 &\le C\EP_{\cL_{\bm A,\bm H_a}}(\bm\sigma_a^{(\lambda)}\|\bm\rho_{\bm H_a})\\
 &\le C\sum_{\bm A'\in\{\bm X_i,\bm Y_i,\bm Z_i\}}
 \EP_{\cL_{\bm A',\bm H_a}}(\bm\sigma_a^{(\lambda)}\|\bm\rho_{\bm H_a})\\
 &=-C\Tr\bigl(\cL_{i,\bm H_a}[\bm\sigma_a^{(\lambda)}](\log\bm\sigma_a^{(\lambda)}+\beta\bm H_a)\bigr)\\
 &\le C M\left\{\zeta+\poly(\beta)e^{-c(L-b)/\beta}\right\}.
\end{aligned}
\end{equation}
where the last inequality uses \eqref{app:patch-stationarity}, the logarithm bound \eqref{meta:log-bound}, and trace-norm duality. Taking the maximum over sites and Pauli choices proves \eqref{meta:adb-local}.

\paragraph{Logarithmic commutators.}
Fix $a$ and write $\bm H=\bm H_a$, $\bm\sigma=\bm\sigma_a^{(\lambda)}$, $\eta=\eta_a$, and $\bm X=\log\bm\sigma+\beta\bm H$. Then $\norm{\bm X}_\infty\le M=C\{\log(1/\lambda)+\beta L^D\}$ by \eqref{meta:log-bound}. 
Write $\bm P=\bm A_1\cdots\bm A_s$, where each single-site Pauli factor satisfies $\mathsf{ADB}_{\bm H,\bm A_j}(\bm\sigma)\le\eta$ by \eqref{meta:adb-local}.

Let $\bm B$ be a Hermitian approximation to $\bm P$ with $\norm{\bm B}_\infty\le2$, to be constructed below. We will bound its logarithmic commutator using ADB, while controlling $\bm P-\bm B$ directly in operator norm.
Fourier inversion and $\int f_\beta(t)^2dt=1$ give the reconstruction identity
\begin{equation}\label{app:gaussian-reconstruction}
 \bm B=\frac1{\sqrt{2\pi}}\iint f_\beta(t)e^{-i\nu t}\bm J_{\bm B,\bm H}(\nu,t)\,dt\,d\nu.
\end{equation}
Let $\bm B^{(1)}$ denote the same integral restricted to $|\nu|\le\Omega_1$.
For $0<\delta<1/2$, first separate the interior interval from the endpoints. The triangle inequality and $\norm{[\bm Z,\bm X]}_{\bm\sigma,v}\le2\norm{\bm X}_\infty\norm{\bm Z}_\infty$ give
\begin{equation}\label{app:commutator-error-decomposition}
\begin{aligned}
 \left(\int_0^1\norm{[\bm P,\bm X]}_{\bm\sigma,v}^2\,dv\right)^{1/2}
 \le{}&2\norm{\bm X}_\infty\norm{\bm P-\bm B}_\infty
 +2\norm{\bm X}_\infty\norm{\bm B-\bm B^{(1)}}_\infty\\
 &+\left(\int_\delta^{1-\delta}\norm{[\bm B^{(1)},\bm X]}_{\bm\sigma,v}^2\,dv\right)^{1/2}
 +4\norm{\bm X}_\infty\sqrt{2\delta}.
\end{aligned}
\end{equation}
The last term uses $\norm{\bm B}_\infty\le2$ on endpoint intervals of total length $2\delta$.

To estimate the interior term, put $v=1/2+z$ and apply weighted Cauchy--Schwarz to the commutator of the truncated reconstruction formula:
\begin{equation}\label{app:reconstruction-cauchy-schwarz}
\begin{aligned}
 \norm{[\bm B^{(1)},\bm X]}_{\bm\sigma,1/2+z}^2
 &\quad\le\frac1{2\pi}
 \left(\int_{|\nu|\le\Omega_1}\int
 \frac{|f_\beta(t)|^2}{h_z(\nu)g_z^{\rm ADB}(t)}\,dt\,d\nu\right)\\
 &\qquad\times\left(\int_{|\nu|\le\Omega_1}\int
 h_z(\nu)g_z^{\rm ADB}(t)
 \norm{[\bm J_{\bm B,\bm H}(\nu,t),\bm X]}_{\bm\sigma,1/2+z}^2\,dt\,d\nu\right).
\end{aligned}
\end{equation}
On $|z|\le1/2-\delta$ and $|\nu|\le\Omega_1$,
\begin{equation}\label{app:interior-kernel-bounds}
 g_z^{\rm ADB}(t)\ge c\delta g(t),\qquad
 h_z(\nu)^{-1}\le e^{C+C\beta\Omega_1},\qquad
 \int\frac{|f_\beta(t)|^2}{g(t)}\,dt=C\beta.
\end{equation}
Thus the first parenthesis is at most $C\delta^{-1}\beta\Omega_1e^{C+C\beta\Omega_1}$.
The integral of the second parenthesis over this $z$ interval is at most
$\mathsf{ADB}_{\bm H,\bm B}(\bm\sigma)$ by \eqref{app:adb-gradient}.
Taking square roots and absorbing the polynomial factors into the exponential gives
\begin{equation}\label{app:interior-adb-bound}
 \left(\int_\delta^{1-\delta}\norm{[\bm B^{(1)},\bm X]}_{\bm\sigma,v}^2\,dv\right)^{1/2}
 \le C\delta^{-1/2}e^{C\beta(\Omega_1+1)}
 \sqrt{\mathsf{ADB}_{\bm H,\bm B}(\bm\sigma)}.
\end{equation}

It therefore remains to construct $\bm B$ with three properties: a small approximation error $\norm{\bm P-\bm B}_\infty$, a small reconstruction tail $\norm{\bm B-\bm B^{(1)}}_\infty$, and an ADB bound inherited from the single-site factors of $\bm P$. The frequency windows below give the first and third properties. Their uniform imaginary-time bound gives the second without a loss depending on the first cutoff $\Omega_2$.

\paragraph{Frequency windows and a uniform imaginary-time bound.}
Choose a fixed real, even $\vartheta\in C_c^\infty(\mathbb R)$ with $\sum_{k\in\mathbb Z}\vartheta(\omega-k)=1$, and define
\begin{equation}\label{app:frequency-window-definition}
 \bm F_{j,k}:=c_\beta\int\vartheta(\omega-k)\bm J_{\bm A_j,\bm H}(\omega)\,d\omega,
 \qquad c_\beta:=\frac{1}{\sqrt{2\pi}f_\beta(0)}=\Theta(\sqrt\beta).
\end{equation}
Fourier inversion gives $\sum_k\bm F_{j,k}=\bm A_j$, and $\bm F_{j,k}^*=\bm F_{j,-k}$. For some fixed $y_0,c>0$,
\begin{equation}\label{app:window-strip}
 \sum_{k\in\mathbb Z}e^{c|k|}
 \sup_{|y|\le y_0}\norm{e^{-y\bm H}\bm F_{j,k}e^{y\bm H}}_\infty\le C.
\end{equation}
To prove this, bounded interaction degree and range give
$\norm{\ad_{\bm H}^{r}(\bm A_j)}_\infty\le C^r r!$, since each successive commutator adds only a bounded number of sites. Thus imaginary-time evolution is bounded for $|y|\le2y_0$.
The time kernel of $\bm F_{j,k}$ is a fixed constant times
$e^{-ikt}\widehat\vartheta(t)e^{-t^2/\beta^2}$: the normalization $c_\beta f_\beta(0)$ is constant.
Shifting the contour a fixed distance within this strip gives exponential decay in $|k|$; the Gaussian is uniformly bounded and $\widehat\vartheta$ is rapidly decaying along horizontal lines. Decreasing $c$ then gives \eqref{app:window-strip}.

For a cutoff $\Omega_2$, put
\begin{equation}\label{app:retained-operator-definition}
 \bm P_{\Omega_2}:=\sum_{\sum_j|k_j|\le\Omega_2}\bm F_{1,k_1}\cdots\bm F_{s,k_s},
 \qquad \bm B:=\tfrac12(\bm P_{\Omega_2}+\bm P_{\Omega_2}^*).
\end{equation}
Since $\bm P$ is Hermitian, \eqref{app:window-strip} implies
\begin{equation}\label{app:retained-strip}
 \norm{\bm P-\bm B}_\infty\le e^{Cs-c\Omega_2},\qquad
 \sup_{|y|\le y_0}\norm{e^{-y\bm H}\bm B e^{y\bm H}}_\infty\le e^{Cs}.
\end{equation}
The second bound is uniform in $\Omega_2$, because the retained sum is bounded by the sum over all tuples. We take $\Omega_2\ge Cs$ large enough that also $\norm{\bm B}_\infty\le2$.

\paragraph{ADB of the retained operator.}
Introduce the time-integrated defect
\begin{equation}\label{app:integrated-defect-definition}
 \mathcal E(\bm Z)^2:=\int g(t)\norm{\mathcal D_{\bm\sigma}^{\bm H}[\mathcal U_t^{\bm H}[\bm Z]]}_2^2\,dt.
\end{equation}
Because $\gamma\le1$, Parseval and $(g*|f_\beta|^2)(t)\le Cg(t)$ give
$\sqrt{\mathsf{ADB}_{\bm H,\bm Z}(\bm\sigma)}\le C\mathcal E(\bm Z)$, where the same quadratic expression defines ADB for non-Hermitian intermediate operators. Cauchy--Schwarz on the fixed-width frequency window, using $\gamma(\omega)^{-1}\le e^{C\beta(|k|+1)}$ there, and Lemma~\ref{lem:gaussian-conjugation} give
\begin{equation}\label{app:window-defect-bounds}
 \mathcal E(\bm F_{j,k})\le C\sqrt\beta\,e^{C\beta(|k|+1)}\sqrt\eta,
 \qquad
 \norm{e^{\beta\bm H/2}\bm F_{j,k}e^{-\beta\bm H/2}}_\infty\le C\beta e^{C\beta(|k|+1)}.
\end{equation}
Apply the product rule
\begin{equation}\label{app:window-defect-product-rule}
 \mathcal D_{\bm\sigma}^{\bm H}[\bm Y\bm Z]
 =\bm Y\mathcal D_{\bm\sigma}^{\bm H}[\bm Z]
 +\mathcal D_{\bm\sigma}^{\bm H}[\bm Y]e^{\beta\bm H/2}\bm Z e^{-\beta\bm H/2}
\end{equation}
to the real-time-evolved factors and use the triangle inequality in $L^2(g(t)dt)$. Real-time evolution preserves the operator norms, so
\begin{equation}\label{app:product-integrated-defect}
 \mathcal E(\bm F_{1,k_1}\cdots\bm F_{s,k_s})
 \le s(C\beta)^s e^{C\beta(s+\sum_j|k_j|)}\sqrt\eta.
\end{equation}
There are at most $2^s\binom{\lfloor\Omega_2\rfloor+s}{s}\le e^{C(s+\Omega_2)}$ retained tuples. The adjoint has the same bound by reversing the factors and replacing $k_j$ by $-k_j$. Thus
\begin{equation}\label{app:retained-adb}
 \sqrt{\mathsf{ADB}_{\bm H,\bm B}(\bm\sigma)}
 \le e^{C\beta(s+\Omega_2)}\sqrt\eta.
\end{equation}

\paragraph{Combining the estimates.}
The strip bound in \eqref{app:retained-strip}, applied to the Gaussian Fourier integral, gives
\begin{equation}\label{app:retained-frequency-decay}
 \norm{\bm J_{\bm B,\bm H}(\nu,t)}_\infty\le C\sqrt\beta\,e^{Cs-c|\nu|},
\end{equation}
uniformly in real $t$ and in $\Omega_2$. Integrating the discarded frequencies in \eqref{app:gaussian-reconstruction} yields
\begin{equation}\label{app:reconstruction-tail}
 \norm{\bm B-\bm B^{(1)}}_\infty\le C\beta e^{Cs-c\Omega_1}.
\end{equation}
The other approximation error satisfies $\norm{\bm P-\bm B}_\infty\le e^{Cs-c\Omega_2}$ by \eqref{app:retained-strip}, and the ADB bound is \eqref{app:retained-adb}.
Combining these estimates with the initial decomposition and the interior ADB bound gives
\begin{equation}\label{app:two-cutoff-errors}
\begin{aligned}
 \left(\int_0^1\norm{[\bm P,\bm X]}_{\bm\sigma,v}^2\,dv\right)^{1/2}\le{}&C\delta^{-1/2}e^{C\beta(s+\Omega_1+\Omega_2+1)}\sqrt\eta\\
 &{}+C\norm{\bm X}_\infty\left[\sqrt\delta+\beta e^{Cs-c\Omega_1}+e^{Cs-c\Omega_2}\right].
\end{aligned}
\end{equation}
Choose $\delta=e^{-2u}$ and $\Omega_i=C_i(s+u)$ with sufficiently large fixed $C_i$. Since $u\ge1+\log(\beta+1)$, all additive terms are bounded by $C\norm{\bm X}_\infty e^{-u}\le CM e^{-u}$, proving \eqref{app:ordered-cutoff-bound}. 
\end{proof}

\subsection{The stationarity-test comparison}\label{app:continuity-proof}
\begin{proof}[Proof of Lemma~\ref{lem:meta-continuity}]
We follow the proof of \citep[Lemma 6.14]{Bergamaschi2026Stationarity}, keeping the $\beta$-dependence explicit and using Lemma~\ref{lem:meta-regularized-adb} for the final logarithmic-commutator estimate.
Suppress the patch index, writing $\bm H=\bm H_a$, $\bm K=\bm H+\bm U$, $\bm\sigma=\bm\sigma_a^{(\lambda)}$, and $\bm J_{\bm G}=\bm J_{\bm A_a,\bm G}(\omega,t)$. Let
\begin{equation}\label{app:comparison-notation}
 \bm X=\log\bm\sigma+\beta\bm H,\qquad
 d\nu:=g(t)h_0(\omega)\,dt\,d\omega\,dz,
\end{equation}
where $h_0(\omega)=e^{-1/8-\beta|\omega|/2}$ and $-1/2\le z\le1/2$. Inner products below are those associated with $\norm{\cdot}_{\bm\sigma,1/2+z}$.
The gradient-flow identity \citep[Eq.~(B7)]{bergamaschi2025structuraltheoryquantummetastability} and the Fisher form \eqref{meta:Q-definition} give the expansion \citep[Eqs.~(183)--(185)]{Bergamaschi2026Stationarity}
\begin{equation}\label{app:comparison-expansion}
\begin{aligned}
 &T_a(\bm U,\bm K;\bm\sigma)-T_a(\bm U,\bm H;\bm\sigma)+\beta\mathcal Q_a(\bm U,\bm K;\bm\sigma)\\
 &=-\int\langle[\bm J_{\bm K}-\bm J_{\bm H},\bm U],[\mathcal U_{-iz\beta}^{\bm H}[\bm J_{\bm H}],\bm X]\rangle_{\bm\sigma,1/2+z}\,d\nu\\
 &\quad-\int\langle[\bm J_{\bm K},\bm U],[\mathcal U_{-iz\beta}^{\bm K}[\bm J_{\bm K}]-\mathcal U_{-iz\beta}^{\bm H}[\bm J_{\bm H}],\bm X]\rangle_{\bm\sigma,1/2+z}\,d\nu.
\end{aligned}
\end{equation}
For the first integral, put $\bm W=[\bm J_{\bm K}-\bm J_{\bm H},\bm U]$. We first perform the $z$ integration explicitly. Since $\bm X=\log\bm\sigma+\beta\bm H$,
\begin{equation}\label{app:comparison-z-derivative}
 \frac{d}{dz}\left(\bm\sigma^{1/2+z}\mathcal U_{-iz\beta}^{\bm H}[\bm J_{\bm H}]\bm\sigma^{1/2-z}\right)
 =-\bm\sigma^{1/2+z}[\mathcal U_{-iz\beta}^{\bm H}[\bm J_{\bm H}],\bm X]\bm\sigma^{1/2-z}.
\end{equation}
Consequently,
\begin{equation}\label{app:comparison-z-integral}
 \int_{-1/2}^{1/2}\bm\sigma^{1/2+z}[\mathcal U_{-iz\beta}^{\bm H}[\bm J_{\bm H}],\bm X]\bm\sigma^{1/2-z}\,dz
 =\bm J_{\bm H}^{[-1/2]}\bm\sigma-\bm\sigma\bm J_{\bm H}^{[1/2]}.
\end{equation}
Recall that the map in \eqref{meta:adb-defect} is explicitly
$\mathcal D_{\bm\sigma}^{\bm H}[\bm Z]=\bm Z\sqrt{\bm\sigma}-\sqrt{\bm\sigma}e^{\beta\bm H/2}\bm Ze^{-\beta\bm H/2}$.
Adding and subtracting $\sqrt{\bm\sigma}\bm J_{\bm H}\sqrt{\bm\sigma}$ gives
\begin{equation}\label{app:comparison-endpoint-split}
 \bm J_{\bm H}^{[-1/2]}\bm\sigma-\bm\sigma\bm J_{\bm H}^{[1/2]}
 =\mathcal D_{\bm\sigma}^{\bm H}[\bm J_{\bm H}^{[-1/2]}]\sqrt{\bm\sigma}
 +\sqrt{\bm\sigma}\mathcal D_{\bm\sigma}^{\bm H}[\bm J_{\bm H}].
\end{equation}
The assumption $\mathsf{ADB}_{\bm H,\bm A_a}(\bm\sigma)\le\eta$ controls the second term through \eqref{meta:ADB}. For the first, use
$\mathcal D_{\bm\sigma}^{\bm H}[\bm J_{\bm H}^{[-1/2]}]=-\mathcal D_{\bm\sigma}^{\bm H}[\bm J_{\bm H}^*]^*$ and
$\bm J_{\bm A_a,\bm H}(\omega,t)^*=\bm J_{\bm A_a,\bm H}(-\omega,t)$.
Changing $\omega$ to $-\omega$ therefore gives the two bounds
\begin{equation}\label{app:comparison-integrated-bounds}
\begin{aligned}
 \iint g(t)\gamma(\omega)\norm{\mathcal D_{\bm\sigma}^{\bm H}[\bm J_{\bm H}]}_2^2\,dt\,d\omega&\le\eta,\\
 \iint g(t)\gamma(-\omega)\norm{\mathcal D_{\bm\sigma}^{\bm H}[\bm J_{\bm H}^{[-1/2]}]}_2^2\,dt\,d\omega&\le\eta.
\end{aligned}
\end{equation}
Applying Hilbert--Schmidt Cauchy--Schwarz to the trace pairing with $\bm W$, and then Cauchy--Schwarz over $(t,\omega)$, yields
\begin{equation}\label{app:comparison-weighted-cauchy-schwarz}
\begin{aligned}
 |\text{first integral}|
 &\le\iint gh_0\left(
 \norm{\bm W\sqrt{\bm\sigma}}_2\norm{\mathcal D_{\bm\sigma}^{\bm H}[\bm J_{\bm H}^{[-1/2]}]}_2
 +\norm{\sqrt{\bm\sigma}\bm W}_2\norm{\mathcal D_{\bm\sigma}^{\bm H}[\bm J_{\bm H}]}_2\right)dt\,d\omega\\
 &\le\sqrt\eta\left(\iint\frac{g(t)h_0(\omega)^2}{\gamma(-\omega)}
 \norm{\bm W\sqrt{\bm\sigma}}_2^2\,dt\,d\omega\right)^{1/2}+\sqrt\eta\left(\iint\frac{g(t)h_0(\omega)^2}{\gamma(\omega)}
 \norm{\sqrt{\bm\sigma}\bm W}_2^2\,dt\,d\omega\right)^{1/2}.
\end{aligned}
\end{equation}
Since $h_0(\omega)^2/\gamma(\pm\omega)\le C$, this implies the first inequality below:
\begin{equation}\label{app:comparison-single-adb}
 |\text{first integral}|\le C\sqrt\eta
 \left(\iint g(t)\{\norm{\bm W\sqrt{\bm\sigma}}_2^2+\norm{\sqrt{\bm\sigma}\bm W}_2^2\}\,dt\,d\omega\right)^{1/2}
 \le C\beta\norm{\bm U}_\infty^2\sqrt\eta.
\end{equation}
To prove the second inequality, Duhamel's formula and $\bm K-\bm H=\bm U$ give
\begin{equation}\label{app:comparison-duhamel-commutator}
\begin{aligned}
 \norm{\mathcal U_{t+r}^{\bm K}[\bm A_a]-\mathcal U_{t+r}^{\bm H}[\bm A_a]}_\infty
 &\le2|t+r|\norm{\bm U}_\infty,\\
 \norm{[\mathcal U_{t+r}^{\bm K}[\bm A_a]-\mathcal U_{t+r}^{\bm H}[\bm A_a],\bm U]}_\infty
 &\le4|t+r|\norm{\bm U}_\infty^2.
\end{aligned}
\end{equation}
The definition of the filtered jump operators gives explicitly
\begin{equation}\label{app:comparison-W-fourier}
 \bm W=\frac{1}{\sqrt{2\pi}}\int_{\mathbb R}e^{-i\omega r}f_\beta(r)
 [\mathcal U_{t+r}^{\bm K}[\bm A_a]-\mathcal U_{t+r}^{\bm H}[\bm A_a],\bm U]\,dr.
\end{equation}
Parseval in Hilbert--Schmidt norm, applied after multiplication by $\sqrt{\bm\sigma}$ on either side, and $\norm{\sqrt{\bm\sigma}}_2=1$ now give
\begin{equation}\label{app:comparison-parseval-moments}
\begin{aligned}
 \iint g(t)\left(\norm{\bm W\sqrt{\bm\sigma}}_2^2+
 \norm{\sqrt{\bm\sigma}\bm W}_2^2\right)dt\,d\omega
 &\le32\norm{\bm U}_\infty^4
 \iint g(t)|f_\beta(r)|^2(t+r)^2\,dr\,dt
 \\
 &\le C\beta^2\norm{\bm U}_\infty^4.
\end{aligned}
\end{equation}
The last step uses $(t+r)^2\le2t^2+2r^2$: both kernels have bounded total mass and second moments $O(\beta^2)$. Taking the square root proves the second inequality in \eqref{app:comparison-single-adb}.

For the second integral, we decompose the operator difference into terms supported on successively larger balls. For $\bm G=\bm H,\bm K$, let $\bm G_r$ retain only the interactions supported entirely in $B(a,r)$, and set $r_0=\lceil\ell+R\rceil$. Since $\bm U$ is supported in $B(a,\ell)$, $\bm K_r=\bm H_r+\bm U$ for $r\ge r_0$. Define
\begin{equation}\label{app:comparison-shell-definition}
\begin{aligned}
 \bm Z_{r_0}
 &=\mathcal U_{-iz\beta}^{\bm K_{r_0}}[\bm J_{\bm K_{r_0}}]-\mathcal U_{-iz\beta}^{\bm H_{r_0}}[\bm J_{\bm H_{r_0}}],\\
 \bm Z_r
 &=\bigl(\mathcal U_{-iz\beta}^{\bm K_r}[\bm J_{\bm K_r}]-\mathcal U_{-iz\beta}^{\bm H_r}[\bm J_{\bm H_r}]\bigr)
-\bigl(\mathcal U_{-iz\beta}^{\bm K_{r-1}}[\bm J_{\bm K_{r-1}}]-\mathcal U_{-iz\beta}^{\bm H_{r-1}}[\bm J_{\bm H_{r-1}}]\bigr),\qquad r>r_0.
\end{aligned}
\end{equation}
All jump operators here have arguments $(\omega,t)$, and each conjugation uses its own truncated Hamiltonian:
$\mathcal U_{-iz\beta}^{\bm G_r}[\bm J_{\bm G_r}]=e^{z\beta\bm G_r}\bm J_{\bm A_a,\bm G_r}(\omega,t)e^{-z\beta\bm G_r}$.
Adding these differences cancels the intermediate terms. Once the ball contains the full patch, the truncations equal the original Hamiltonians, so
\begin{equation}\label{app:comparison-shells}
 \mathcal U_{-iz\beta}^{\bm K}[\bm J_{\bm K}]-\mathcal U_{-iz\beta}^{\bm H}[\bm J_{\bm H}]=\sum_{r\ge r_0}\bm Z_r,
 \qquad \supp(\bm Z_r)\subseteq B(a,r).
\end{equation}
We will show uniformly for $|z|\le1/2$ that
\begin{equation}\label{app:comparison-shell-bound}
 \norm{\bm Z_r}_\infty\le\poly(\beta,\ell)\norm{\bm U}_\infty
 e^{z\beta\omega+\alpha|t|/\beta-c(r-\ell)/\beta},\qquad \alpha<\pi.
\end{equation}
First consider the corresponding differences before the Gaussian time integration. Write
$\bm Y_r(v)=\mathcal U_v^{\bm K_r}[\bm A_a]-\mathcal U_v^{\bm H_r}[\bm A_a]$.
For $r>r_0$, the quantity to bound is $\bm Y_r(v)-\bm Y_{r-1}(v)$. Its dependence on $\bm U$ can be made explicit by
\begin{equation}\label{app:comparison-interpolation}
 \bm Y_r(v)=\int_0^1\frac{d}{dx}\mathcal U_v^{\bm H_r+x\bm U}[\bm A_a]\,dx,
 \qquad
 \frac{d}{dx}\mathcal U_v^{\bm H_r+x\bm U}[\bm A_a]
 =i\int_0^v\mathcal U_{v-u}^{\bm H_r+x\bm U}
 [[\bm U,\mathcal U_u^{\bm H_r+x\bm U}[\bm A_a]]]\,du.
\end{equation}
To compare the two integrands, use that time evolution preserves products:
\begin{equation}\label{app:comparison-evolved-commutator}
 \mathcal U_{v-u}^{\bm H_r+x\bm U}
 [[\bm U,\mathcal U_u^{\bm H_r+x\bm U}[\bm A_a]]]
 =[\mathcal U_{v-u}^{\bm H_r+x\bm U}[\bm U],
   \mathcal U_v^{\bm H_r+x\bm U}[\bm A_a]].
\end{equation}
The two Hamiltonians differ only near the boundary, at distance at least $r-\ell-R$ from $B(a,\ell)$. The truncation bound \eqref{eq:capped-lr}, applied separately to $\bm U$ and $\bm A_a$, gives uniformly in $x\in[0,1]$
\begin{equation}\label{app:comparison-two-locality-bounds}
\begin{aligned}
 \norm{\mathcal U_s^{\bm H_r+x\bm U}[\bm U]
       -\mathcal U_s^{\bm H_{r-1}+x\bm U}[\bm U]}_\infty
 &\le C(1+\ell)^D\norm{\bm U}_\infty
 e^{v_{\rm LR}|s|-c_{\rm LR}(r-\ell-R)},\\
 \norm{\mathcal U_s^{\bm H_r+x\bm U}[\bm A_a]
       -\mathcal U_s^{\bm H_{r-1}+x\bm U}[\bm A_a]}_\infty
 &\le C e^{v_{\rm LR}|s|-c_{\rm LR}(r-\ell-R)}.
\end{aligned}
\end{equation}
Using $[\bm B,\bm C]-[\bm B',\bm C']=[\bm B-\bm B',\bm C]+[\bm B',\bm C-\bm C']$ and preservation of operator norms, the difference of the integrands in \eqref{app:comparison-interpolation} is at most
$C(1+\ell)^D\norm{\bm U}_\infty e^{v_{\rm LR}|v|-c_{\rm LR}(r-\ell-R)}$:
for $u$ between $0$ and $v$, both $|u|$ and $|v-u|$ are at most $|v|$.
Integration over $u$ contributes $|v|$, and integration over $x\in[0,1]$ contributes no additional factor. Combining this estimate with the direct bound $\norm{\bm Y_r(v)-\bm Y_{r-1}(v)}_\infty\le4|v|\norm{\bm U}_\infty$ gives
\begin{equation}\label{app:comparison-unfiltered-shell}
 \norm{\bm Y_r(v)-\bm Y_{r-1}(v)}_\infty
 \le C(1+\ell)^C\norm{\bm U}_\infty|v|
 e^{\theta v_{\rm LR}|v|-\theta c_{\rm LR}(r-\ell-R)},
 \qquad 0<\theta\le1.
\end{equation}

Next integrate at $v=t+s$ against the Gaussian filter. The frequency factor has absolute value one, and
\begin{equation}\label{app:comparison-gaussian-integral}
 \int_{\mathbb R}|f_\beta(s)|\,|t+s|e^{\theta v_{\rm LR}|t+s|}\,ds
 \le C\sqrt\beta\,(|t|+\beta)
 e^{(\theta v_{\rm LR}\beta)^2/2+\theta v_{\rm LR}|t|}.
\end{equation}
Choose $\theta=c_*/\beta$ with $0<c_*\le1$ sufficiently small. The Gaussian exponential is then bounded independently of $\beta$, and
$|t|+\beta\le C\beta e^{c_*|t|/\beta}$.
Thus the time dependence is bounded by $e^{\alpha|t|/\beta}$ with $\alpha=c_*(v_{\rm LR}+1)<\pi$, while the spatial factor is $e^{-c(r-\ell)/\beta}$, up to a constant depending on $R$.

Finally, conjugation by $e^{z\beta\bm G}$ can be evaluated exactly for a Gaussian filter:
\begin{equation}\label{app:comparison-modular-shift}
 e^{z\beta\bm G}\bm J_{\bm A_a,\bm G}(\omega,t)e^{-z\beta\bm G}
 =e^{z\beta\omega+z^2}\bm J_{\bm A_a,\bm G}(\omega+2z/\beta,t).
\end{equation}
Applying \eqref{app:comparison-modular-shift} to the differences in \eqref{app:comparison-shell-definition}, with $e^{z^2}\le e^{1/4}$, proves \eqref{app:comparison-shell-bound}.

For each term in the second integral, Cauchy--Schwarz in the weighted inner product gives
\begin{equation}\label{app:comparison-weighted-pairing}
 \left|\left\langle[\bm J_{\bm K},\bm U],[\bm Z_r,\bm X]\right\rangle_{\bm\sigma,1/2+z}\right|
 \le2\norm{\bm J_{\bm K}}_\infty\norm{\bm U}_\infty
 \norm{[\bm Z_r,\bm X]}_{\bm\sigma,1/2+z}.
\end{equation}
For $r>b$, use $\norm{[\bm Z_r,\bm X]}_{\bm\sigma,1/2+z}\le2M\norm{\bm Z_r}_\infty$. The single-site imaginary-time estimate used in \eqref{app:window-strip} also gives
$\norm{\bm J_{\bm K}(\omega,t)}_\infty\le C\sqrt\beta e^{-c|\omega|}$.
Since $h_0(\omega)e^{z\beta\omega}\le1$ and $\int g(t)e^{\alpha|t|/\beta}dt\le C$, summing these shells bounds their contribution by
\begin{equation}\label{app:comparison-far-shells}
 \poly(\beta,\ell)\norm{\bm U}_\infty^2 M e^{-c(b-\ell)/\beta}.
\end{equation}
For $r\le b$, expand $\bm Z_{\le b}=\sum_{\bm P\in\mathcal P(B(a,b))}c_{\bm P}\bm P$. Here $\mathcal P(B(a,b))$ consists of all Pauli strings supported in $B(a,b)$, without additional phases.
Pauli orthogonality gives $\sum_{\bm P}|c_{\bm P}|^2\le\norm{\bm Z_{\le b}}_\infty^2$.
Apply Cauchy--Schwarz in the coefficients and then in $z$, and use \eqref{app:comparison-shell-bound} and the same scalar integrals. Since there are at most $4^V$ strings, combining the near and far contributions with \eqref{app:comparison-single-adb} yields
\begin{equation}\label{app:comparison-before-reconstruction}
\begin{aligned}
 &|T_a(\bm U,\bm K;\bm\sigma)-T_a(\bm U,\bm H;\bm\sigma)+\beta\mathcal Q_a(\bm U,\bm K;\bm\sigma)|\\
 &\quad\le\poly(\beta,\ell,b)\norm{\bm U}_\infty^2
 \Biggl\{\sqrt\eta+M e^{-c(b-\ell)/\beta}+e^{CV}\max_{\bm P}\left(\int_0^1\norm{[\bm P,\bm X]}_{\bm\sigma,v}^2\,dv\right)^{1/2}\Biggr\}.
\end{aligned}
\end{equation}
Every Pauli string is supported on at most $V$ sites. Applying \eqref{app:ordered-cutoff-bound} therefore gives Lemma~\ref{lem:meta-continuity}.

\end{proof}

\section{Preservation of Coercivity under Short-Time Evolution}\label{app:short-time-coercivity}
We prove the finite-time bounds used in Lemma~\ref{lem:oracle}. 
\paragraph{Local generator bounds.}
For the sum of the generators associated with $\bm X_i,\bm Y_i,\bm Z_i$, \citep[Lemma 2.18 and Lemma A.4]{Bergamaschi2026Stationarity} give, for $\beta\ge1$,
\begin{equation}\label{app:site-generator-bounds}
 \|\cL_{i,\bm K}\|_{1\to1}\le C\beta,
 \qquad
 \|\cL_{i,\bm K}-\cL_{i,\bm G}\|_{1\to1}
 \le C\beta\sum_\ell|k_\ell-g_\ell|e^{-c\,\operatorname{dist}(i,\supp\bm P_\ell)/\beta}.
\end{equation}
Summing over the $O(r^D)$ sites in the patch and dropping the exponential weights gives
\begin{equation}\label{app:patch-generator-bounds}
\begin{aligned}
 \|\sum_{i\in B_j(r/2)}\cL_{i,\bm K_{B_j(r)}}\|_{1\to1}&\le C\beta r^D,
 \\
 \|\sum_{i\in B_j(r/2)}\cL_{i,\bm K_{B_j(r)}}-\sum_{i\in B_j(r/2)}\cL_{i,\bm G_{B_j(r)}}\|_{1\to1}
 &\le C\beta r^D\sum_{\ell:\supp(\bm P_\ell)\subseteq B_j(r)}|k_\ell-g_\ell|.
\end{aligned}
\end{equation}

\paragraph{Finite-time correction.}
For any two Lindblad generators $\mathcal L,\mathcal M$, with $\id$ denoting the identity superoperator, channel contractivity and Duhamel's formula imply
\begin{equation}\label{app:semigroup-difference}
 e^{s\mathcal L}-e^{s\mathcal M}
 =\int_0^s e^{(s-u)\mathcal L}(\mathcal L-\mathcal M)e^{u\mathcal M}\,du,
 \qquad
 \|e^{s\mathcal L}-e^{s\mathcal M}\|_{1\to1}
 \le s\|\mathcal L-\mathcal M\|_{1\to1}.
\end{equation}
Also $(e^{t\mathcal L}-\id)/t-\mathcal L=t^{-1}\int_0^t(e^{s\mathcal L}-\id)\mathcal L\,ds$. Subtracting the corresponding expression for $\mathcal M$ and using $\|e^{s\mathcal L}-\id\|_{1\to1}\le s\|\mathcal L\|_{1\to1}$ yields
\begin{equation}\label{app:semigroup-remainder}
\begin{aligned}
 \left\|\left(\frac{e^{t\mathcal L}-\id}{t}-\mathcal L\right)
       -\left(\frac{e^{t\mathcal M}-\id}{t}-\mathcal M\right)\right\|_{1\to1}&\le\frac1t\int_0^t s\bigl(\|\mathcal L\|_{1\to1}+\|\mathcal M\|_{1\to1}\bigr)
       \|\mathcal L-\mathcal M\|_{1\to1}\,ds\\
 &=\frac t2\bigl(\|\mathcal L\|_{1\to1}+\|\mathcal M\|_{1\to1}\bigr)
       \|\mathcal L-\mathcal M\|_{1\to1}.
\end{aligned}
\end{equation}
Apply these inequalities to the two patch generators associated with coefficient vectors $\mathbf x,\mathbf y$. Using $\|\bm P_j\|_\infty=\|\bm\tau\|_1=1$, we obtain
\begin{equation}\label{app:coordinate-finite-time-bounds}
\begin{aligned}
 |F_{\nu,j}(\mathbf x)-F_{\nu,j}(\mathbf y)|
 &\le C\beta r^D\sum_{\ell:\supp(\bm P_\ell)\subseteq B_j(r)}|x_\ell-y_\ell|,\\
 |(F_{\nu,j}-R^{(r)}_{\bm\tau,j})(\mathbf x)
   -(F_{\nu,j}-R^{(r)}_{\bm\tau,j})(\mathbf y)|
 &\le Ct_{\rm ev}\beta^2r^{2D}\sum_{\ell:\supp(\bm P_\ell)\subseteq B_j(r)}|x_\ell-y_\ell|.
\end{aligned}
\end{equation}
Each patch retains at most $Cr^D$ coefficients, and each coefficient is retained in at most $Cr^D$ patches. Cauchy--Schwarz therefore gives
\begin{equation}\label{app:patch-cauchy-schwarz}
 \sum_j\left(\sum_{\ell:\supp(\bm P_\ell)\subseteq B_j(r)}|x_\ell-y_\ell|\right)^2
 \le Cr^D\sum_j\sum_{\ell:\supp(\bm P_\ell)\subseteq B_j(r)}|x_\ell-y_\ell|^2
 \le C^2r^{2D}\|\mathbf x-\mathbf y\|_2^2.
\end{equation}
Combining \eqref{app:coordinate-finite-time-bounds} and \eqref{app:patch-cauchy-schwarz} proves \eqref{cutoff:polynomial-oracle} and \eqref{oracle:finite-time-correction}.

\paragraph{Radial coercivity and the baseline.}
Writing $\mathbf e=\mathbf k-\mathbf h$, the assumed coercivity of the full residual, the spatial error \eqref{cutoff:bias}, and \eqref{oracle:finite-time-correction} imply
\begin{equation}\label{app:centered-finite-time-coercivity}
\begin{aligned}
 \langle\mathbf e,\mathbf F_\nu(\mathbf k)-\mathbf F_\nu(\mathbf h)\rangle
 &\ge (2\mu-Ct_{\rm ev}\beta^2r^{3D})\|\mathbf e\|_2^2-\tfrac\nu2\|\mathbf e\|_1\\
 &\ge\mu\|\mathbf e\|_2^2-\tfrac\nu2\|\mathbf e\|_1.
\end{aligned}
\end{equation}
At $\mathbf k=\mathbf h$, the identity $e^{t\mathcal L}-\id=\int_0^t e^{s\mathcal L}\mathcal L\,ds$, trace-norm contractivity, and \eqref{app:patch-stationarity-error} give
\begin{equation}\label{app:finite-time-baseline}
\begin{aligned}
 |F_{\nu,j}(\mathbf h)|
 &\le\frac1{t_{\rm ev}}\int_0^{t_{\rm ev}}
 \|e^{s\sum_{i\in B_j(r/2)}\cL_{i,\bm H_{B_j(r)}}}\sum_{i\in B_j(r/2)}\cL_{i,\bm H_{B_j(r)}}[\bm\tau]\|_1\,ds\\
 &\le\|\sum_{i\in B_j(r/2)}\cL_{i,\bm H_{B_j(r)}}[\bm\tau]\|_1
 \le\sum_{i\in B_j(r/2)}\|\cL_{i,\bm H}[\bm\tau]\|_1+\nu/4
 \le Cr^D\zeta+\nu/4.
\end{aligned}
\end{equation}
For Gibbs inputs, take $\zeta=0$. Combining \eqref{app:centered-finite-time-coercivity} and \eqref{app:finite-time-baseline} gives \eqref{oracle:finite-time-coercivity}, completing the proof.

\bibliographystyle{alphaurl}
\bibliography{bibilography}
\end{document}